\documentclass[aps,prd,superscriptaddress,reprint,nofootinbib]{revtex4-1}
\usepackage{graphicx}
\usepackage{amsmath}
\usepackage{amssymb}
\usepackage{siunitx}
\usepackage[acronym]{glossaries}
\usepackage{hyperref}
\usepackage{cleveref}
\usepackage{ulem}
\usepackage{xcolor}
\usepackage{float}

\usepackage{todonotes}
\DeclareMathOperator{\evsym}{E}
\newcommand\ev[1]{\evsym\left\langle#1\right\rangle}

\usepackage{hyperref}
\hypersetup{
    colorlinks=true,
    linkcolor=blue,
    filecolor=magenta,      
    urlcolor=cyan,
    pdftitle={Overleaf Example},
    pdfpagemode=FullScreen,
    }

\begin{document}

\title{Impact of the noise knowledge uncertainty for the science exploitation of cosmological and astrophysical stochastic gravitational wave background with LISA }

\author{Martina~Muratore}\email{contact: martina.muratore@aei.mpg.de}
\affiliation{\addressii}
\author{Jonathan Gair}
\affiliation{\addressii}
\author{Lorenzo~Speri}
\affiliation{\addressii}

\def\addressii{Max Planck Institute for Gravitational Physics (Albert Einstein Institute), D-14476 Potsdam, Germany}
\date{\today}

% \hfill \vspace{.35cm} \hrule \vspace{1cm}

\begin{abstract}
This paper investigates the impact of a lack of knowledge of the instrumental noise on the characterisation of stochastic gravitational wave backgrounds with the Laser Interferometer Space Antenna (LISA). We focus on constraints on modelled backgrounds that represent the possible backgrounds from the mergers of binary black holes of stellar origin, from  primordial black hole generation, from non-standard inflation, and from sound wave production during cosmic fluid phase transitions. We use splines to model generic, slowly varying, uncertainties in the auto and cross-spectral densities of the LISA time delay interferometry channels. We find that allowing for noise knowledge uncertainty in this way leads to one to two orders of magnitude degradation in our ability to constrain stochastic backgrounds, and a corresponding increase in the background energy density required for a confident detection. We also find that to avoid this degradation, the LISA noise would have to be known at the sub-percent level, which is unlikely to be achievable in practice. %The paper also discusses the practical challenges associated with setting reasonable noise requirements for LISA. 
\end{abstract}
 
\maketitle

\section{Introduction}
The Laser Interferometer Space Antenna (LISA) is part of the European Space Agency Cosmic Vision program and is due to be launched in the mid-2030s. LISA will be the first observatory in space to study gravitational waves (GWs) at mHz frequencies. It will consist of a constellation of three satellites forming a quasi equilateral triangle and continuously exchanging laser beams \cite{LISA:2017pwj}. LISA is expected to observe a large variety of sources, such as galactic binaries (GBs), massive black hole binaries (MBHBs) \citep{Klein:2015hvg}, stellar-origin black hole binaries (SOBHB) \citep{PhysRevLett.116.231102,Gerosa:2019dbe,Moore:2019pke}, extreme-mass-ratio inspirals (EMRIs) \citep{Babak:2017tow} and possibly stochastic backgrounds arising from astrophysical and cosmological processes \cite{Caprini_2019}.\\

When considering the science that can be done with LISA, it is typical to assume a known model for the instrumental noise in the detector data channels. However, these noise levels will not be known in practice. This is also true for ground-based gravitational wave detectors, but in that context spectral density estimation is easier because signals are rare and short-lived, allowing the spectral density to be estimated from data in the vicinity of observed events. LISA signals, by contrast, are typically long lived, which means that noise and signal properties must be simultaneously estimated by fitting a suitable model. While such methods and models are still under development, it is expected that the characterisation of deterministic signals will not be significantly affected by lack of instrumental noise knowledge (see Appendix~\ref{sec:detsrc_FM}). The case of stochastic GW backgrounds (SGWB) is different, however, as these are intrinsically of the same character as the stochastic instrumental noise. Searches for stochastic signals in ground-based interferometers rely on the cross-correlation of data from independent detectors~\cite{1999PhRvD..59j2001A}. This would only be possible if there is another space-based interferometer in operation concurrently, such as Taiji~\cite{2018arXiv180709495R}, but this is not certain at the moment. Here we explore the challenge of distinguishing between the stochastic instrumental noise and a stochastic GW signal.\\ 

% It is expected to be a signal dominated detector where massive black hole binaries (MBHB) and galactic binaries (GB) are going to be present in the LISA band for the entire observation time with very high signal to noise ratio (SNR) \cite{}. Moreover, other gravitational wave (GW) sources such as extreme mass ratio inspirals (EMRIs) or LIGO type binary black holes  are expected to be present and detection techniques as well as parameter estimation algorithm are under development. All these GW sources will have smaller or bigger SNR relative to each other, but they will be all above the LISA noise.\\
% Both cosmological and astrophysical backgrounds are predicted in the LISA band. 

One approach is to use a model for the instrumental noise. It is possible to derive analytical models that describe how different known noise sources propagate into the LISA data stream. However, not all noise sources will be known in advance, so we will not be able to strictly rely on the models as we will not be able to perform full tests and directly measure the noise. In the LISAPathfinder mission \cite{PhysRevLett.116.231101} it was seen that at low frequency the analytical models couldn't fully explain the measured noise. Therefore, when we plan for LISA data analysis, we must be prepared for uncertainty in the noise models.

The goal of this paper is to assess the impact of lacking a noise model for LISA in parameter estimation of SGWBs. We consider four different models of cosmological and astrophysical SGWBs: a power law to model signals from stellar origin binary black hole inspirals, a Gaussian bump to model a background from primordial black hole generation, a power law with running to model a background from non-standard inflation and finally a first order phase transition model, representing GW production from sound waves in the cosmic fluid generated by colliding phase transition bubbles \cite{Caprini_2019}. For each model, we will take a reference amplitude that corresponds to a relatively low signal to noise ratio (SNR) that is close to the boundary for detection. These are the backgrounds that will be most difficult to distinguish from instrumental noise. We will also explore what happens as the background energy density is varied in each model.

We represent our lack of knowledge of the LISA instrumental noise by multiplying a set of reference auto- and cross-spectral densities with cubic splines. For the reference spectral densities we use the noise model from \cite{LISA:2017pwj}, which includes only the so-called secondary noises~\cite{Muratore:2022nbh}, the test mass (TM) acceleration and optical metrology noise (OMS). This noise model assumes that the laser noise \cite{Armstrong_1999}, clock noise \cite{Tinto:2018kij}, tilt to length coupling \cite{Hartig_2022,PhysRevD.106.042005} have been suppressed by the initial noise reduction pipeline \cite{Hartwig:2021dlc,Muratore:2021rwq}. To represent the fact that we will have some amount of information from noise modelling before launch, we place a Gaussian prior on the weights of the cubic spline. By varying the Gaussian variance we explore the effect of having more or less knowledge of the noise. \\

Several previous studies have tackled the problem of detecting a SGWB with LISA and distinguishing it from the noise, but these have used different methods than the one we employ in this paper. In \cite{Flauger:2020qyi} it was shown that SGWB reconstruction was possible for generic SGWB models, if the LISA instrumental noise can be represented by just two parameters, representing the level of TM and OMS noise, assumed equal for all arms of the interferometer. The authors of \cite{Adams:2010vc,Hartwig:2023pft} allowed the TM and OMS noises to differ from arm to arm, but still assumed that these noises had a known spectral shape as a function of frequency. In \cite{Baghi:2023qnq} an arbitrary noise shape was allowed, described by a spline, but using a simplified noise model for the single link. Finally, \cite{Muratore:2022nbh} derived an upper bound on the detectable SGWB amplitude when being agnostic on both the signal and noise shape and discussed limitations of the utility of the null channel for distinguishing between instrumental noise and a stochastic GW background.\\

The paper is organised as follows: in Section~\ref{sec:first} we introduce the general data model that we use in the analysis and we describe the Fisher matrix formalism that will be used for this work. In Section~\ref{sec:splines} we describe the spline model that we use to represent the uncertainties in the power spectral density (PSD) and cross-spectral density (CSD) of the instrumental noise. In Section~\ref{sec:noise_and_tdi} we give the analytical noise model for a single LISA link that is used as the reference model, and the corresponding PSDs and CSDs for the time delay interferometry (TDI) channels, $A$, $E$ and $\zeta$. In Section~\ref{sec:signal_response} we describe how a stochastic signal appears in the three TDI channels and their cross-correlations, while in Section~\ref{sec:sgwbmods} we describe the models for the cosmological and astrophysical SGWBs that we use in this paper. %We compare the SNR for all the models versus their value of the energy density $\Omega h^2$ evaluated at 1mHz. 
In Section~\ref{sec:ampdep} we show how well we can estimate the parameters of the different SGWB models when we allow for uncertainty in our knowledge of the instrumental noise. For each model, we compare the precision of parameter estimation to that when noise knowledge is perfect and show how the parameter precisions vary as a function of the background energy density, $\Omega$, evaluated at 1mHz. In Section~\ref{sec:priordep} we show how the results change as we vary our priors uncertainty on the instrumental noise. We conclude our results in Section~\ref{sec:sigrecon} by showing how well the signal, noise and galactic foreground can be reconstructed for a power law SGWB background. Section~\ref{sec:nine} summarises our conclusions and future perspectives.

\section{Methods}\label{sec:first}
\subsection{Likelihood}

We assume that the output of a gravitational wave detector, $s(t)$, is expressed as a linear combination of a signal, $h(t| \vec{\mu})$, determined by a finite set of (unknown) parameters, $\vec{\mu}$, and instrumental noise, $n(t)$. If we ignore the presence of calibration errors \cite{Savalle_2022}, the content of a single data stream, i.e., one output channel from one detector, can be written in the frequency domain as:
\begin{equation}
\tilde{s}(f) = \tilde{h}(f|\vec{\mu})+ \tilde{n}(f),
\end{equation}
where the tilde indicates the Fourier transform.
The likelihood for the observed data can be written as $p(\tilde{s}(f) | \vec\mu) = p(\tilde{n}(f) = \tilde{s}(f) - \tilde{h}(f|\vec{\mu}))$. In a gravitational wave context it is usual to further assume that the instrumental noise follows a Gaussian distribution characterized by a one-sided PSD, $S_n(f)$, defined such that
\begin{equation}\label{eq:psd_definition}
 \mathbb{E} [\tilde{n}^*(f) \tilde{n}(f')] = \frac{1}{2} S_n(f) \delta(f-f'),
\end{equation}
for $f$, $f'>0$, where the expectation value $\mathbb{E}$ is taken over the data generating process. The delta function in the previous equation implies that different frequencies are not correlated.

In reality the noise model is not known perfectly and could vary from the assumption above in a number of ways. For example, the PSD might have a different shape from the reference one \cite{babak2021lisa}, the probability distribution of the noise might not be Gaussian, or the noise might not be stationary, leading to correlations between frequencies. 

In this work we will continue to assume that the noise is Gaussian and stationary, but we will allow the power spectral density to vary using a parametrized spectral density, $S_n(f) \rightarrow S_n(f|\vec\lambda)$, described by parameters $\vec\lambda$. Then the log-likelihood depends on both sets of parameters, $\vec\mu$ and $\vec\lambda$, and can be written as:
\begin{align}
   l := \ln p(\tilde{s}|\vec\mu,\vec\lambda) = - \sum_{k = 1}^{n} \ln\left[2 \pi \frac{S_n(f_k|\vec\lambda)}{4 \Delta f} \right] -\\ \frac{1}{2} \sum_{k = 1}^{n} \frac{|\tilde{s}(f_k) -  \tilde{h}(f_k|\vec\mu) |^2}{\frac{1}{4\Delta f} S_n(f_k|\vec\lambda)}
\end{align}
where the sum is performed over $n$ frequencies and $\tilde{n}(f_k) = \tilde{s}(f_k) -  \tilde{h}(f_k|\vec\mu)$ are the discrete Fourier component at frequency $f_k = k \Delta f$, of the data minus signal model. The frequency bin width, $\Delta f$, is related to the total observation time as $T =1 /\Delta f$. The first term does not include the $1/2$ factor because the real and imaginary parts of $\tilde{n}(f_k)$ are independent random variables. This follows from the fact that, for a real time series, $\tilde{n}^*(f) = \tilde{n}(-f)$, which combined with Eq.~(\ref{eq:psd_definition}) means that $\langle \tilde{n}(f) \tilde{n}(f') \rangle = 0$ for $f$, $f'>0$. This allows equation~\ref{eq:psd_definition} to be rewritten as
\begin{equation}
\langle \Re[\tilde{n}(f_k)]^2\rangle = \langle \Im[\tilde{n}(f_k)]^2\rangle = \frac{\langle | \tilde{n}(f_k)|^2\rangle}{2}= \frac{S_n(f_k| \vec\lambda)}{4 \Delta f} 
\end{equation}
for a discrete set of frequencies.

Stochastic gravitational wave backgrounds are not deterministic signals and can be treated on the same footing as the instrumental noise by defining the total variance at frequency $f_k$ as
\begin{equation}\label{eq:st_sn_sgw}
S_{ t}(f_k|\vec\theta,\vec\lambda) = S_{\rm GW}(f_k|\vec\theta) + S_{\rm n}(f_k|\vec\lambda) \, .
\end{equation}
If we assume that all the deterministic sources have been correctly subtracted from the datastream $s$ the log-likelihood becomes:
\begin{align}
l(\vec\theta,\vec\lambda) &=  - \sum_{k = 1}^{n} \ln\left[T \pi \frac{S_t(f_k|\vec\theta,\vec\lambda)}{2 } \right] \nonumber \\ & -\frac{1}{2} \sum_{k = 1}^{n} \frac{|\tilde{s}(f_k)|^2}{\frac{T}{4} S_t(f_k|\vec\theta,\vec\lambda)} \label{eq:loglik}
 \end{align}
The derivation of this likelihood can be found in Appendix~\ref{sec:appendix}.

\subsection{Fisher matrix}\label{sec: fisher}

We are interested in understanding the impact of noise knowledge uncertainties on the parameter measurement precision of stochastic gravitational wave backgrounds. The Fisher information matrix provides a lower bound on the covariance of an unbiased estimator of the model parameters and provides a good approximation to the precision of parameter estimation in the high signal-to-noise ratio limit. We will therefore use it to quantify our ability to measure both the noise parameters, $\vec{\lambda}$, and the background parameters, $\vec \theta$.

In a general context the Fisher matrix is defined by
\begin{equation}\label{eq:fisher}
\Gamma_{ij} = \mathbb{E} \left[ \frac{\partial l}{\partial \upsilon^i} \frac{\partial l}{\partial \upsilon^j} \right] = - \mathbb{E} \left[ \frac{\partial^2 l}{\partial \upsilon^i \partial \upsilon^j} \right] 
\end{equation}
where the expectation value $\mathbb{E}$ is taken over the data generating process, and the partial derivatives are taken with respect to the parameters, $\vec\upsilon$, on which the likelihood depends. We want to compute the Fisher matrix on the extended parameter space $\vec\upsilon = \{\vec\theta,\vec\lambda\}$.

It can be shown that the expectation value of the product between the derivative of the log-likelihood with respect to deterministic and stochastic parameters is zero. Therefore, at the level of the Fisher matrix approximation it can be shown that the estimation of the noise and deterministic signal parameters is independent (see Appendix~\ref{sec:detsrc_FM}).

For SGWBs, we can compute the Fisher matrix in continuous domain as:
\begin{equation}\label{log_likelihood}
\Gamma_{ij}  =T \int_{0} ^\infty  (\Sigma^{-1})_{lr}  \frac{\partial \Sigma ^{rp}   }{\partial \upsilon^i}( \Sigma ^{-1}  )_{pm}\frac{\partial \Sigma^{ml} }{\partial \upsilon^j} \,\rm{d}f\, .
\end{equation}
with 
\begin{align}
\Sigma(f|\vec\upsilon = \{\vec\theta,\vec\lambda\})= \frac{1}{2}\begin{pmatrix}
 S^{AA} _t  &  S^{AE} _t  &S^{A\zeta} _t  \\
 S^{AE*} _t  &  S^{EE} _t  & S^{E\zeta} _t  \\
 S^{A\zeta *} _t & S^{E\zeta *} _t & S^{\zeta \zeta} _t  
\end{pmatrix}
\, .
\end{align}
where each element of the matrix can be written as a sum of an instrumental noise component and a stochastic gravitational wave component as indicated in Eq.~\ref{eq:st_sn_sgw}. A complete derivation of this formula can be found in Appendix~\ref{app:fisher_derivation}.

Prior knowledge on the noise can be incorporated by imposing a prior on the instrumental parameters, $\vec{\lambda}$. When doing numerical marginalisation any prior can be imposed, but in the Fisher matrix formalism it is easiest to work with a Gaussian prior \cite{Savalle_2022}. The posterior covariance is then given by the inverse of the modified Fisher matrix:
\begin{equation}
\Gamma = \begin{pmatrix}
 \Gamma^{\theta \theta} &  \Gamma^{\theta \lambda}  \\
( \Gamma^{\theta \lambda})^T &  \Gamma^{\lambda \lambda}+\Theta ^{\lambda \lambda} \label{eq:fisher_prior}
\end{pmatrix}
\end{equation}
with normal prior on the instrumental noise parameters with zero mean and covariance given by $(\Theta^{\lambda \lambda})^{-1}$. The diagonal elements of the inverse of this matrix provide estimates for the precision with which the corresponding parameters can be measured. The estimated precision of measurement of the SGWB parameters accounting for noise model uncertainty is thus given by the diagonal elements of the matrix: 
\begin{equation}
\sigma_{\theta } = \sqrt{{\rm diag}[(\Gamma^{\theta \theta} - \Gamma^{\theta \lambda} (\Gamma^{\lambda \lambda}+ \Theta^{\lambda \lambda})^{-1}(\Gamma^{\theta \lambda})^T)^{-1}]}
\end{equation}
Note that in the limit in which the instrumental noise parameters are perfectly known $\Theta \rightarrow \infty$ and the measurement precision of the  SGWB parameters is given by $\sigma_{\theta } = \sqrt{{\rm diag}[(\Gamma^{\theta \theta})^{-1}]}$.

\subsection{Modeling noise knowledge uncertainties}\label{sec:splines}
\begin{figure}
\includegraphics[width=0.45\textwidth]{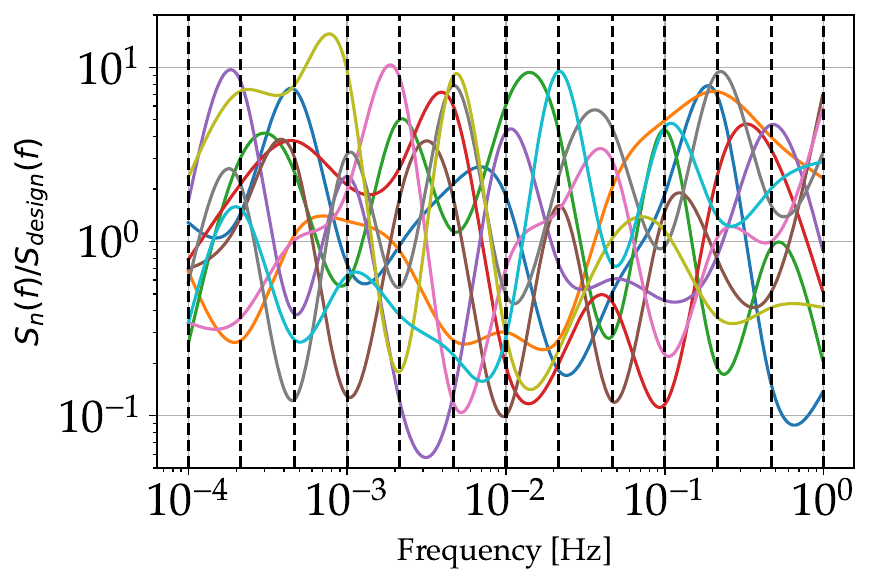}
\caption{
Deviations from the design power spectral density obtained using the cubic spline model, with $\lambda_i \sim U[-1,1]$, and with knots equally spaced between $\log _{10}(f)=-4$ and $\log_{10}(f)=0$. The plot shows the ratio of the total PSD and the design one for different parameter realizations $\lambda_i$.}
\label{fig:splinemod_unbound}
\end{figure}

To model noise uncertainties, we allow the PSD and CSD of the instrumental channels to deviate from the design specification. However, we assume that such deviations vary smoothly over a relatively wide range of frequency and model the noise uncertainties as fractional deviations from the design PSD/CSD that are described by natural cubic splines. We write the PSD of the instrumental noise in each channel as:
\begin{equation}
    S_n(f|\lambda) = S_{\rm des}(f) \, 10^{C(f|\vec\lambda) } \, ,
\end{equation}
where $C(f|\vec\lambda)$ is a natural cubic spline. The parameters $\vec\lambda$ specify the values of the spline at the knots, labelled by $i$. In this study we use knots evenly spaced in $\log_{10}(f)$ between $\log_{10}(f) = -4$ and $\log_{10}(f) = 0$ and we fix the number of knots to 13. %and evenly log-spaced in frequency . The weights have a range of $\lambda_i \in [-\infty, \infty]$ with $i=1,...,13$.
%\LS{The following is the old text}
% To construct a cubic spline we need to specify the spline knots $\{\log_{10}(f_i)\}$ and the value of the weights at those points $\{w_i\}$
%\begin{itemize}
%\item Natural cubic spline model evaluated at $\log_{10}(f)$: $C(f | \{\log_{10}(f_i)\}, \{w_i\})$;  C represents a cubic spline function used to model a cross spectral density (CSD) or power spectral density function. The function takes as input the logarithm (base 10) of a frequency value f, and computes the CSD or PSD of f given a set of input variables $ \{\log_{10}(f_i)\}$ and associated weights $\{ w_i\}$ which are used to define the cubic spline 
%\item Set of spline knots: $\{ \log_{10}f_i: i= 1, \ldots, N_{\rm knots} \}$. We use 13 knots per spline equally spaced between $10^{-4}$ Hz and 1 Hz. 
%\item{Weights at the spline points: $w_i \in [-\infty, \infty]$. These could be related to a bounded parameter via a logit transform, $w_i = \log(p_i/(1-p_i)$, where $p_i$ are parameters defined to lie in the range $[0,1]$ or allowed to freely vary. \LS{are we actually using this?} We characterise our spline directly via the $w_i$'s. }
%\end{itemize}
Noise curves corresponding to this model, with the weights at each knot drawn randomly from a $\log_{10}(f) \sim U[-1,1]$ distribution, are shown in \cref{fig:splinemod_unbound}. We note that this choice of prior means we are allowing approximately one order of magnitude variation in the PSD. 
%Therefore, our model for the PSD is given by
%\begin{equation}
%S_n(f|\{w_i\}) = S(f)_{\rm des.}  10^{C(f | \{\log_{10}(f_i)\}, \{w_i\})};
%\end{equation}
When we evaluate the Fisher matrix we will always do so at the reference point where the weights of the spline are zero, i.e., where the PSD is equal to the reference value shown in Fig. \ref{fig:noises}.\\

\begin{figure}
\centering
\includegraphics[width=0.5\textwidth]{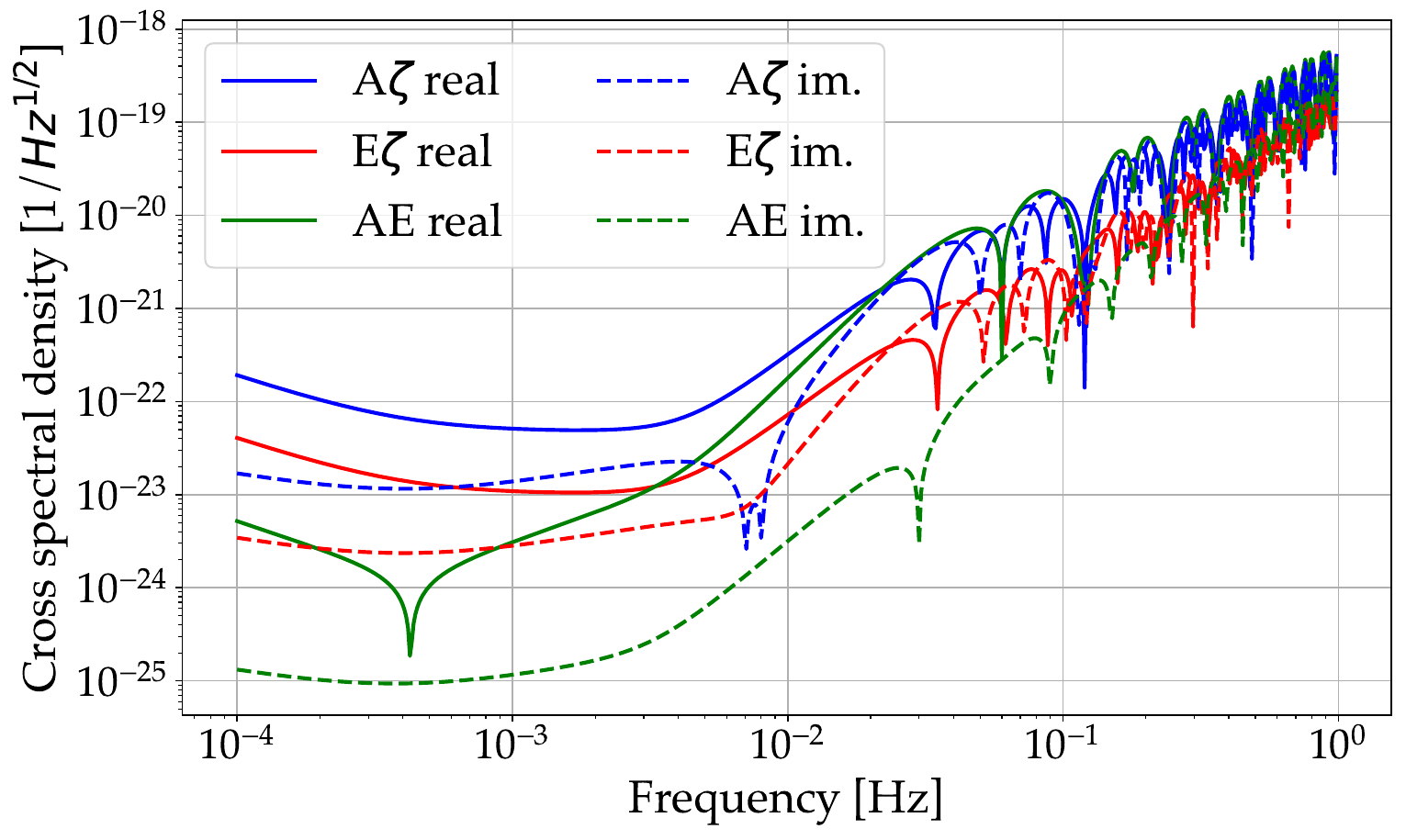}
\caption{\label{fig:noises_csd} Real and imaginary part of the reference square root of the cross spectral density for the time delay interferometry channels AE, E$\zeta$ and A$\zeta$ considering only TM acceleration and OMS noise and assuming a constellation of three fixed unequal arm-lengths.
}
\end{figure}

We cannot follow the same procedure for specifying the CSD, because the reference model is smaller by 1 or even 2-3 order of magnitude a low frequency with respect to the PSD (see Fig.~\ref{fig:noises_csd} and Fig. \ref{fig:noises}).
%~\ref{fig:GWcsd})
It was shown in~\cite{Hartwig:2023pft} that when the LISA response is constructed allowing for unequal noises in the different laser links, the CSD can be much larger and become comparable to the PSD.
%The choice of our model for the CSD has been driven by two factors: the first factor is that in case of unequal arm-length the cross-correlation terms between the TDI channels AE, A$\zeta$ and E$\zeta$ are smaller than the PSD but non-negligible, as can be seen in the lower panel of Fig. \ref{fig:noises}; the second factor is that in case we consider a more inclusive noise model where the six OMS and TMs can have different PSDs  among each other then the CSD noise contribution starts to become relevant as shown in  where the CSD can be of comparable size to the PSD. 
Since our goal is to allow the splines to vary in such a way to mimic un-expected and un-modelled noise components with respect to the simplified scenario (three unequal but fixed-length arms) we model the CSD as\footnote{ In principle our model does not force the matrix to be positive definite. We are forcing the reference spectral density matrix to be positive definite, but in principle we could have a factor of 10 variation in the CSD while the PSD is unchanged. It doesn't matter for the Fisher matrix because this is a local approximation and we are evaluating it at a point where the matrix is positive definite. The CSD at the central point is $0.1$ of its maximum value, so in an open set around that point it will be positive definite and thus all derivatives are well defined. The conclusion is that the model used here is fair for what we want to demonstrate but would not be a suitable model to use when analysing the data. %because it won't be imposing positive definiteness
}:
\begin{align}
S_n(f|\{\lambda_i\}) = \nonumber \\  & \sqrt{S_{\rm des,i}(f) S_{\rm des,i}(f)} \, \sigma_R \, 10^{C(f| \{\log_{10}(f_i)\}, \{\lambda_i\})}  \nonumber \\  + & \sqrt{S_{\rm des,i}(f) S_{\rm des,i}(f)} \,i  \sigma_I \, 10^{C(f | \{\log_{10}(f_i)\}, \{\lambda_i\}) } \,  \label{eq:csd_splines}
\end{align}
where we fix $\sigma_R = 0.1, \sigma_I =0.8 \sigma_R$. We do not expect that varying the relationship between $\sigma_I$ and $ \sigma_R$ to significantly change the conclusions; although we fix $\sigma_I$ slightly smaller than $\sigma_R$ accordingly to Fig. \ref{fig:noises_csd}. There at lower frequencies the imaginary components are about 1 order of magnitude smaller than the real components. The indexes $i$ and $j$ run over the number of detectors or channels with $i \neq j$. The additional factors $\sigma_I$ and  $\sigma_R$ are used to limit the amplitude, and allow us to model the CSD as a sum of splines times the geometric mean of the square root-PSDs. Using the (scaled-)geometric mean of the PSDs as a reference for the CSD rather than the CSD of the reference equal-noise configuration, allows for much larger CSD variations. This is consistent with the results presented in~\cite{Hartwig:2023pft}.

It is important to state that our model is not completely general since we are imposing a certain amount of smoothness in the PSD variation, and consequently in the CSD, when we specify the number and spacing of the knots. Thus we are not able to fit for all possible noise scenarios.
In particular, this model does not attempt to reproduce the zeros of the TDI transfer functions faithfully. This will become important above $f \sim 0.05$Hz, 
%is not informative. Note that even if we are not properly capturing the zeros of the CSD at high frequency, 
but this should not affect our results as the SGWBs we consider do not have much power at those frequencies, as can be seen from Fig. \ref{fig:sgwbnoisepw}. Other models could be considered, for example by imposing the spline variations at the level of the noise in individual laser links (generalising the approach taken in~\cite{Baghi:2023qnq}), before applying the TDI transfer function. This should be explored in the future, but this would increase the number of parameters further so we might expect there to be additional degeneracies, which would lead to practical difficulties in fitting noise and signal simultaneously. However, for the purpose of the current study, the model we use is adequate to represent generic, slowly varying, fluctuations in the PSD and CSD.\\

%The equivalent expression when we have multiple detectors, represented by covariance matrices at each frequency $f$, which we denote by $C(f)$, is
%\begin{equation}
%\Gamma_{ij} = \frac{1}{2} \int \frac{\partial C_{ij} (f)}{\partial \pi^i}\frac{\partial C^{-1}_{ij} (f)}{\partial \pi^j} \,{\rm d}f
%\label{eq:SGWBfish}
%\end{equation}
%where the indices $i,j$ run over the number of detectors included in the measurement.

%Noting that
%\begin{align}
%\frac{\partial}{\partial \pi^i} \left( C_{jk} C^{-1}_{kl} \right) &= 0 \nonumber \\
%\Rightarrow \qquad \frac{\partial C^{-1}_{jk} }{\partial \pi^i} &= - C^{-1}_{jl} \frac{\partial C_{lm} }{\partial \pi^i}  C^{-1}_{mk},
%\end{align}
%we can rewrite Eq.~\eqref{eq:SGWBfish} as
%\begin{equation}
%\Gamma_{ij} = \frac{1}{2} \int C^{-1}_{kl} \frac{\partial C_{lm} (f)}{\partial \pi^i} C^{-1}_{mn} \frac{\partial C_{nk} (f)}{\partial \pi^j} \,{\rm d}f.
%\label{eq:SGWBfish}
%\end{equation}

\subsection{Noise at the TDI input and outputs }\label{sec:noise_and_tdi}
Here, we present the instrumental noise model used to define the reference PSD in this work.
Among the different noise sources for LISA, the laser noise is the main source of noise, which must be reduced by eight orders of magnitude by applying a post-processing technique called time delay interferometry (TDI) \cite{tintoD}. TDI synthesises an equal arm-length interferometer by appropriately delaying and combining the interferometric measurements in many different ways to form TDI channels free from laser noise. The standard second generation TDI channels (unequal and time varying arm-length) are the Michelson interferometer channels, $X$, $Y$ and $Z$,  from which we form the more GW sensitive channels, $A$ and $E$~\cite{PhysRevD.66.122002}. Together with the GW sensitive channels we consider a null channel, the $\zeta$ channel~\cite{Muratore:2021uqj} that is less sensitive to GWs and can in principle be used as a noise monitor. \\

In the current work we will assume that laser noise has already been reduced thus we can work directly with the first generation TDI \cite{PhysRevD.105.062006}, allowing us to  consider three unequal but fixed arm-lengths. This choice should not significantly affect the conclusions of the analysis but simplifies the transfer function derivations. \\

%In this section, we describe the main limiting noise sources after the post-processing TDI technique has been applied. 
We also assume that all known calibrated and measured instrumental noise sources have been subtracted, such as the optical tilt to length cross-coupling to spacecraft motion and clock noise (\cite{Hartig_2022}, \cite{PhysRevApplied.14.014030} and \cite{PhysRevD.103.123027}). The remaining noises, for which we have neither a measurement for coherent subtraction nor a high precision a priori model \cite{Muratore:2022nbh}, fall into two broad categories, the acceleration noise of each individual test-mass (TM) and an overall optical metrology system (OMS) noise term for each single link measurement (see \cite{nam2022tdi} for the case of multiple OMS noise terms).  \\ 

We represent the TM acceleration noise PSD of a single TM by $S_{g_{ij}}$. To directly compare the OMS and TM contributions we can directly convert the acceleration noise of a single TM to an equivalent displacement, whose PSD is given by
\begin{equation}
	S_{g_{ij}}^{\rm disp} =S_{g_{ij}} /(2 \pi f)^4 \,
\end{equation}
where $f$ is the Fourier frequency.
We denote the time series associated with this displacement as $x^g_{ij}(t)$. We also define the PSD of the OMS noise as $S_{{\rm oms}_{ij}}(f)$ and we denote the time series of the single OMS as $x^m_{ij}(t)$.  
All TDI combinations can be constructed from a combination of single link TM to TM measurements. Such measurements are represented by the intermediary variables~\cite{Hartwig:2021dlc}:
\begin{equation}
	\tilde{\eta}^N_{ij}(\omega) = \tilde{x}(\omega)^g_{ji}e^{-i \omega L_{ji}}+ \tilde{x}(\omega)^g_{ij} + \tilde{x}(\omega)^m_{ij},\label{eq:link}
\end{equation}
where $\tilde{\eta}^N_{ij}(\omega)$ is the noise in a single link measurement, the first index $i$ indicates the spacecraft where the measurement is performed at time $t$, and the second index $j$ indicates the distant spacecraft from which light was emitted at time $t - \tau$, and $ \omega = 2\pi f $. Equation \ref{eq:link} implies that each single link measurement contains TM noise terms from the distant and local spacecraft, such that the TM noise appearing in the measurements on the two ends of the same arm is correlated (between the two links): 
\begin{equation}
\langle\tilde \eta^N_{ij}(\omega) \tilde \eta^N_{ji}(\omega)\rangle \neq 0
\end{equation}
From these measurements it is possible to build any TDI channels \cite{Armstrong_1999,muratore2020revisitation} and therefore the corresponding first generation orthogonal channels $A_1$ and $E_1$ \cite{PhysRevD.105.062006} that will be used in this work:
\begin{equation}
{\rm A_1} = \frac{{\rm Z_1} - {\rm X_1}}{\sqrt{2}}\;, \qquad  {\rm E_1} = \frac{{\rm X_1} - 2 {\rm Y_1} + {\rm Z_1}}{\sqrt{6}}.
\end{equation}
The $X_1$ variable is defined as:%
\begin{align}
{\rm X_1}  = & (D_{13}D_{31}-1)(\eta_{12} +  D_{12} \eta_{21}) \nonumber
\\ & + (1-D_{12}D_{21} )(\eta_{13} + D_{13} \eta_{31}) \; ,
\end{align}\label{eq:tdi-definition}
where the delays $D_{ij}$ corresponds to a constant time shift and thus in frequency to $\mathcal{F}\{D_{ij}\} = e^{-i \omega L_{ij}}$. The $Y_1$ and $Z_1$ are given from $X_1$ by cyclic permutations of the three satellites. The fully symmetric channel, $\zeta_1$, is defined by:
\begin{equation}
\zeta_1 = D_{12}(\eta_{31} - \eta_{32}) + D_{23}(\eta_{12} - \eta_{13}) + D_{31}(\eta_{23} - \eta_{21}) \; .
\label{eq:zeta-1.5}
\end{equation}
\\
%\LS{ and directly restart here:}

The assumed model for the TM acceleration noise is
\begin{align}
&\ev{\tilde{x}^g_{ij}(f) \tilde{x}^{g*}_{lm}(f')} = \frac{1}{2} \delta_{il}\delta_{jm} \delta(f-f') S_{g_{ij}}(f)\nonumber \\
 &S_{g_{ij}}(f) = \left(3 \times 10^{-15}~\frac{\textrm{m}}{\textrm{s}^2~\sqrt{\textrm{Hz}}}\right)^2  \\ & \times \left(1 + \left(\frac{0.4 ~ \textrm{mHz}}{f}\right)^2\right)\left(1 + \left(\frac{f}{8 ~ \textrm{mHz}}\right)^4\right),\nonumber
\end{align}
and for the  OMS noise 
\begin{align}
&\ev{\tilde{x}^m_{ij}(f) \tilde{x}^{m*}_{lm}(f')} = \frac{1}{2} \delta_{il}\delta_{jm} \delta(f-f') S_{{\rm oms}_{ij}}(f)\nonumber \\
&	S_{{\rm oms}_{ij}}(f)  =\left(15 ~\textrm{pm}/\sqrt{\textrm{Hz}}\right)^2 \times \left(1 + \left(\frac{2~\textrm{mHz}}{f}\right)^4\right)\label{eq:readout},
\end{align}
This model assumes that individual noise components are uncorrelated. In reality the test masses in the same satellite will share environmental noise, such as temperature fluctuations, so this assumption might not hold. However, this model serves as a reference one, and any variation is captured by the flexible spline model previously presented.

\begin{figure}
\centering
\includegraphics[width=0.5\textwidth]{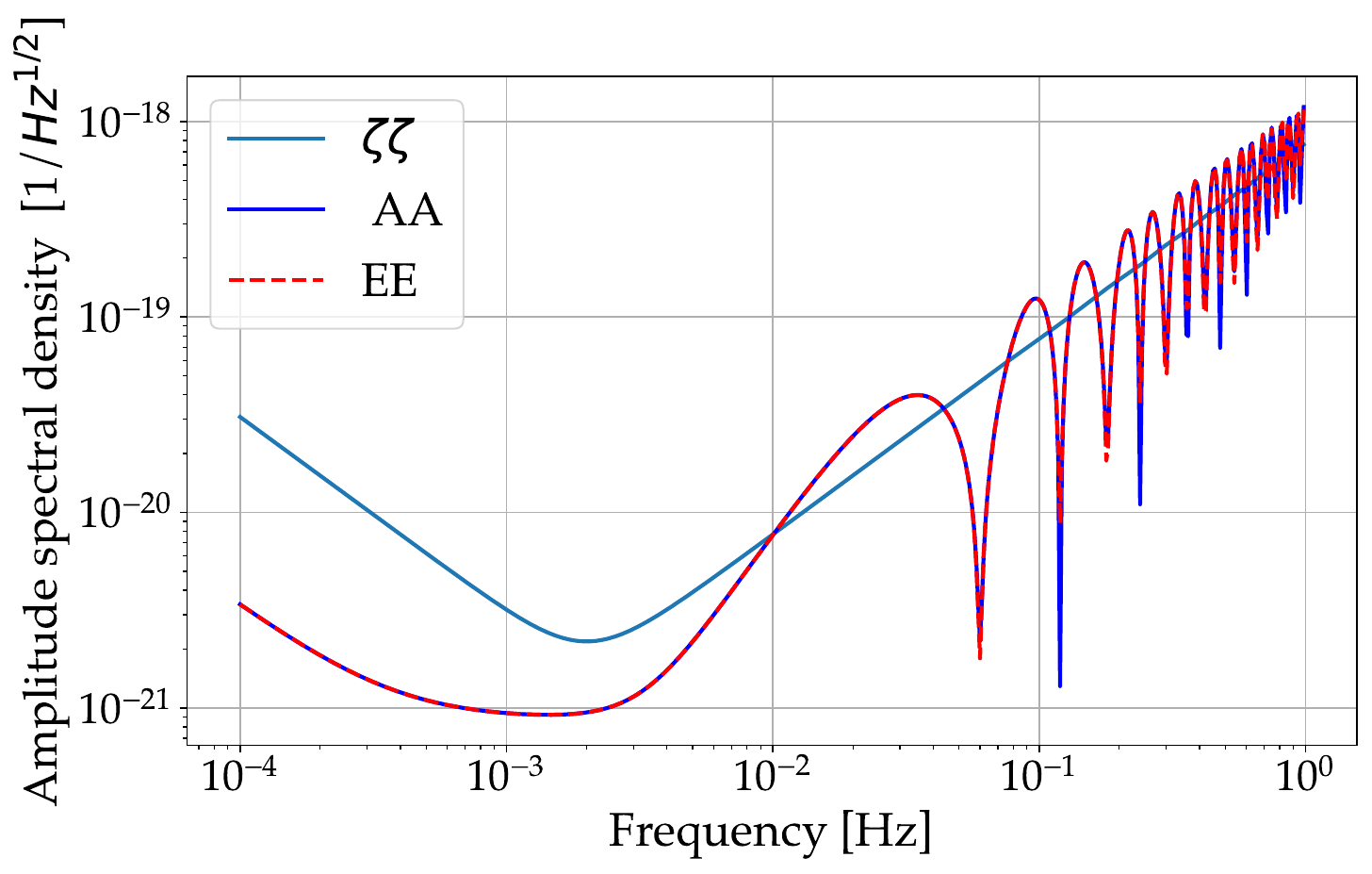}
\caption{Reference Amplitude spectral density for the Time delay interferometry channels A, E and $\zeta$ considering only test mass acceleration and optical metrology noise and assuming a constellation of three fixed unequal arm-lengths. }\label{fig:noises}
\end{figure}

To derive the PSDs and CSDs of the TDI channels $A$, $E$ and $\zeta$, as illustrated in \cite{Muratore:2021uqj}, one can express the arm-lengths $L_{ij}$ in terms of the breathing modes of the LISA triangle, $\delta_a$ and $\delta_b$  as:
\begin{subequations}\label{Lterm2}
\begin{align}
L_{12}(t)&= L \left[1 + \frac{1}{2}\left(\sqrt{3}\,\delta_a - \delta_b\right)\right] \; ,\\
L_{23}(t)& = L \left(1 + \delta_b\right) \; ,\\
L_{31}(t) &= L \left[1 - \frac{1}{2}\left(\sqrt{3}\,\delta_a + \delta_b\right)\right] \; .
\end{align}
\end{subequations}
The full expressions are rather long thus we give them in a separate $Mathematica$ notebook file
(\href{https://github.com/martinaAEI/noise_knowledge_uncertainty.git}{noise-analytical-model}) and we plot them 
%below the expansions at low frequencies, $\omega<<1$, for the PSDs to have a quantitative description of what we see 
in Fig. \ref{fig:noises}.  The Amplitude spectral density (ASD) are computed in terms of $\delta_a$ and $\delta_b$.  Indeed, while $L = (L_{12} + L_{23} + L_{31})/3\approx \SI{8.3}{\second}$ is the average arm-length, the small parameters $\delta_a $ and $\delta_b$ are typically $\sim 0.005 - 0.009$ for realistic ESA orbits \cite{lisa_orbit}. The case $\delta_a = \delta_b = 0$ corresponds to the equal-arm LISA scenario.  
%\begin{align}
%\begin{subequations}   
%S_{AA}(\omega) \approx & 3\omega^2(4S_g^{disp} + S_{oms})(8T^2 + 3T^2\delta_5^2 \nonumber \\ & + 2\sqrt 3T^2\delta_5 \delta_6 +  T^2\delta_6^2)\\
  % S_{EE}(\omega) \approx   &   3\omega^2(4S_g^{disp} + S_{oms})(8T^2 + T^2\delta_5^2 %\nonumber \\  & - 2\sqrt{3}T^2\delta_5 \delta_6 +   3T^2\delta_6^2)\\
%S_{\zeta\zeta}(\omega) \approx  &    6 (S_{oms}+T^2 \omega^2 S_g^{disp} (1+2 \delta_5^2+2 \delta_6^2))
%\end{subequations}
%\end{align}

%As is possible to see from the low frequency expansion the CSD terms given by arm-length inequality are proportional to the $\delta$s and thus few orders of magnitude smaller than the PSD so in principle can be discarded. Looking at the PSD we see that these $\delta$s represent a correction, at the low frequencies, to the equal arm-lengh case. 

\begin{figure}
\centering
\includegraphics[width=0.5\textwidth]{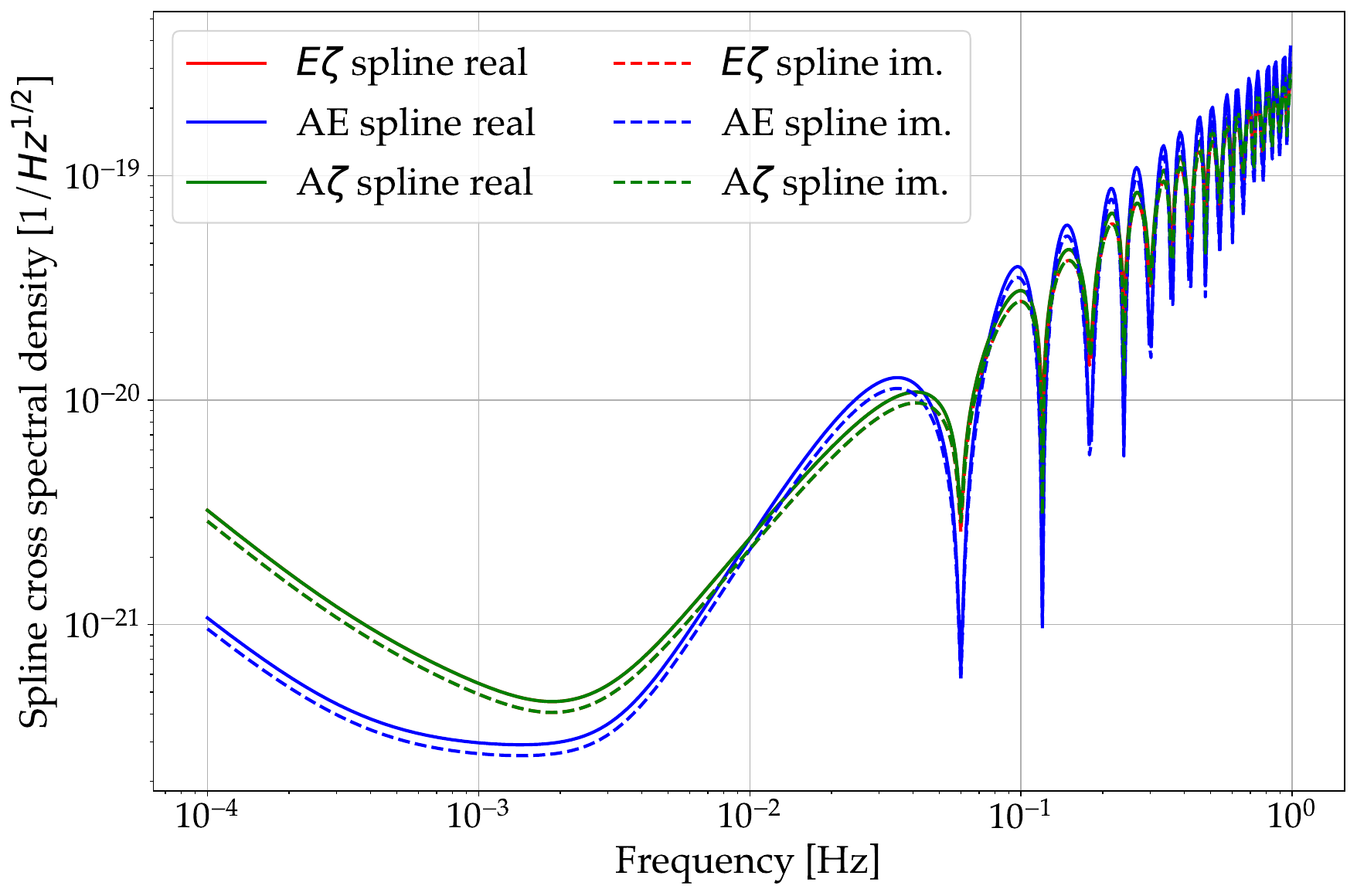}
\caption{\label{fig:noises2} Real and Imaginary part of splines-square root cross spectral density for the time delay interferometry channels AE, E$\zeta$ and A$\zeta$ considering test mass acceleration and optical metrology noise assuming a constellation of three fix unequal arm-lengths
}
\end{figure}

We consider that the six TMs have the same PSD as well as the six OMS noise terms, but this suppresses the contribution in the CSD. It was shown in \cite{Hartwig:2023pft} that if the levels of the noises differ by $20\%$ then the CSD can be $10\%$ of the PSD at low frequencies and several tens of percent at high frequency. %up to $50\%$ (or even $100\%$ at the zeros).
This motivates the particular choice of flexible CSD model that we introduced in Eq.~(\ref{eq:csd_splines}) and is illustrated in Fig.~\ref{fig:noises2}.
%Therefore, since our goal is to be as general as possible on the noise model we take into account the CSD terms when considering the total covariance matrix.  %To account for it we then used a different CSD model with respect to \ref{fig:noises}  (see \cref{eq:csd_splines}) which is figuratively reported in Fig. \ref{fig:noises2}. \LS{I do not understand what you are trying to say here. }

\subsection{Signal transfer function} \label{sec:signal_response}
The detector response to a stochastic background can be computed by expressing a GW signal as a superposition of plane waves, and by assuming that the LISA constellation has static arm lengths and is in a flat background spacetime. Following \cite{Hartwig:2023pft}, it is possible to show that the component of the single link measurement $\eta_{ij}(t)$ due to a GW is given by:
% The response of the various TDI channels to a stochastic background can be understood by considering the response of a single link in the constellation to an incident gravitational wave~\cite{Hartwig:2023pft}. 
% The component of the single link measurement $\eta_{ij}(t)$ due to a GW is given by:
\begin{align}
    \eta^{GW}_{ij}(t) &= i \int_{-\infty}^{\infty} \left\{ \frac{f}{f_{ij}} {\rm e}^{2\pi i f (t-L_{ij})} \right. \nonumber \\
    & 
\left.    \int  \left[ {\rm e}^{-2\pi i f \hat{k} \cdot \vec{x}_i} \sum_{\mathcal{A}} \xi_{ij}^\mathcal{A}(f,\hat{k}) \tilde{h}_\mathcal{A} (f,\hat{k}) \right] {\rm d}\Omega_{\hat{k}} \right\} {\rm d}f,
\end{align}
where $i$ stands for imaginary component, $f_{ij}=(2\pi L_{ij})^{-1}$, $\vec{x}_i$ denotes the position of satellite $i$, $\mathcal{A}= +, \, \times$ denotes the GW polarization, $\tilde{h}_\mathcal{A}(f,\hat{k})$ is the Fourier transform of the GW signal, $f$ is the GW frequency, $\hat{k}$ the outward vector in the direction of the incoming GW and ${\rm d}\Omega_{\hat{k}}$ is the infinitesimal solid angle.\\ The above expression quantifies the fractional frequency shift due to a superposition of plane waves coming from different directions $\hat k$. \\ The term $\xi^\mathcal{A}_{ij}$ projects the incoming wave with polarization $\mathcal{A}$ onto the detector, and its functional dependence is given by:
\begin{equation}
\xi^\mathcal{A}_{ij}\left(f, \hat k\right)= e^{-2\pi i f \hat k\cdot \vec L_{ij}}\mathcal M_{ij}(f, \hat k) \; \mathcal{G}^\mathcal{A}(\hat k,\hat l_{ij}) \, ,
\end{equation}
where
\begin{equation}
\mathcal{M}_{ij}(f, \hat k) \equiv \mathrm{e}^{\pi i f L_{ij} ( 1 + \hat{k}\cdot \hat l_{ij})}  \; \mathrm{sinc}\left(\pi f L_{ij} ( 1 + \hat{k}\cdot \hat l_{ij})\right)\\
\end{equation}
and
\begin{equation}
\mathcal{G}^\mathcal{A}(\hat k, \hat l_{ij}) \equiv \frac{\hat l^a_{ij}\hat l^b_{ij}}{2}e^\mathcal{A}_{ab}(\hat k) \, ,
\end{equation}
where $\hat l_{ij} = (\vec x_j - \vec x_i)/|\vec x_j - \vec x_i|$ is a unit vector pointing from spacecraft $i$ to $j$ and $e^\mathcal{A}_{ab}(\hat{k})$ denotes the GW polarization tensors.

For an a homogeneous, isotropic and non-chiral, stochastic background, the GW signal is only specified statistically
$$\langle \tilde h_\mathcal{A}(f,\hat k) \, \tilde h_B^*(f', \hat k')\rangle = \delta(f - f')\delta(\hat k -\hat k')\delta_{\mathcal{A}B}\frac{P_{h}^{\mathcal{A}B}(f)}{16\pi} $$ and 
$$\langle \tilde h_\mathcal{A}(f,\hat k) \, \tilde h_B(f',\hat k')\rangle = 0 \, .$$
Homogeneity and isotropy implies that $P_{h}^{\mathcal{A}B}(f)$ is diagonal, whereas the non-chirality implies $P^{\times \times}_h=P^{++} _h$, so that we can define $ P _h := \sum_\mathcal{A} P^{\mathcal{A}\mathcal{A}} _h$.

We characterise the response of the individual links to a stochastic background statistically
\begin{equation}
\langle \tilde \eta^{GW}_{ij} \, \tilde \eta^{GW}_{mn}\rangle = \frac{1}{2} S^{\eta,\mathrm{GW}}_{ij,mn}(f) \delta(f-f') \, ,
\end{equation}
where spectral densities for the link measurements are given by
\begin{equation}
S^{\eta,\mathrm{GW}}_{ij,mn}(f)  = \frac{f^2}{f_{ij}f_{mn}} e^{-2\pi i f(L_{ij}-L_{mn})}\sum_\mathcal{A} P_h^{\mathcal{A}\mathcal{A}}(f)  \; \Upsilon_{ij,mn}^{\mathcal{A}}(f) \; ,
\label{eq:CSDsignal}
\end{equation}
with:
\begin{equation}
\Upsilon_{ij,mn}^{\mathcal{A}}(f)=\int \frac{\rm{d} \Omega_{\hat{k}}}{4 \pi}  \; \textrm{e}^{-2\pi i f \hat{k}\cdot (\vec x_i - \vec x_m)}  \;  \xi^\mathcal{A}_{ij}(f, \hat k)  \, \xi^\mathcal{A}_{mn}(f, \hat k)^* \, .
\label{eq:Upsilon}
\end{equation}

The power  spectral densities of the signal in the TDI variables described in sec. \ref{sec:noise_and_tdi} can then be computed from
\begin{align}
\langle \tilde{U}(f) \tilde{V}^*(f')\rangle &= \frac{1}{2} S^{\rm GW}_{UV}(f) \delta(f-f') \nonumber \\
S^{\rm GW}_{UV}(f)&=\sum_{ij,mn\in \mathcal{I}} c^U_{ij}(f)  c^{V*}_{mn}(f) S^{\eta,\mathrm{GW}}_{ij,mn}(f)\; ,
\label{eq:sign_psd}
\end{align}

where $\tilde{U}$ and $\tilde{V}$ denote any two TDI variables, which in our case are TDI A, E and $\zeta$, and $\mathcal{I} = \{12, 13, 23, 21, 31, 32\}$ denotes the set of pairs of indices that define the six inter-satellite links. The coefficients $c^{U}_{ij/mn}$ map the single-link measurements onto the TDI variable $U$. Refer to the Mathematica code for the computation of such coefficients (\href{https://github.com/martinaAEI/noise_knowledge_uncertainty.git}{noise-analytical-model}).\\ 

\begin{figure}
\centering
\includegraphics[width=0.5\textwidth]{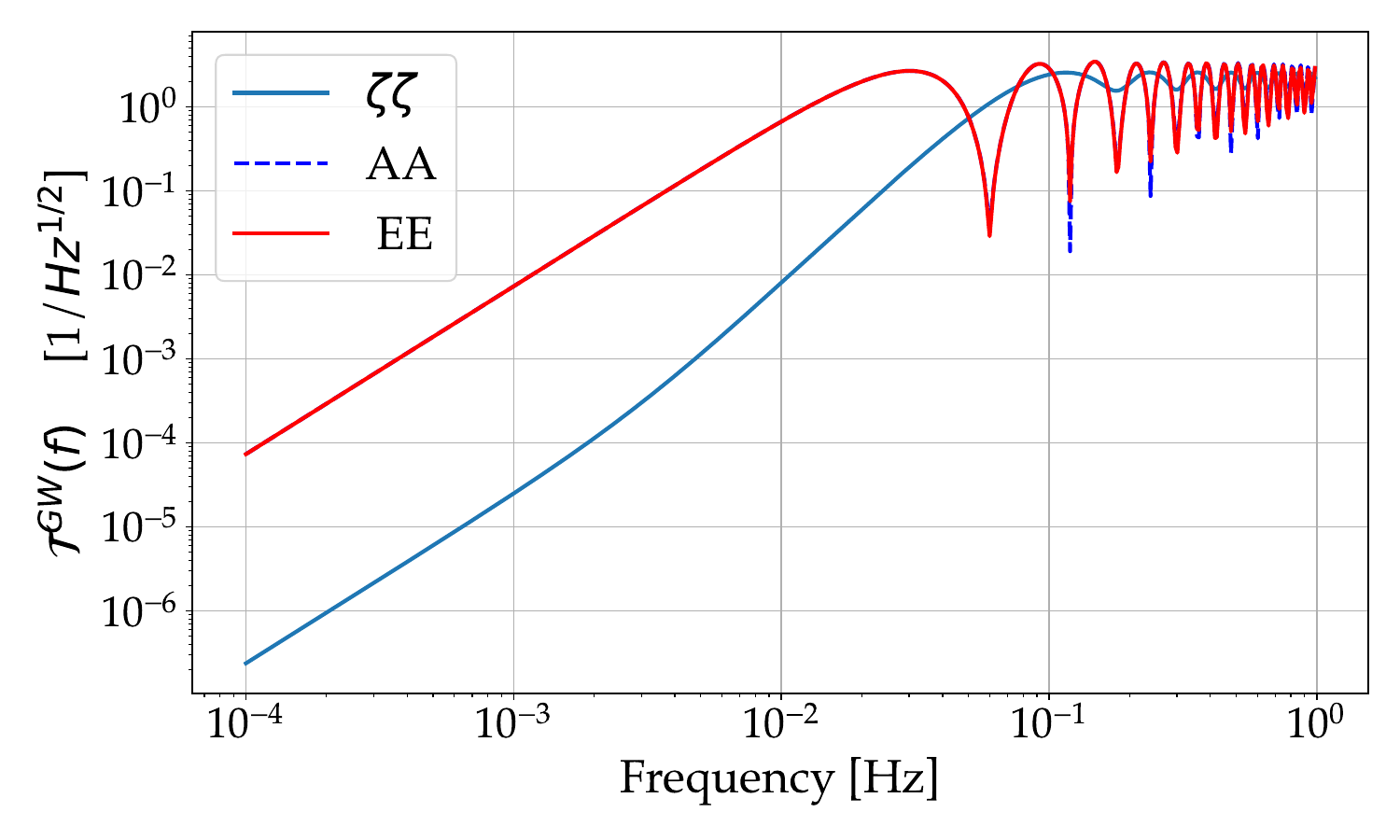}
\includegraphics[width=0.5\textwidth]{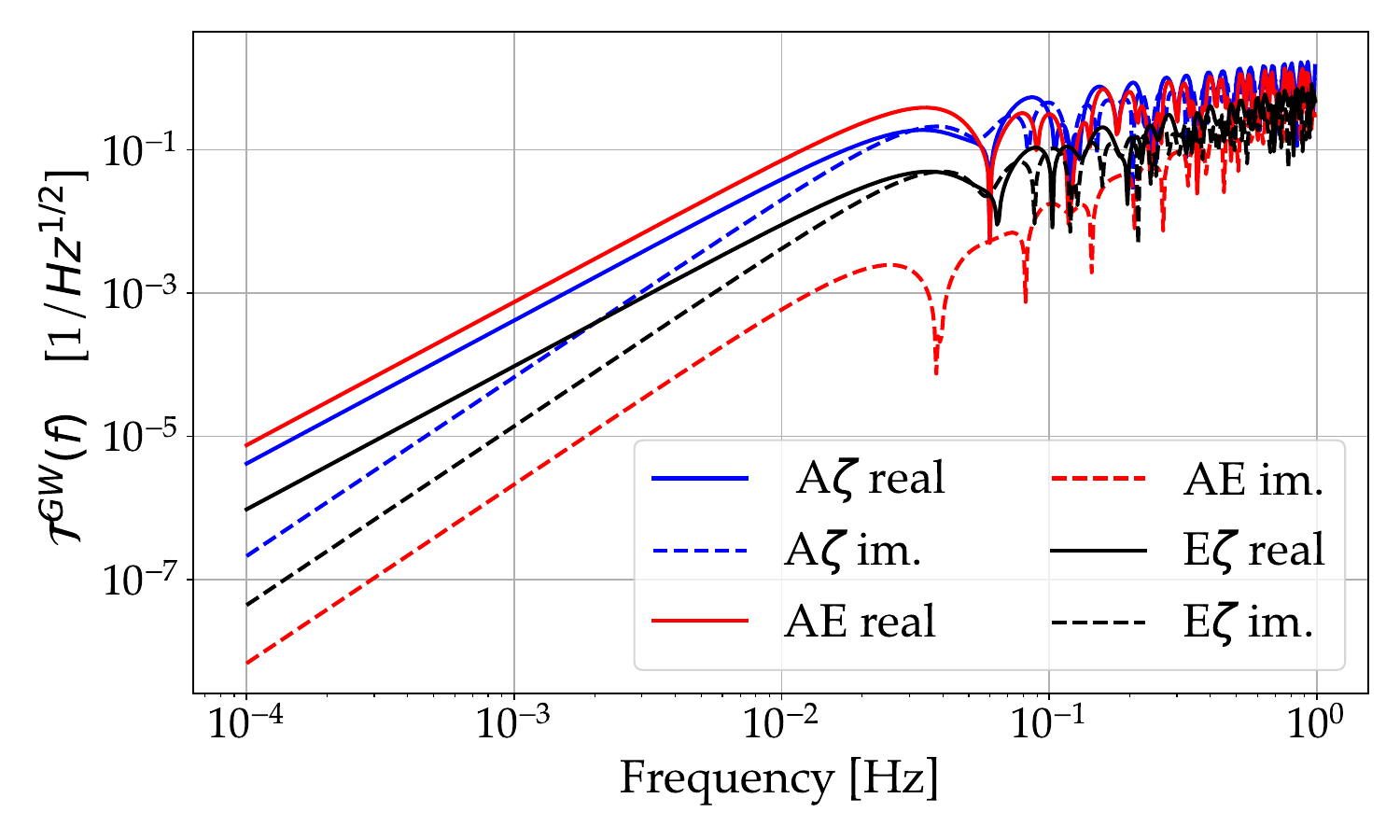}
\caption{Upper panel: Gravitational wave transfer functions $\mathcal{T}^{GW}(f)$ of the three time delay interferometry channels A,E and $\zeta$ assuming a constellation of three fixed unequal arm-lengths; lower panel: real and imaginary components of the gravitational wave transfer functions $\mathcal{T}^{GW}(f)$ of the time delay interferometry channels AE, E$\zeta$,A$\zeta$.}
\label{fig:GWcsd}
\end{figure}

Note that considering each polarization of the SGWB contributes equally to the background, i.e. $P^{\times \times} _h=P^{++} _h$, %, then we define $ P _h := \sum_\mathcal{A} P^{\mathcal{A}\mathcal{A}} _h$.
we can rewrite Eq.~\eqref{eq:sign_psd} as a product of the SGWB spectral density $P_h(f)$ and a transfer function $\mathcal{T}^{GW}(f)$ which takes into account the LISA detector response, i.e. $S^{\rm GW}_{UV}(f) = \mathcal{T}^{GW}(f) P_h (f)$.  \\
$S^{\rm GW}_{UV}(f)$ would correspond to the first term on the right hand side of Eq.~\eqref{eq:st_sn_sgw}. The transfer functions for the three TDI channels and their cross correlation are shown in Fig.\ref{fig:GWcsd}.
% \begin{align}
% \mathcal{T}_{UV}(f)=\sum_{ij,mn\in \mathcal{I}} c^U_{ij}(f)  c^{V*}_{mn}(f)
%  \frac{f^2}{f_{ij}f_{mn}} \\
%  e^{-2\pi i f(L_{ij}-L_{mn})}\sum_\mathcal{A} \Upsilon_{ij,mn}^{\mathcal{A}}(f) 
% \end{align}

%The coefficients can be computed by replacing each delay in the time domain, $D_{ij}$, with the corresponding frequency domain expression. \LS{This last paragraph is not clear to me. What do you mean by "the corresponding frequency domain expression"?}

%Assuming that an isotropic GW background is made of superposition of many GW sources coming from different directions and with different polarizations, we can consider that the output of a TDI$_j$ as given superpositions of $n$ plane waves is:
%\begin{equation}
%S_{{j}_h} = \sum_{i}^n T_{j_h}^i(\omega) S_{h_i}(\omega)\label{sh},
%\end{equation}
%where $S_{h_i}(\omega)$ is the PSD/CSD of the $i$'th GW source expressed as dimensionless strain, and $T_{j_h}^i(\omega)$ is the absolute squared value transfer function for the $j$'th TDI.

\subsection{SGWB Signal models}
\label{sec:sgwbmods}
There are a large variety of models for stochastic gravitational wave backgrounds that might manifest in the LISA band \cite{red_b}. In this work we focus on four models, which can be described by their energy density, $ h^2 \Omega_{GW}$ \cite{Caprini:2018mtu}, which is a function of some parameters, $\theta$:
\begin{itemize}
\item{\textbf{Power law}: \\
\begin{equation} 
h^2 \Omega_{GW}(f) \approx  A 
%6.9 \times 10^a 
\left(\frac{f}{f_p} \right)^n,
\end{equation}
where $f_p$ is the pivot frequency, defined as the geometrical mean of the LISA frequency interval ($10^{-4}$ Hz, $0.1$ Hz), $f_p = 3$mHz. The model parameters are the log-amplitude, $A$, and slope, $n$ \cite{babak2023stochastic, lehoucq2023astrophysical}. We use reference values of $n = 2/3$ and $A = 7.87  \times 10^{-13}$, representing a SGWB from stellar origin black hole binaries that has energy density at 1mHz of $h^2 \Omega_{GW}(1mHz) = 3.78 \times 10^{-13} $. This value was chosen to be compatible with LIGO-VIRGO-KAGRA constraints \cite{babak2023stochastic}}
\item{\textbf{Gaussian bump}: \\
\begin{equation} 
h^2 \Omega_{GW} = A e^{- \frac{1}{2 \sigma^2} \ln(\frac{f}{f_p})^2},
\end{equation}
where $f_p$ is the pivot frequency as before. The model parameters are the log-amplitude, $A$, and width, $\sigma$. We use reference values of $A = 10^{-12.48}$ and $\sigma = 0.3$ whose energy density at 1mHz is $h^2 \Omega_{GW}(1mHz) = 4.05 \times 10^{-16} $.   This signal is chosen as a simple way to mimic the one which might arise from particle production taking place for a limited number of e-folds during inflation (as, for instance, required by some models of primordial BH generation). (See e.g.~ \cite{Bartolo_2019,Bartolo:2018evs,franciolini2021primordial} )}
\item{\textbf{Power law with running:} \\
\begin{equation} 
h^2 \Omega_{GW} = A \left(\frac{f}{f_p}\right)^{n + \alpha  \ln(\frac{f}{f_p})},
\end{equation}
where $f_p$ is the pivot frequency as before. The model parameters are the log-amplitude, $A$, the slope, $n$, and the running index, $\alpha$. We use reference values of $A = 10^{ -12.65}$, $n = 1$ and  $\alpha = -0.1$.  This signal is motivated by non-standard inflationary models. For example, gravitational wave generation can be enhanced by sustained particle production during inflation, leading to a power law stochastic GW background, which would deviate from a simple power law at higher frequency when back-reaction kicks in (see e.g. \cite{Bartolo:2016ami}). The energy density at 1mHz is $h^2 \Omega_{GW}(1mHz) = 6.61 \times 10^{-14} $ }
\item{\textbf{First Order Phase Transition}:
\begin{equation}
 h^2\Omega_{GW}(f) = h^2\Omega_p  \left(\frac{f}{f_p}\right)^3 \left(\frac{7}{4 + 3 \big(\frac{f}{f_p}\big)^2}\right)^{n} \label{eq:PT_SGWB},
\end{equation} 
where $f_p=2\cdot 10^{-4}  $ Hz (note this is different to the reference frequency in the previous models). The model parameters are the energy density, $h^2\Omega_p$, and spectral index,  $n$. We use  reference values of $A\equiv h^2\Omega_p= 10^{-10}$ and $n = 7/2$ whose energy density at 1mHz is $h^2 \Omega_{GW}(1mHz) = 2.59 \times 10^{-12} $. This signal is motivated by the production of sound waves in the cosmic fluid from colliding phase transition bubbles \cite{Caprini_2016,PhysRevD.96.103520}.}
\end{itemize}

In our analysis, we also include the contribution to the spectral density from the foreground of galactic binaries (GB). 
%We also compare, for all the models, the SNR and amplitude that we would need in order to be able to detect and characterise them in the case we take into account the galactic foreground which peaks at 1mHz. 
We use the following model for the foreground \cite{red_b}:
\begin{itemize}
\item{\textbf{Foreground of Galactic Binaries}: 
\begin{align} 
S_{GB}(f) = & A_{GB}\left(\frac{f}{Hz}\right)^{-\frac{7}{3}} e^{- (f/f_1)^{\alpha}} \nonumber \\  & \times \frac{1}{2} \left[1 + \tanh \left(\frac{f_{knee}-f}{f_2}\right)\right]
\end{align} 
with: $$f_1 = 10^{a_1 \log_{10}(T)+b_1} , f_{knee} = 10^{a_k \log_{10}(T)+b_k}$$}.
\end{itemize}
setting $A = 1.15\cdot 10^{-44};~\alpha = 1.56;~a_1 = -0.15;~b_1=-2.72;  ~a_k = -0.37;~b_k=-2.49;~f_2 = 6.7 \times 10^{-4}$Hz. When considering the background in conjunction with other SGWBs we allow the amplitude to vary, but keep the other parameters fixed. \\

The relation between the energy density $\Omega_{GW}$ and the stochastic GW background power spectral density $P_h(f)$ is given by \cite{Caprini:2018mtu}: 
\begin{equation}
\Omega_{GW}(f) = \frac{4 \pi^2}{3 H_0^2} f^3 P_h(f),
\end{equation}
where $H_0$ is the Hubble constant fixed to be 67.8 km/s/Mpc, as a consequence $h=0.678$. The conversion between the energy density $\Omega_{GW}(f) $ and gravitational power spectral density $P_h(f)$ used to compute Eq.\eqref{eq:sign_psd}  is then \cite{ Caprini:2018mtu}:
 \begin{equation} 
 P_h(f)= 7.98 \times 10^{-37} \left(\frac{Hz}{f}\right)^3 h^2 \Omega_{GW}(f) \frac{1}{Hz},
 \end{equation}.\\

\begin{figure}
\includegraphics[width=0.45\textwidth]{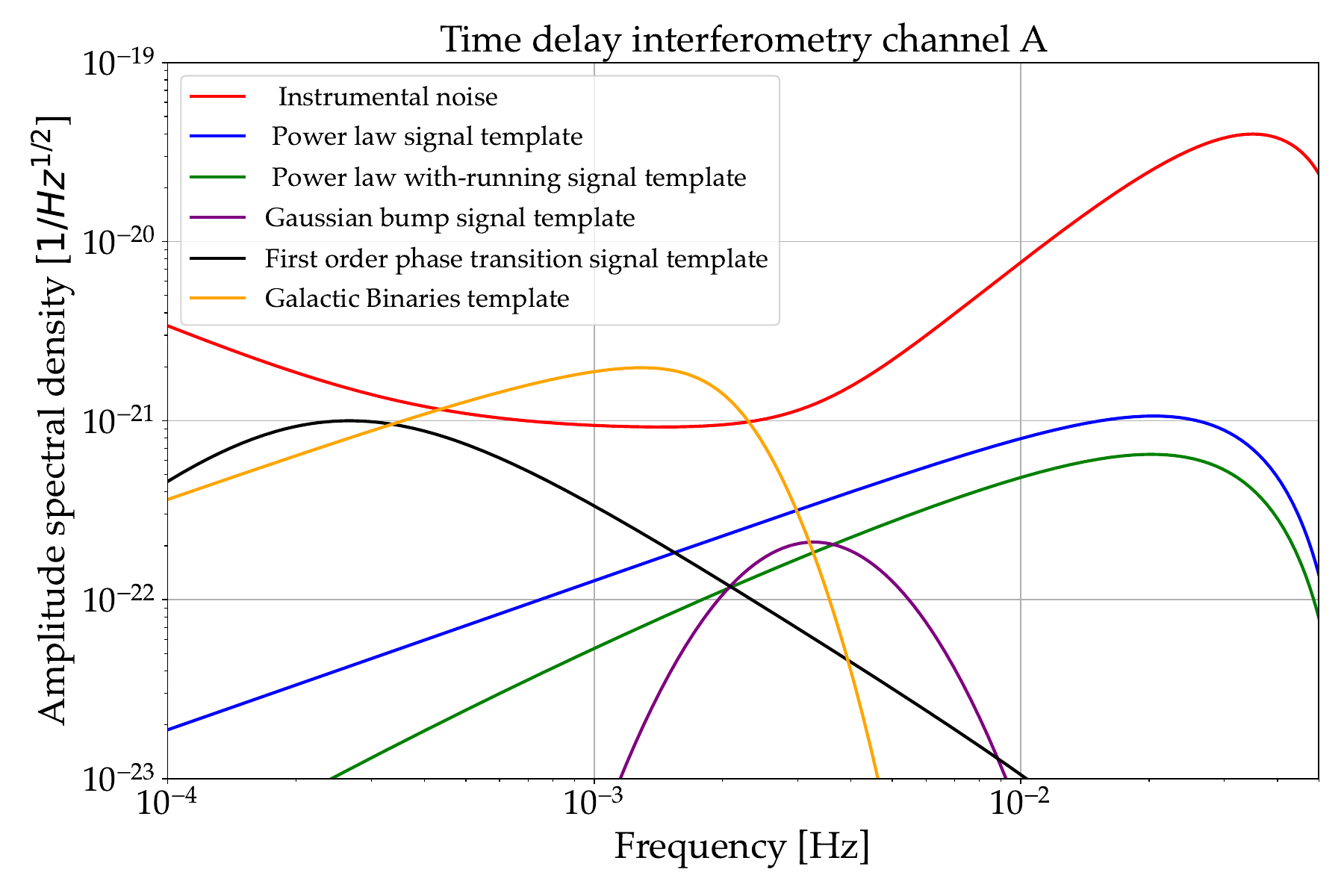}
\caption{\label{fig:sgwbnoisepw} Amplitude spectral density of the stochastic GW background models and the amplitude spectral density of the reference test mass and optical metrology noise in the time delay interferometry channel A.}
\end{figure}

We report in Fig. \ref{fig:sgwbnoisepw} the ASD of the four SGWB models together with the ASD of the reference instrumental noise in TDI channel A. \cite{babak2021lisa} 
\\ 

%The conversion  between amplitude and Energy densitity we consider:
%\begin{equation} 
%\Omega_{GW}= 4/3 \pi^2/ H_0^2  f^3  S_h(f),
%\end{equation}
We also provide the computation of the SNRs of these different backgrounds in the TDI channel A using the following formula \cite{Thrane:2013oya}:
\begin{equation} 
%SNR_j= \sqrt{T}\left[ \int_0^\infty  \frac{T_{j_h}(f)^2 S_{h}(f)^2 }{S_n(f)^2} {\rm d}f \right]^{1/2}
SNR_A= \sqrt{T}\left[ \int_0^\infty  \frac{S^{GW}_{AA}(f)^2 }{S^A_n(f)^2} {\rm d}f \right]^{1/2}
\label{eq:SNRdef}
\end{equation}
with an observation time span of $T= 4$ years. Here $S^{GW}_{AA}(f)$ is the spectral density in channel $A$ that can be computed from Eq.~(\ref{eq:sign_psd}) and $S^A_n$ is the PSD of the A channel. The results\footnote{We note that this formula is derived assuming that we have access to two independent channels that have uncorrelated noise and perfectly correlated signals. This is not a good approximation to LISA, so the SNR is not directly interpretable. However, it still provides an indication of the relative detectability of different backgrounds. }  are shown in Table \ref{tab:snr}.
\begin{table}
\begin{tabular}{|c||c|c |}
\hline
 SGWB Model                   &   SNR w/o GB &   SNR w/ GB \\ \hline \hline
 Power law with running       &     14.54 & 13.35 \\
 Power law                    &     48.70 & 42.89 \\
 Gaussian bump                &     13.51  & 11.65 \\
 First order phase transition &    118.68 & 64.18 \\ \hline
\end{tabular} 
\caption{Signal to noise ratio in TDI channel A for the four SGWB models, with and without the presence of the galactic foreground as an additional noise component. The galactic foreground here considered has an SNR of 1627.39}
\label{tab:snr}
\end{table}\\ 
It is possible to notice that including the foreground as part of the noise ($S^A_n(f) := S^A_{GB}(f)+S^A_{n}(f)$) leads to a substantial decrease of the SNR for the FOPT background but the SNR does not change very much for the other models. 
%\todo[inline]{I checked the SNR for power with running, gaussian bump and they are same order of magnitude of chiara's notes, for the others not really but we use different models with respect to Chiara's note}

\begin{figure}
\centering
\includegraphics[width=0.45\textwidth]{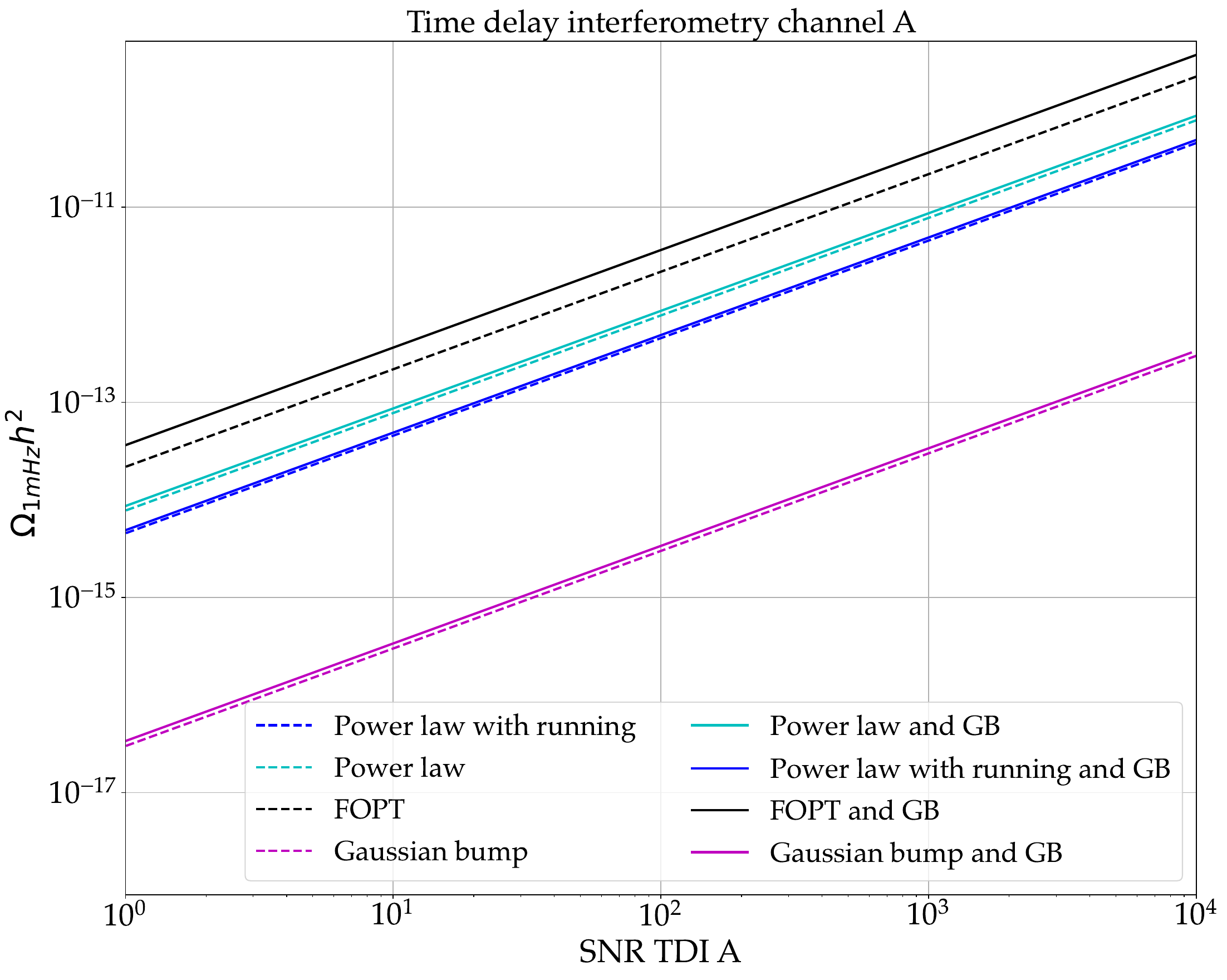}
\caption{Signal to noise ratio of four different SGWB signals in the TDI channel A,  Power law, Power law-with-running, Gaussian bump and First order phase transition, versus the energy density at 1mHz, both considering (continuous lines) or not (dashed lines) the presence foreground\label{fig:snr_omega} }
\end{figure}

We plot in \cref{fig:snr_omega} the value of the SNR in channel $A$ vs. the energy density at 1mHz for the different models. The dotted lines assume no presence of the foreground, whereas the continuous lines include the presence of the foreground. As expected there is a direct correlation between increasing the energy density and an increase in the SNR. Moreover, the presence of the foreground mostly affects the SNR of the FOPT. In fact, in the presence of the foreground the energy density must be two times larger to have the same SNR as it would in the absence of a foreground.

\section{Results}\label{sec:results}
%\LS{Maybe we can have a unique section called results, which is subdivided into Measurement precision in the presence of nosie, and Impact of noise knowledge uncertainty on SGWB precision. We can discuss about the sections and subsection titles.}

\subsection{Impact of instrumental noise knowledge uncertainty on SGWB recovery}
\label{sec:ampdep}
Here, we explore how the measurement precision of the SGWB parameters changes in the presence of instrumental noise knowledge uncertainty, for each of the SGWB models described in Section~\ref{sec:sgwbmods}. We use the Fisher matrix formalism described in Section~\ref{sec: fisher}, which assumes that the noise is uncorrelated at different frequencies. We assume we use three TDI channels in our analysis, A, E and $\zeta$, as described in Section~\ref{sec:noise_and_tdi}. We model uncertainties in the PSD and CSD at each frequency following the model described in Section~\ref{sec:splines}. To build the Fisher matrix we need the following elements:
%the above models using the Fisher matrix formalism. We consider three TDI channels whose noise properties are described by a symmetric $3 \times 3$ matrix defined per each frequency. We assume that each frequency is independent and that the instrumental and GW components are also independent. To build the Fisher matrix then we compute the following elements:
\begin{enumerate}
  \item The derivatives of the PSD and CSD at each frequency with respect to the parameters of the SGWB model.
  \item The derivatives of the PSD and CSD at each frequency with respect to the parameter (amplitude) of the Galactic Binaries. The addition of this parameter extends the dimension of the Fisher matrix by one. 
  \item The derivatives of the PSD and CSD at each frequency with respect to the parameters of the instrumental noise model. The instrumental noise model is based on 9 different splines: 3 splines to model the PSD of A, E and $\zeta$, and 3 splines each for real and imaginary parts of the CSDs for AE, A$\zeta$ and E$\zeta$. Each spline has a number of parameters equal to the umber of knots, which we take to be $13$. The total number of noise parameters is therefore 9 x 13  = 117. 
  \item The evaluation of the Fisher matrix from these elements using Eq.~(\ref{log_likelihood}), which is summed over frequency.
  \item The choice of a prior on the instrumental noise parameters. We use a Gaussian prior, which is implemented in the Fisher matrix formalism by adding the prior matrix to the Fisher matrix before computing its inverse (see Eq.~\eqref{eq:fisher_prior}). For this first study we take the priors on each noise parameter to be independent, with zero mean and equal variance, $\sigma_{\rm inst}$. In this section we fix $\sigma_{\rm inst}=1$, which means we are allowing for up to an order of magnitude uncertainty in the instrumental noise at each frequency.
  \item We compute the inverse of the Fisher matrix after adding the prior to obtain an estimate of the measurement uncertainty, from the square root of the diagonal elements of the inverse as explained in Sec. \ref{sec: fisher}. We also compute the inverse of the SGWB-parameter only sub-matrix of the Fisher matrix, which represents the expected uncertainty in the absence of instrumental noise uncertainties.
  %At each frequency we compute the inverse of noise plus signal covariance matrix
%\item  The prior inverse covariance matrix. Note that we consider that the prior uncertainty $\sigma$ is the uncertainty in the log10 amplitude. So $\sigma = 1$ means we allow one order of magnitude variation in the PSD. 
%\LS{The above method should be described}\M{??? which method? It is not clear to what are you referring to}
%For these initial studies we consider a uniform prior with $\sigma = 1$ and we consider the prior just on the instrumental noise 
\end{enumerate}
For each SGWB model we will present the results in two different ways. Firstly, we will show the ratio of the uncertainties in the SGWB parameters in the presence of instrumental noise uncertainties to those uncertainties when perfect knowledge of the instrumental noise is assumed. These results illustrate the impact of lack of noise knowledge on SGWB characterisation. Secondly, we will show the actual uncertainties in the SGWB parameters, as computed from the Fisher matrix. Of particular interest is the uncertainty in the log-energy density of the background. As a rule of thumb, a background will be detectable if the uncertainty $\Delta \ln(A)<1$ ( $\Delta \ln(A) = (\Delta A)/A$). In both cases, we will plot results as a function of the background amplitude/the background energy density at a reference frequency of 1mHz (the logarithm of these quantities are linearly related, so they can be easily represented using bottom/top axes in a single figure). For the second type of plot, solid lines show results in the presence of noise knowledge uncertainty, and dashed lines give results assuming perfect noise knowledge. In both the analysis we consider the foreground amplitude to vary and we consider it as additional source of noise together with the instrumental noise.

%we want to estimate either versus the log background amplitude or versus the background amplitude (but expressed as the energy density at 1 mHz).  The ratio of the uncertainties is computed as square root between the diagonal term of the inverse of the fisher matrix, in case of non perfect noise knowledge ($\sigma$), over the diagonal term of the inverse of the fisher matrix in case of perfect noise knowledge ($\sigma_{ref}$). \\ The second plot, shows instead the absolute precision for the parameters we could achieve either versus the log amplitude or versus the background amplitude, again,  expressed as the energy density at 1 mHz. In all the plots, the continuous lines are the diagonal terms of the inverse of the fisher matrix in case of non perfect noise knowledge; whereas the dotted lines shows the case of perfect noise knowledge. We also mark the background energy density, at 1mHz, that corresponds to the observing requirement accordingly to the SGWB models defined. \\

%%%%%%%%%%

%%%%%%%
\subsubsection{Power law}
A power law SGWB is described by two parameters: the slope and the amplitude. The full Fisher matrix, including instrumental noise and foreground parameters, is $120 \times 120$.
%For the case of the Power law the fisher matrix is a $119 \times 119$ as we have 2 parameters we want to estimate: the slope and the amplitude.
\begin{figure}
\includegraphics[width=0.45\textwidth]{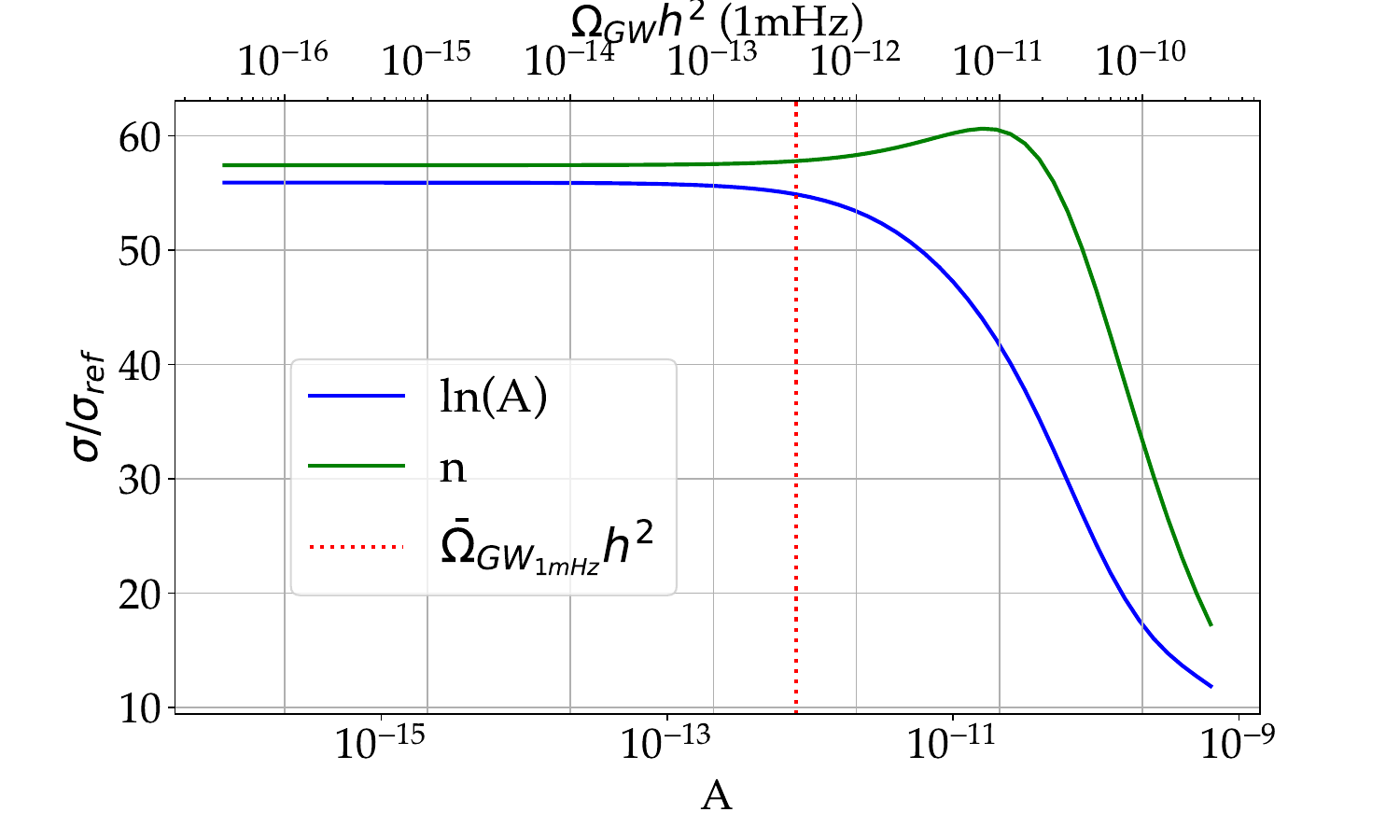}
\includegraphics[width=0.45\textwidth]{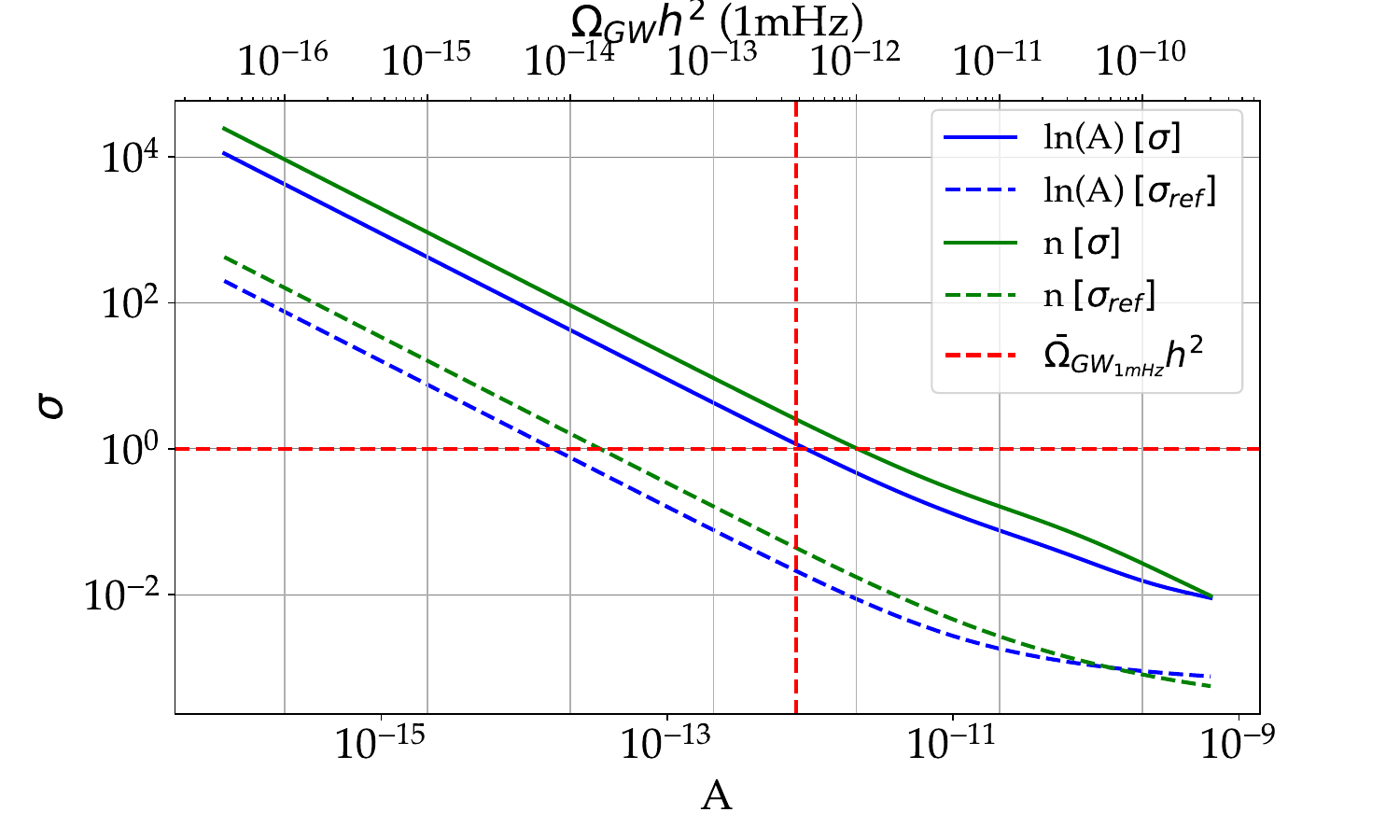}
\caption{\label{fig:power_law2} Results for the power law SGWB model considering the foreground as additional source of noise. The upper panel shows the ratio of the uncertainties of the SGWB parameters (amplitude and slope) when including instrumental noise uncertainties or assuming perfect noise knowledge. This ratio is plotted versus the amplitude (bottom axis) and SGWB energy density at 1mHz (top axis). The lower panel shows the estimated parameter uncertainties for the two cases. Once again this is as a function of amplitude/energy density, but for a restricted range. The horizontal red dashed line corresponds to an uncertainty of one, which is our threshold on the uncertainty in log-energy density for deciding that a background is detectable. The vertical red dashed line indicates the reference SGWB amplitude given in Section~\ref{sec:sgwbmods}. }
\end{figure}
Figure~\ref{fig:power_law2} shows the results computed for this model. We see that in the presence of instrumental noise uncertainties, the uncertainty in the SGWB parameters increases by a factor of $\sim 55$--$60$, with the uncertainty in the slope being slightly more affected than that of the amplitude. The increase is lower for high background amplitudes, as expected, but only when the background is one to two orders of magnitude brighter than the reference value. Considering the raw uncertainties, we see that the uncertainty in the log-energy density is typically a factor of $\sim 50$ larger and the background energy density would have to to be a factor of $\sim 50$ times %a factor of $\sim 20$ times
higher to be characterised with the same measurement precision when there is instrumental noise uncertainty as it could be without those uncertainties. %; and few factors higher to be detected when allowing for lack of knowledge of the instrumental noise.%. 
However, a background with amplitude equal to the reference value should (just) be detectable even allowing for confusion with instrumental noise mismodelling.
%The top plot of  shows that the ratio of the uncertainties in case of non perfect noise knowledge over perfect noise knowledge for the slope is of order 5.4 at the reference value and of order 4.4 for the log amplitude then it decreases for bigger amplitude and slope, as expected, since in these cases the impact of not knowing the noise is less evident. What we see looking at the lower plot of Fig. \cref{fig:power_law} is that there is more than one order of magnitude different between the error we compute when estimating the slope with perfect noise knowledge with respect to the case of non perfect noise knowledge. Similar is for the amplitude estimation. However, we can notice that the error we made in estimating the amplitude, even in case of noise knowledge uncertainties, is lower than 1 $\sigma$ at the reference value this means we will still be able to characterise it.

%%%%%%%%

%%%%%%
\subsubsection{Power law with running}
For the power law with running SGWB, the fisher matrix is $121 \times 121$, as the SGWB model depends on 3 parameters: slope, amplitude and running index, $\alpha$. 
\begin{figure}
\includegraphics[width=0.45\textwidth]{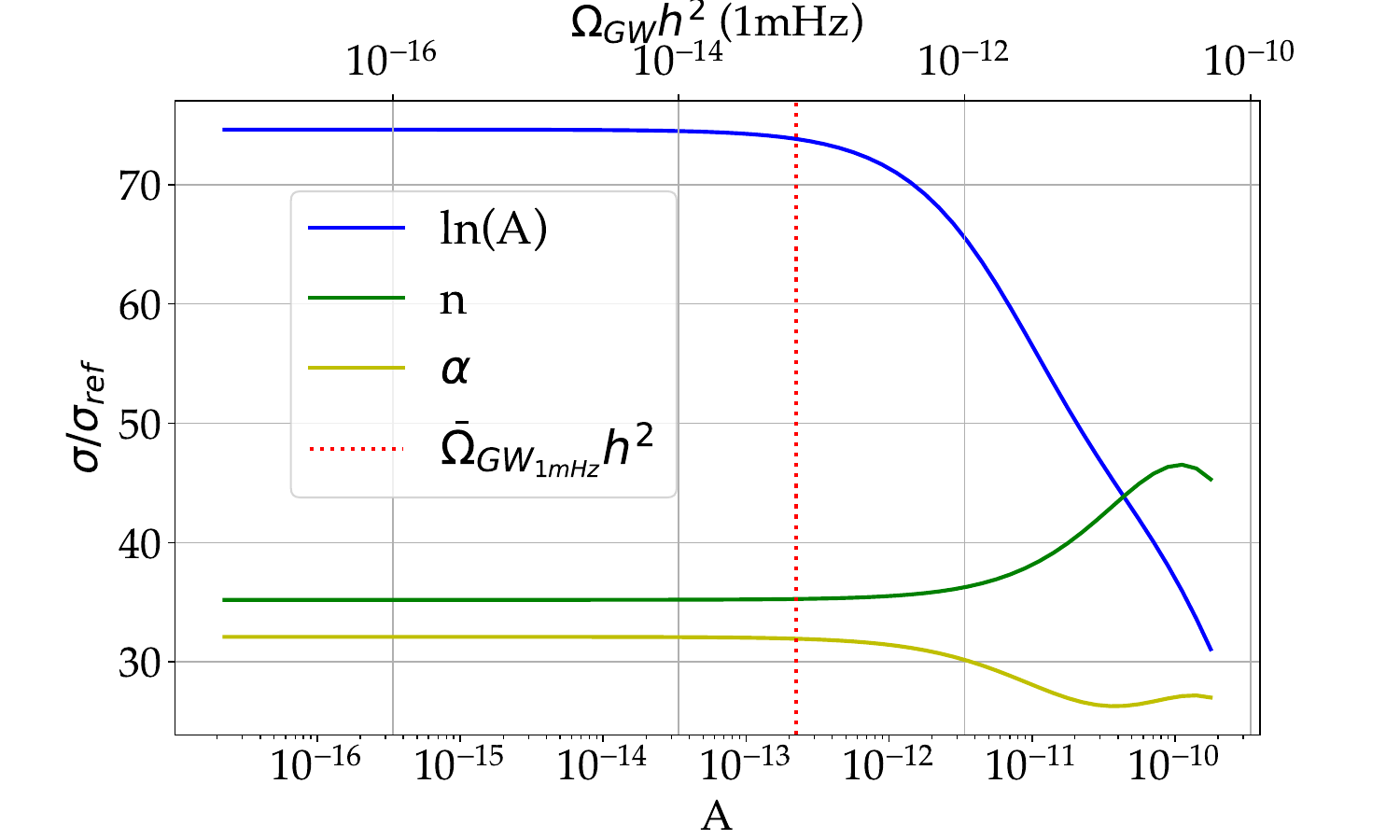}
\includegraphics[width=0.45\textwidth]{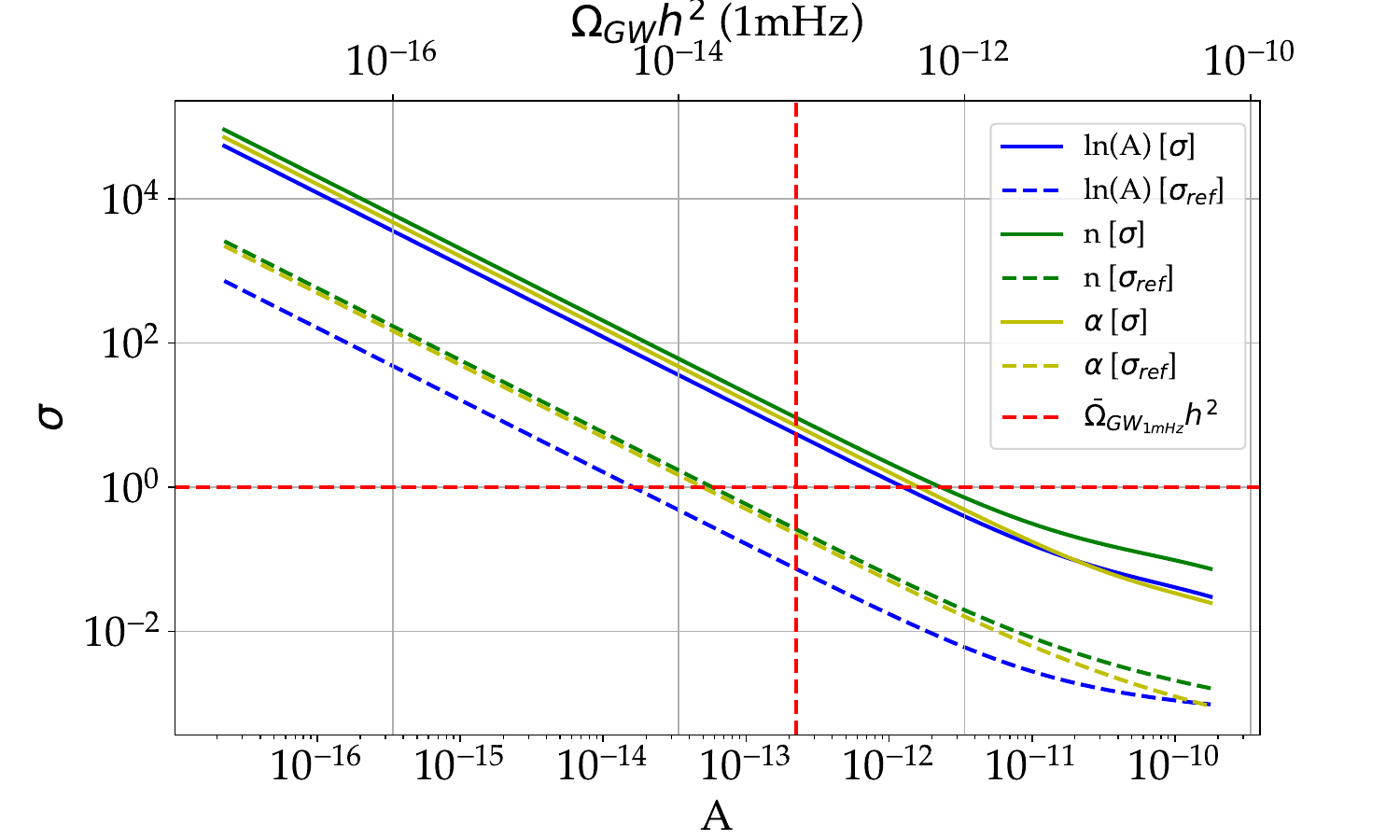}
\caption{\label{fig:power_law_wr2} As Figure~\ref{fig:power_law2}, but now for the power-law with running SGWB model%Power law with running model. Upper plot shows the ratio of the uncertainties of the amplitude, slope and $\alpha$ either versus the background amplitude or versus the background amplitude but expressed as the energy density at 1 mHz. Lower plot shows instead the same uncertainties either versus the  background amplitude or versus the background amplitude but expressed as the energy density at 1 mHz
}
\end{figure}
%%%%%%%%%%%%
The results for this model are shown in \cref{fig:power_law_wr2}. In this case we see that the uncertainties in the SGWB parameters increase by a factor of $\sim 30$--$75$, with the uncertainty on the amplitude being most affected in this case. Once again, the relative increase in the uncertainty is somewhat lower at higher background amplitudes. The lower panel of \cref{fig:power_law_wr2} shows that the background is not detectable at the reference amplitude. An energy density $\sim 20$ times higher would be required for a detection. In general, the background has to have an energy density $\sim 60$ times higher to be characterised with the same measurement precision when there is instrumental noise uncertainty as it could be without those uncertainties and there is a similar increase in the parameter measurement uncertainty at fixed background energy density. 
%The top plot of \cref{fig:power_law_wr} shows that the ratio of the uncertainties in case of non perfect noise knowledge over perfect noise knowledge for the slope is of order 4.8 at the reference value and of order 8.7 for the log amplitude and of order 6.7 for $\alpha$  then it decreases for bigger amplitude, slope and $\alpha$ as expected since in these cases the impact of not knowing the noise is less evident.

%What we see looking at bottom plot of \cref{fig:power_law_wr} is that there are about 2 orders of magnitude between the error we compute when estimating the log amplitude with perfect noise knowledge with respect to the case of non perfect noise knowledge. Similar is the $\alpha$ estimation and a bit better is the slope estimation where the error is about 1 order of magnitude bigger in case of non perfect noise knowledge. 

%Differently from the power law case, we can notice that the error we made in estimating the log amplitude when take into account the noise knowledge uncertainties is bigger than 1 $\sigma$ at the reference value this means we cannot characterise this signal with the same confidence. Also the others parameters such as $n$ and $\alpha$ are poorly characterized. This means that we need a bigger amplitude to be able to characterize this model

%%%%%%
\subsubsection{Gaussian bump}
As for the power law, the Fisher matrix is a $120 \times 120$ as we have 2 signal parameters: the Gaussian width and the amplitude.
\begin{figure}
\includegraphics[width=0.45\textwidth]{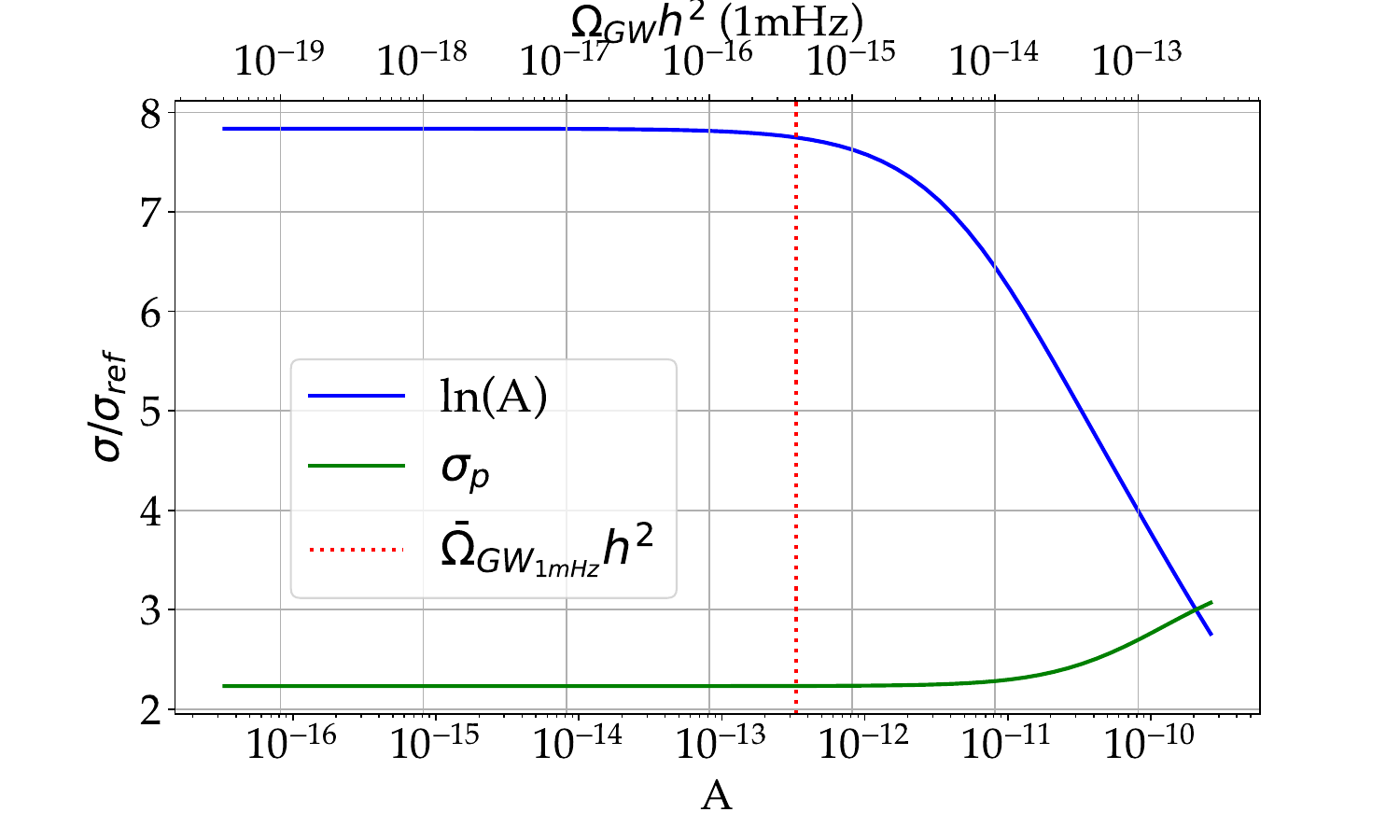}
\includegraphics[width=0.45\textwidth]{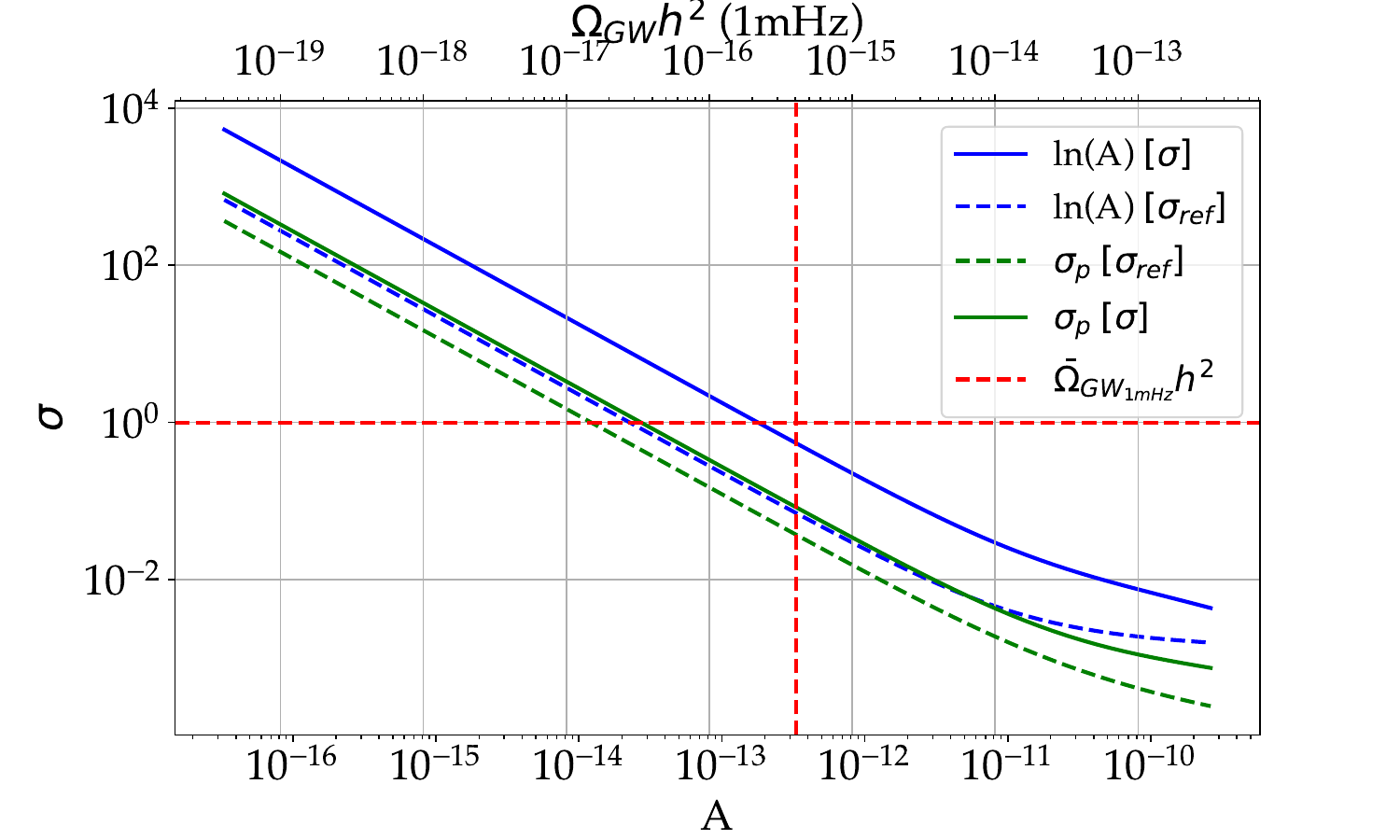}
\caption{\label{fig:gaussian2} As Figure~\ref{fig:power_law2}, but now for the Gaussian bump SGWB model % Gaussian bump model. Upper plot shows the ratio of the uncertainties of the log amplitude and Gaussian amplitude either versus the background amplitude or versus the background amplitude but expressed as the energy density at 1 mHz. Lower plot shows instead the same uncertainties either versus the  background amplitude or versus the background amplitude but expressed as the energy density at 1 mHz
}
\end{figure}
The results for this model are shown in Figure~\ref{fig:gaussian2}. In this case, the degradation in the precision of parameter measurement is a factor of $\sim2$--$8$ when allowing for lack of knowledge of the instrumental noise. This difference in behaviour is related to the different shapes of the SGWBs being considered. A Gaussian is more distinct from the spline model being used to represent the instrumental noise uncertainties than a power law, and hence the degree of confusion between the two models is less in this case. From the lower panel of Figure~\ref{fig:gaussian2}: we see that the energy density in a Gaussian bump SGWB has to be a factor of $\sim 10$ times higher for it to be characterised with the same measurement precision when there is instrumental noise uncertainty as it could be in the absence of those uncertainties. A Gaussian bump background at the reference amplitude would be detectable, and the width of such a Gaussian could be measured to a few tens of percent precision. This measurement precision improves approximately linearly with the background energy density.
%shows that the ratio of the uncertainties in case of non perfect noise knowledge over perfect noise knowledge for the width is of order 1.5 at the reference value and of order 1.8 for the log amplitude then it decreases for bigger amplitude and slope  as expected since in these cases the impact of not knowing the noise is less evident. Looking at this number we could deduce that for the Gaussian bump case not knowing the noise does not have an real impact on parameter estimation. If we then look at the bottom plot of \cref{fig:gaussian} we notice that there is about 1 order of magnitude difference between the error we compute when estimating the log amplitude with perfect noise knowledge with respect to the case of non perfect noise knowledge. Similar is for the slope estimation where the error is even less than 1 order of magnitude in case of non perfect noise knowledge.\\

%Although, similar to the case of power law with running we can notice that the error we made in estimating the log amplitude when take into account the noise knowledge uncertainties is bigger than 1 $\sigma$ at the reference value this means we cannot characterise this signal with enough confidence. This is probably linked with degeneracy with the reference signal model where we can estimate with good accuracy only one of the two parameters; and since we have a good estimate of the slope than we cannot equally well estimate  the amplitude.
%%%%%%%%%%%%%%%
\subsubsection{First order phase transition }
\begin{figure}
\includegraphics[width=0.45\textwidth]{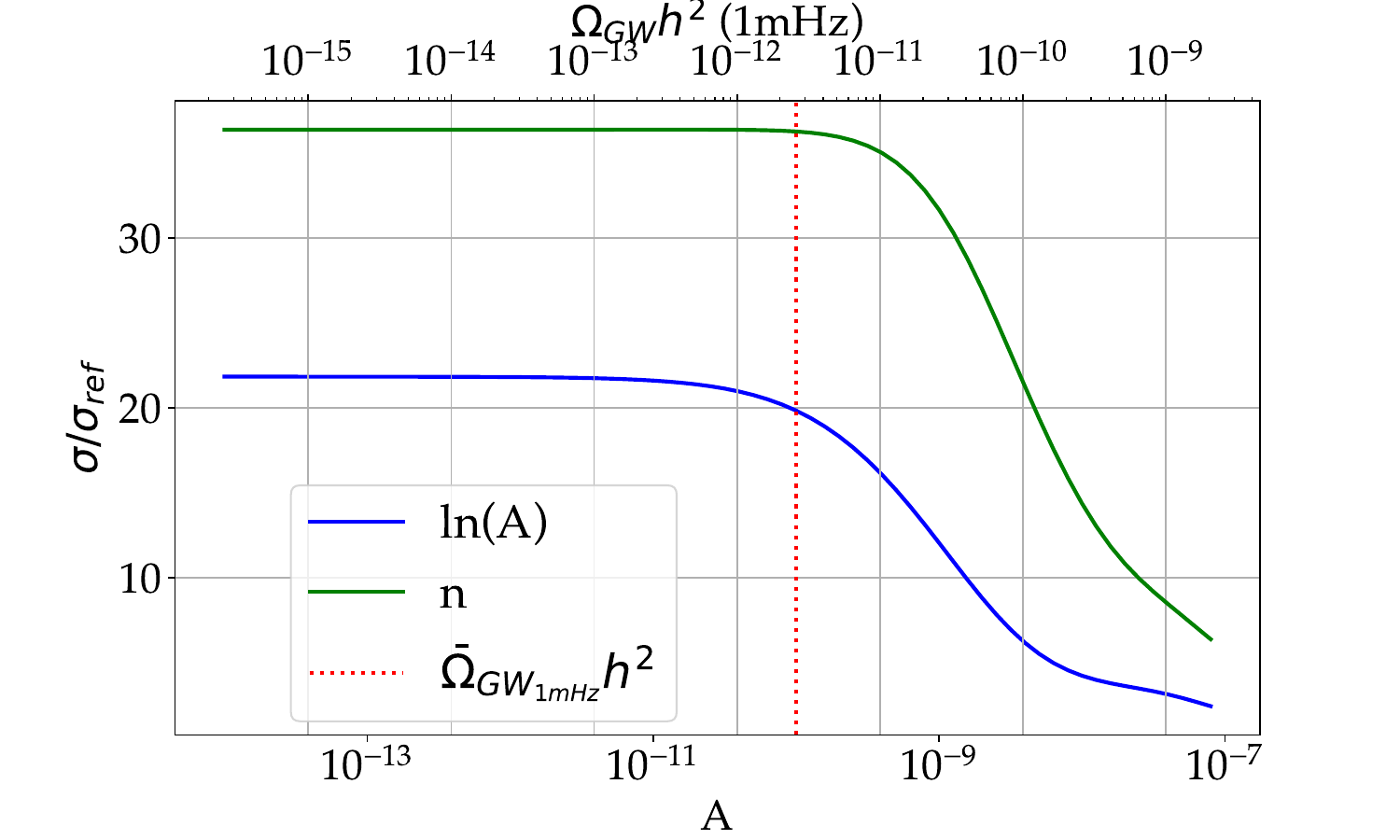}
\includegraphics[width=0.45\textwidth]{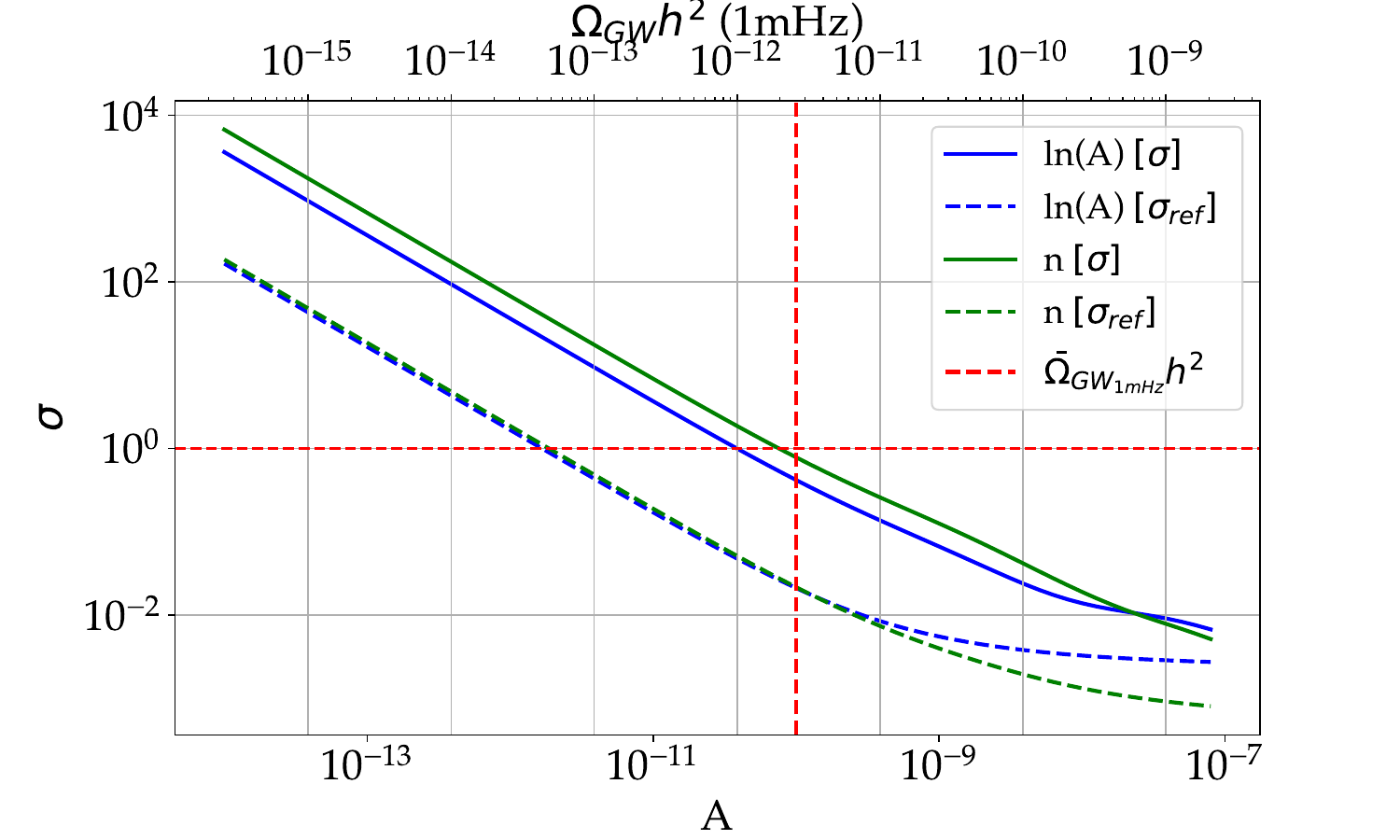}
\caption{\label{fig:FOPT2} As Figure~\ref{fig:power_law2}, but now for the first order phase transition SGWB model}
%First order phase transition model. Upper plot shows the ratio of the uncertainties of the amplitude and sloop either versus the background amplitude or versus the background amplitude but expressed as the energy density at 1 mHz. Lower plot shows instead the same uncertainties either versus the  background amplitude or versus the background amplitude but expressed as the energy density at 1 mHz 

\end{figure}

The FOPT model is again characterised by two parameters, an amplitude and a spectral index, and has a $120 \times 120$ Fisher matrix. The results for this model are shown in Figure~\ref{fig:FOPT2}. When allowing for instrumental noise knowledge uncertainties, the precision with which the SGWB log-energy density can be characterised degrades by a factor of $\sim 20$. The degradation in the determination of the spectral index is even larger, $\sim 35$. Once again, to achieve the same measurement precision, the background energy density would have to be $\sim 20$ times larger than it would need to be in the absence of noise knowledge uncertainties. Nonetheless, a FOPT background at the reference amplitude would still be detectable and provide a measurement of the spectral index at the level of $\sim \pm 0.8$.

%Comparing the behaviour of this model with the previous ones shows that for the FOPT with respect to the power law, including noise knowledge uncertainties, determines a bigger deterioration in estimating the slope of about order 7, whereas for the log amplitude is of order 4. If we then look at the bottom plot of \cref{fig:FOPT} we can confirm the above statement as it is possible to notice that there is about 2 order of magnitude difference between the error we compute when estimating the slope with perfect noise knowledge with respect to the case of non perfect noise knowledge. Similar is for the log amplitude estimation where the error is even less than 1 order of magnitude in case of non perfect noise knowledge. However differently from the Gaussian bump we are still able to characterize this signal at the reference value. \\

%%%%%%%%%%%%%%%%%%%%%%
%\subsubsection{Computation of the signal to noise ratio}\label{subsec:SNR}

\begin{figure}
\centering 
\includegraphics[width=0.45\textwidth]{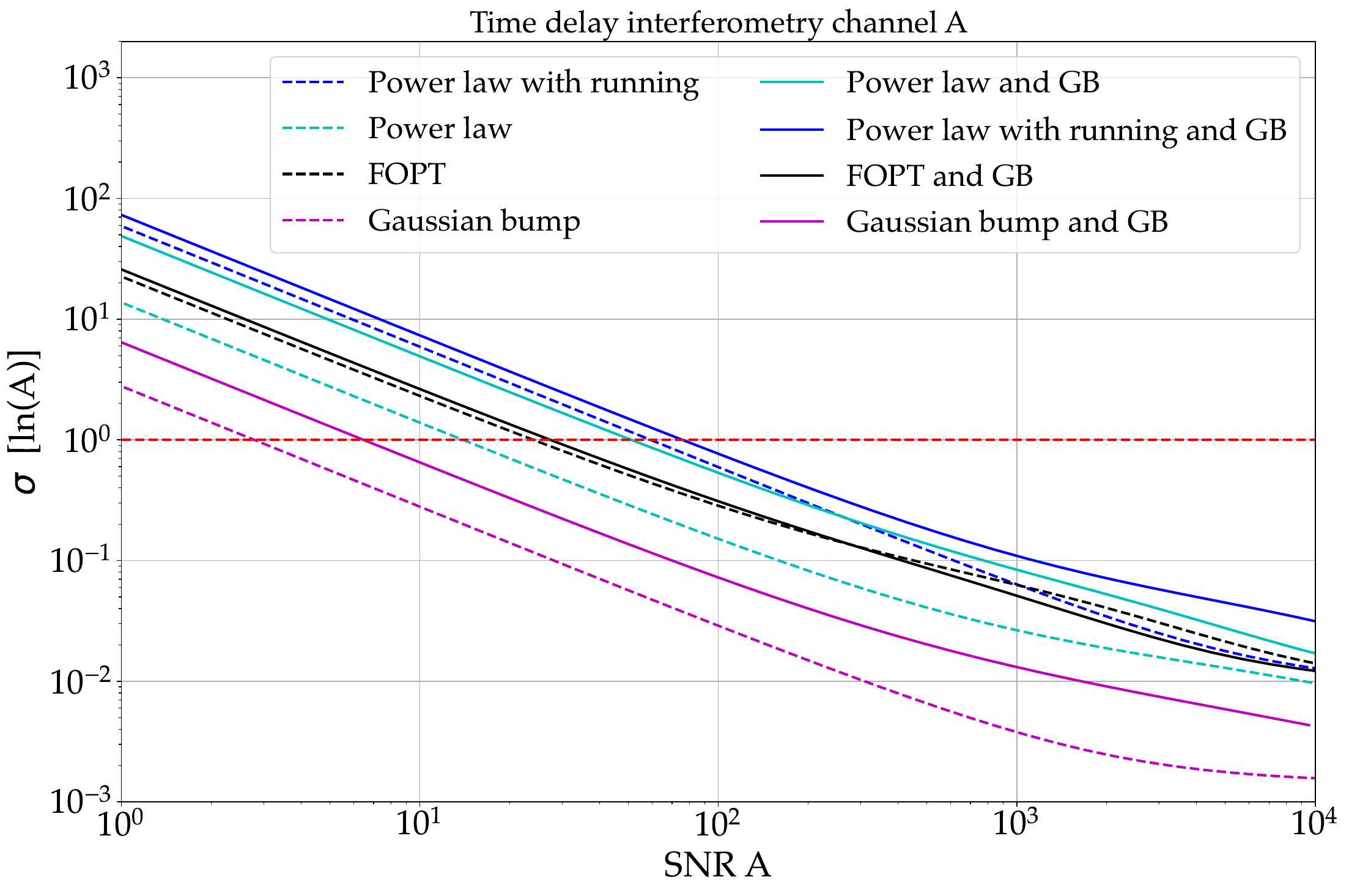}
\caption{Signal to noise ratio of four different SGWB signals in the TDI channel A,  Power law, Power law-with-running, Gaussian bump and First order phase transition, versus the error in log amplitude both considering (continuous lines) or not (dashed lines) the presence foreground \label{fig:snr2} }
\end{figure}

The previous results were computed considering the presence of the galactic foreground. In the appendix \ref{sec:resuls no GB} we report similar results, computed without taking into consideration the Galactic foreground. Redoing these analyses ignoring the foreground we do not see big differences in the uncertainty ratio, nor in the absolute uncertainties, when these are compared at fixed SNR, i.e., when the signal-to-noise ratio is recomputed without the galactic binaries included in the spectral density. To illustrate this, we show in Fig.\ref{fig:snr2} the precision of the measurement of the log-energy density of the background, as a function of the SNR in TDI channel A, for all SGWB models and both including and not including the galactic binary foreground. We see that the uncertainty is typically larger when the foreground is present, but this is typically less than a factor of a few. The Gaussian bump and power law background are most affected, with the uncertainty at fixed SNR and the SNR required for detection both decreasing by a factor of a few when the galactic binary background is removed from the spectral density. For the Gaussian bump, the uncertainty decreases by a factor of a little more than two when the Galactic background is excluded, and the SNR needed to reach the $\Delta \ln(A) < 1$ threshold for detection decreases by a similar factor. For the power law, the uncertainty decreases by about a factor of $4$, and the $\Delta \ln(A) < 1$ threshold required for detection is reached at an SNR that is a factor of $\sim 4$ smaller. For the power law with running and the FOPT backgrounds, the uncertainty at fixed SNR is almost unchanged, and the threshold SNR for detection is within a factor of $1.5$ and $2$, respectively.\\

%The FOPT background is most affected, with the uncertainty at fixed SNR and the SNR required for detection both increasing by about a factor of three.
This behaviour can be understood by looking at the shapes of the various SGWBs in Figure~\ref{fig:sgwbnoisepw}. Figure~\ref{fig:snr_omega} demonstrates that the removal of the foreground does not affect the SNR very much. The only SGWB that shows a significant change is the FOPT, for which most of the power is at frequencies where the foreground is significant. However, in the region around $300 \mu$Hz, where the majority of the SNR is generated, the shape of the FOPT is very different to the foreground. This is also true for the power-law-with running model around $5$mHz, where the majority of its SNR is generated. The power-law model, on the other hand, is quite parallel to the foreground at low frequency, and the Gaussian bump is quite parallel to the foreground at a few mHz. This most likely explains why the latter two backgrounds are more difficult to distinguish from a galactic foreground, and therefore more affected by its inclusion.

%background is mostly at lower frequencies than the GB background (see  \cref{fig:sgwbnoisepw}), at fixed SNR this background is more brighter with respect to the other models, and so the relative contribution from the portion of the background that provides information about the parameters increases.  Indeed if we consider together this plot and the one of  \cref{fig:snr_omega} we see that for a fixed background amplitude, rather than fixed SNR, we would have a worse measurement of the amplitude in the presence of the background. Moreover, we see that the power law with running is less affected than the power law. Looking again at  \cref{fig:sgwbnoisepw}, we can say that the case with running has proportionally more power at higher frequency, and this is probably where the information about the parameters is coming from but this is at frequencies above the GB foreground and this is way it is less affected by it.
 
\subsection{Setting a noise knowledge requirement}\label{sec:priordep}
In this section we will now explore how the amount of uncertainty in the instrumental noise impacts the results. In practice we will not be completely ignorant of the instrumental noise. Measurements on-board the satellites will provide an indication of the size of certain noise components. In principle, it might therefore be possible to place a requirement on how well the instrumental noise must be known in order to not degrade the science output of the mission. To assess this, we will recompute the results while changing the variance of the Gaussian prior on the instrumental noise spline parameters. We will vary the prior on the spline weights from very small values ($\log_{10}(\sigma_{inst}) = -10$), representing near-perfect knowledge of the noise, to very high values $\log_{10}(\sigma_{inst}) = 6$, representing no knowledge of the noise. \\

We fix the amplitude of the background for each SGWB model, so that it corresponds to an SNR of $\sim 120-140$ in each case~\footnote{Specifically the SNR of the power law with running is 135, the SNR of the power law is 138, the SNR of the FOPT is 142 and the SNR of the Gaussian bump is 120.}. This choice of SNR was motivated by Figure~\ref{fig:snr2}, which shows that an SNR greater than 100 is required in order to ensure all types of SGWB are detectable. For the case of the power law we also show results with the amplitude set to the reference energy density, which shows that the exact choice of background amplitude does not make a significant difference to the qualitative behaviour, only to the absolute value of the uncertainty.

\begin{figure}
\includegraphics[width=0.45\textwidth]{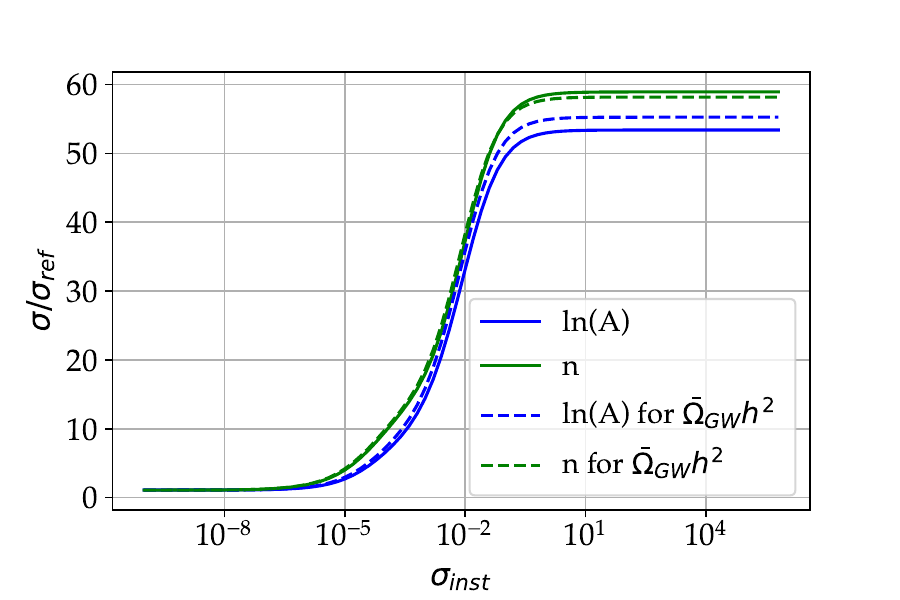}
\includegraphics[width=0.45\textwidth]{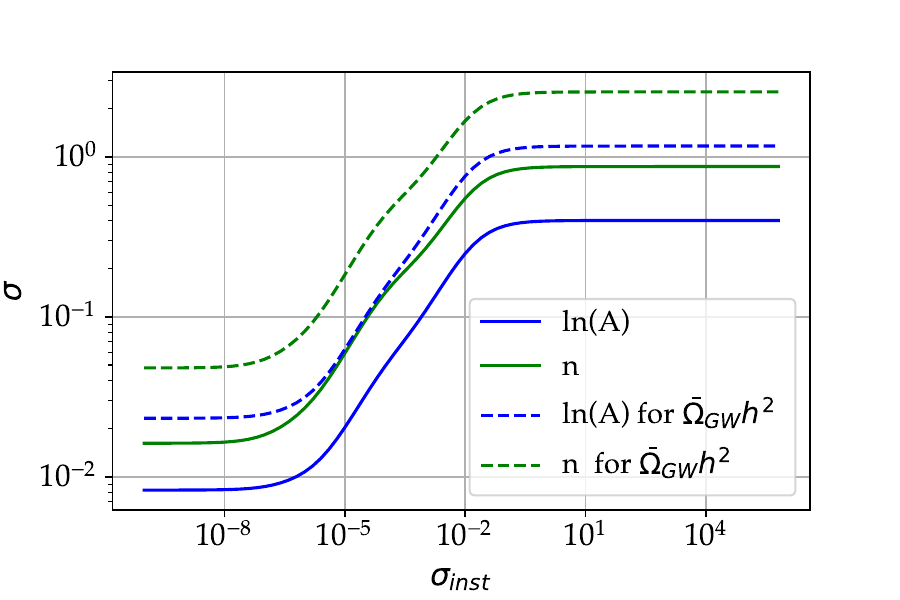}
\caption{\label{fig:PL_prior2} As Figure~\ref{fig:power_law2}, but now for fixed background amplitude and varying the variance of the Gaussian prior on the instrumental noise spline model. This plot is for a power law background, and the amplitude has been fixed such that the SNR in TDI channel A is 138 (continous line) and 43.7 (dashed lines). 
%Results for the power law model. Upper plot shows the ratio of the uncertainties of the amplitude and slope versus prior uncertainty in the instrumental noise. Lower plot shows the uncertainties of the amplitude and slope versus prior uncertainty in the instrumental noise.
}
\end{figure}

\begin{figure}
\includegraphics[width=0.45\textwidth]{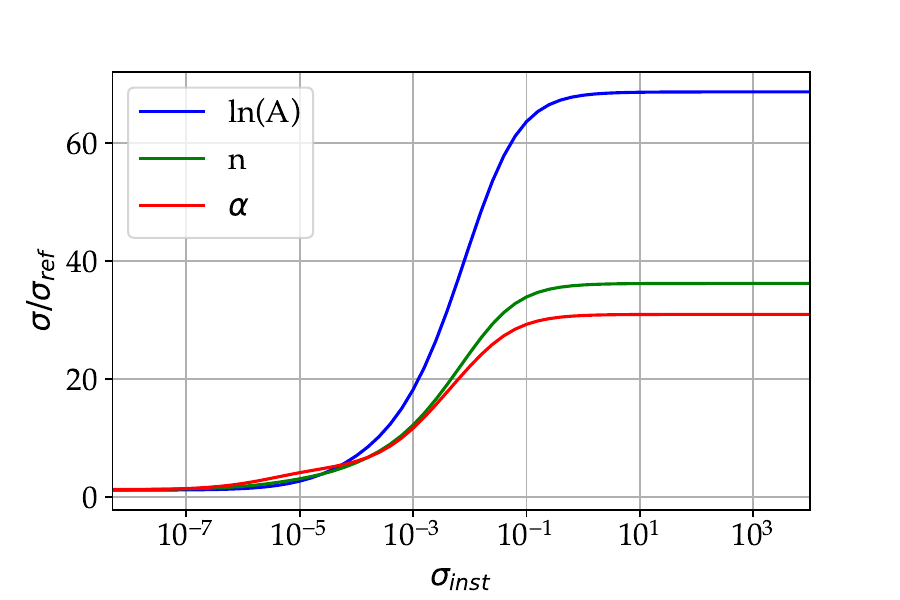}
\includegraphics[width=0.45\textwidth]{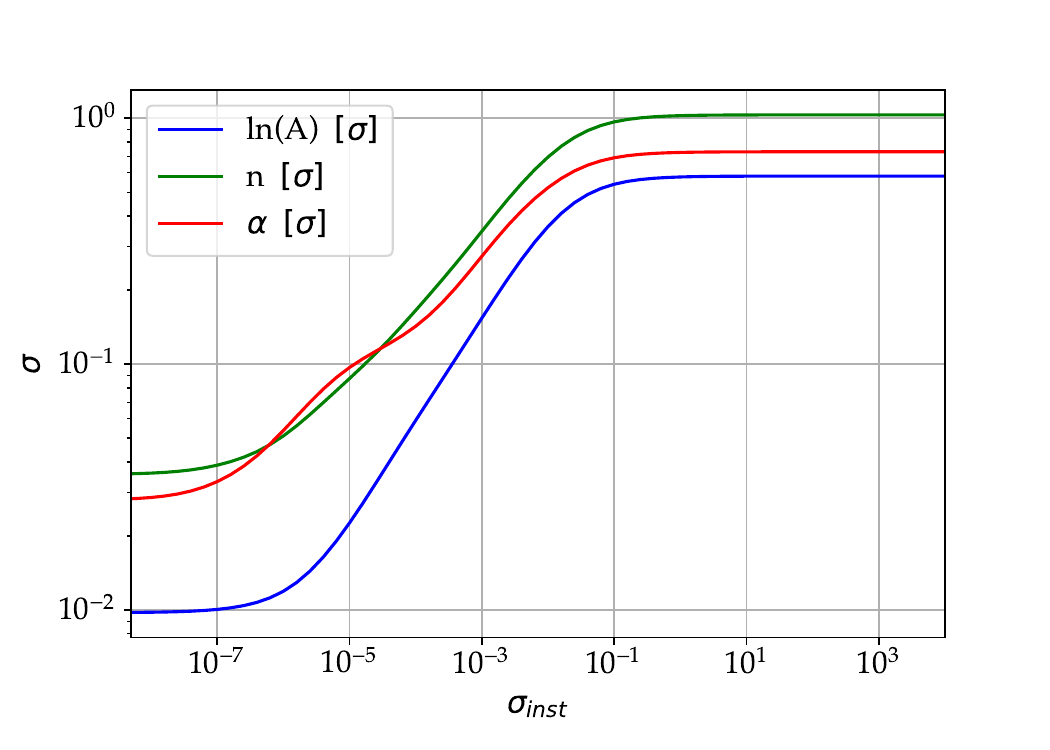}
\caption{\label{fig:PL_wr_prior2} As Figure~\cref{fig:PL_prior2} but now for the power-law-with-running model. The background amplitude has been fixed to give an overall SNR of $135$.
%Power law with running model.  Upper plot shows the ratio of the uncertainties of the amplitude, sloop and $\alpha$ versus prior uncertainty in the instrumental noise. Lower plot shows the uncertainties of the amplitude, sloop and $\alpha$ versus prior uncertainty in the instrumental noise
}
\end{figure}

\begin{figure}
\includegraphics[width=0.45\textwidth]{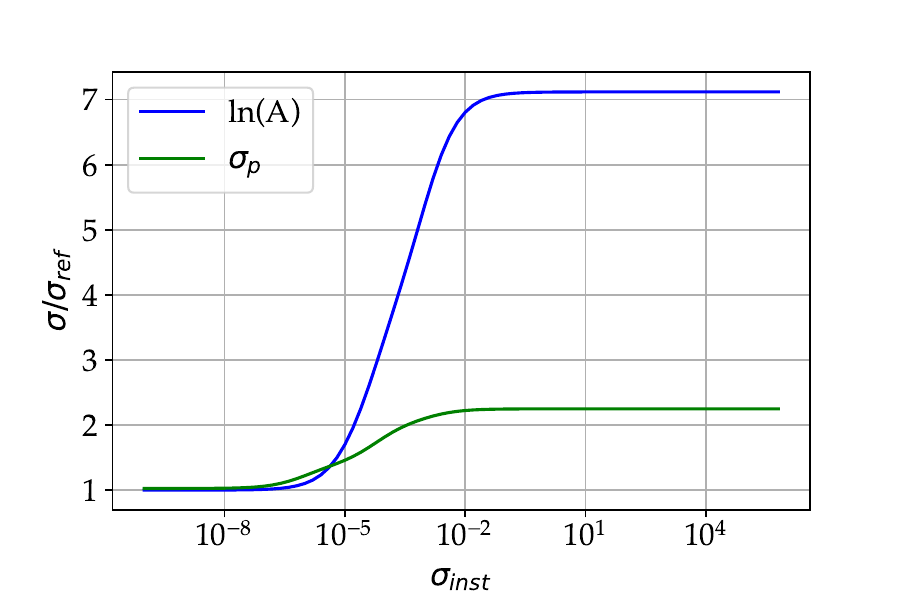}
\includegraphics[width=0.45\textwidth]{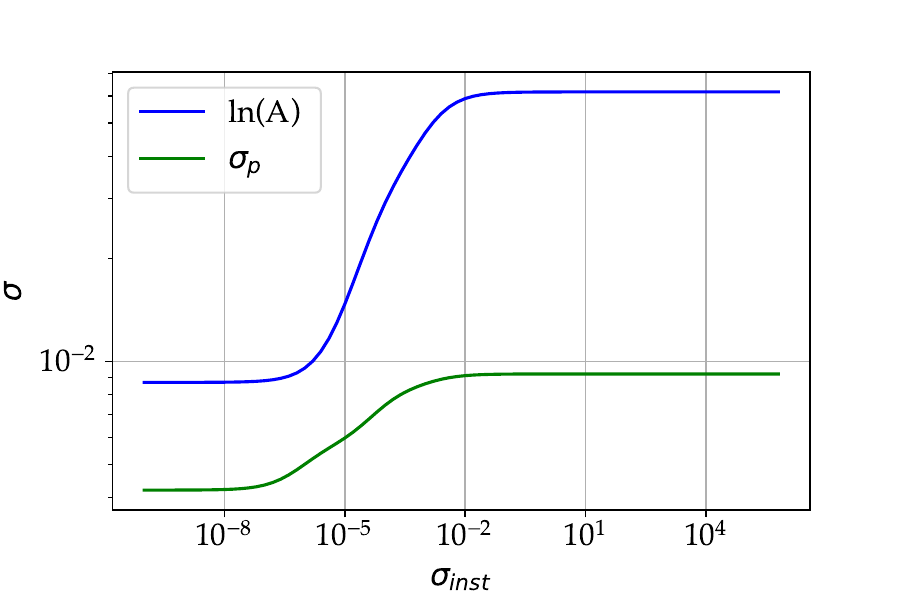}
\caption{\label{fig:gaussian_prior2} As Figure~\cref{fig:PL_prior2} but now for the Gaussian bump model. The background amplitude has been fixed to give an overall SNR of $122$.
%Gaussian bump model.  Upper plot shows the ratio of the uncertainties of the amplitude and Gaussian amplitude versus prior uncertainty in the instrumental noise. Lower plot shows the uncertainties of the amplitude and Gaussian amplitude versus prior uncertainty in the instrumental noise 
}
\end{figure}

\begin{figure}
\includegraphics[width=0.45\textwidth]{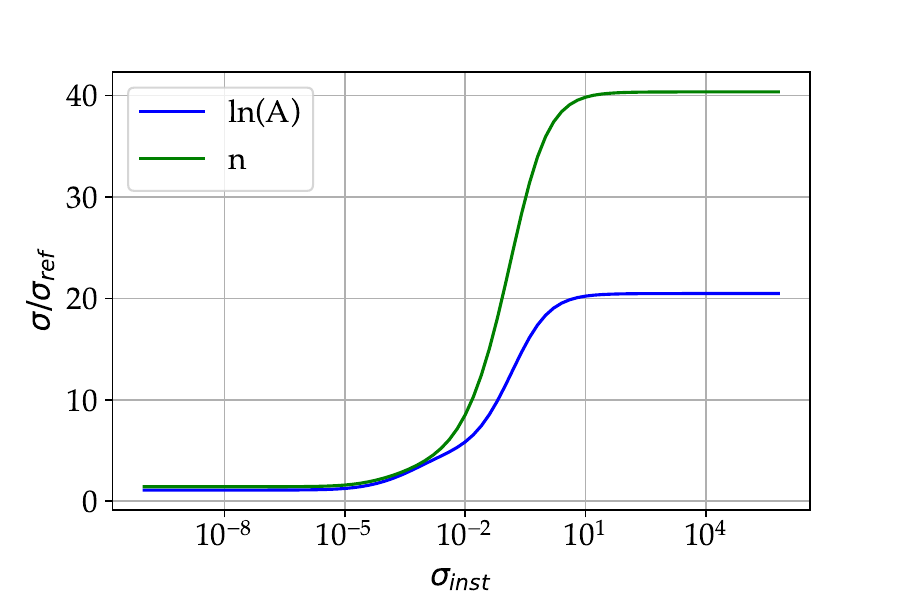}
\includegraphics[width=0.45\textwidth]{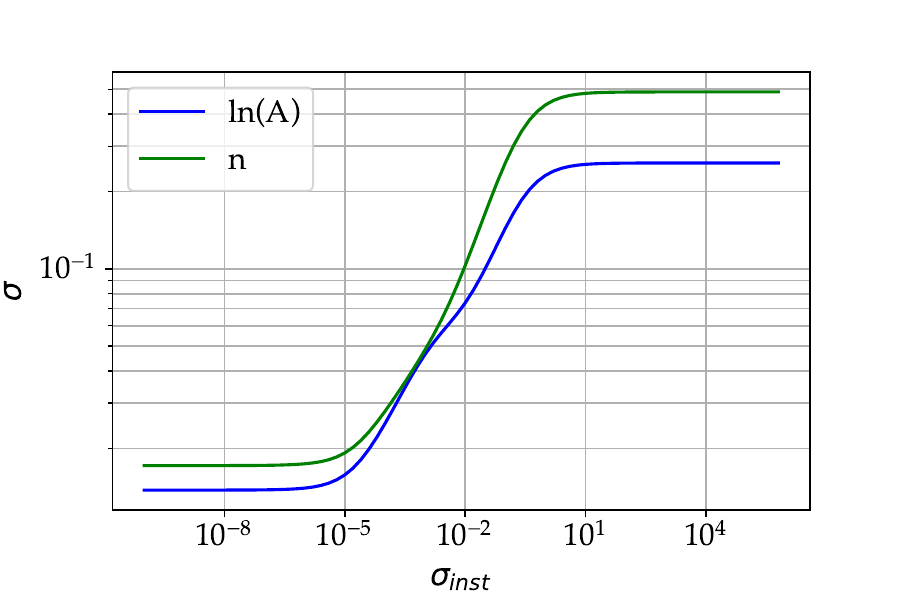}
\caption{\label{fig:FOPT_prior2} As Figure~\cref{fig:PL_prior2} but now for the FOPT model. The background amplitude has been fixed to give an overall SNR of $142$.
%FOPT law model. Upper plot shows the ratio of the uncertainties of the log amplitude and slope versus prior uncertainty in the instrumental noise. Lower plot shows the uncertainties of the log amplitude and slope versus prior uncertainty in the instrumental noise
}
\end{figure}

For all SGWB models we again present the results in two different ways --- as a ratio of the SGWB parameter measurement uncertainties when instrumental noise uncertainties are considered to those assuming perfect noise knowledge, and as the absolute measurement uncertainty. Results for the power law model are shown in Figure~\ref{fig:PL_prior2}, for the Power-law-with-running model in Figure~\ref{fig:PL_wr_prior2}, for the Gaussian bump model in Figure~\ref{fig:gaussian_prior2} and for the FOPT in Figure~\ref{fig:FOPT_prior2}.
%are included on the signal's parameters as a function of the magnitude of the allowed prior uncertainty in the instrumental noise, and the second one shows again the uncertainties on the signal's parameters as a function of the magnitude of the allowed prior uncertainty in the instrumental noise. 
%%%%%%%%%%%%%%%%%%%%%%%%%%
The results for all four SGWB models are qualitatively similar. For very low prior uncertainties the ratio of the uncertainties tends to unity. This is expected as this limit corresponds to the limit in which the instrumental noise is perfectly known. As the prior uncertainty is increased beyond $\sim 10^{-6}$ the measurement precision in the presence of noise knowledge uncertainties starts to increase. When the noise knowledge uncertainty reaches $\sim 10^{-2}$ for power law and Gaussian bump, $\sim10^{-1}$ for power law-with-running and 10 for FOPT, the measurement precision ratio saturates. This final value reflects the expected uncertainty in the absence of any noise knowledge. The results given in Section~\ref{sec:ampdep} were all computed in this regime.

The main conclusion from these results is that if we wanted to ensure that there was no degradation in LISA science due to lack of noise knowledge, the necessary requirement on the noise knowledge would be $<<10\%$.  In the LISA Pathfinder mission, which was designed to accurately characterize the free-fall performance of test masses in a space-based environment, the observed noise could only be explained within some margin: the physical origin of the measured sub-mHz acceleration is only partially understood, as more than 50\% of its PSD is still unmodeled \cite{LorePhD,lpf_noiseperf3}.\\

It is therefore unrealistic to expect that a noise requirement at the $\sim1$--$10\%$ level could be met. At noise uncertainties above this threshold, there is little difference between some and no noise-knowledge, at least within the model for instrumental noise variations considered here. We conclude that no useful and achievable noise knowledge requirement could be implemented in practice.

While we will not be able to achieve the precision that would be possible under ideal circumstances, it is important to emphasise that this does not mean we will not be able to detect and characterise modelled SGWBs. In all cases, at SNR of $\sim >100$, the amplitude can be constrained to a few tens of percent, even without any knowledge of the  instrumental noise. 

%Looking at \cref{fig:PL_prior} we see that when we have a good knowledge of our instrument thus $\sigma_{inst}$ is small we have smaller error in the estimation of the slope and amplitude. The errors increase with the increasing of uncertainties in the noise knowledge up to $\sigma_{inst} = 10^{-3}$ where we see it becomes constant to $\sigma \approx \times 10^{-1}$ for the slope and to $\sigma = 5 \times 10^{-1}$ for the log amplitude. This means that we need noise uncertainty accuracy smaller than $0.001\%$ to be able to estimate a power law signal with SNR of 136 with  $\sigma <  10^{-1}$ error in the amplitude. Also if we consider the reference energy density at $\sigma_{inst} > 10^{-3}$ we reach a constant error both for the amplitude and for the slope, but as expected given the law SNR we are able to estimate the power law signal amplitude with less accuracy.
%%%%%%%%%%%%
%%%%%%%%%%%
%Looking at \cref{fig:PL_wr_prior} we again see that when we have a good knowledge of our instrument thus $\sigma_{inst}$ is small we have smaller error in the estimation of the slope and log amplitude. The errors increase with the increasing of uncertainties in the noise knowledge up to $\sigma_{inst} = 10^{-2}$ where we see it becomes constant to $\sigma = 4 \times 10^{1}$ for the  log amplitude and  to $\sigma = 7 \times 10^{-1}$  for $\alpha$ , and up to $\sigma_{inst} = 10^{-3}$ for the slope where it becomes constant to  $4 \times 10^{2}$ . This means that we should know the noise with an accuracy better than $0.01\%$ to be able to estimate a power law with running signal with SNR of 145 with  $\sigma < 5 \times 10^{-1}$ error in the log amplitude.

%%%%%%%%%%%%
%%%%%%%%%%%%%
%Looking at \cref{fig:gaussian_prior} we again see that when we have a good knowledge of our instrument thus $\sigma_{inst}$ is small we have smaller error in the estimation of the Gaussian amplitude and log amplitude. The errors increase with the increasing of uncertainties in the noise knowledge up to $\sigma_{inst} = 10^{-3}$ where we see it becomes constant to $\sigma = 2 \times 10^{-2}$ for the amplitude and to $\sigma = 6 \times 10^{-3}$ for the $\sigma$. This means that we should know the noise with an accuracy better than $0.001\%$ to be able to estimate a Gaussian bump signal with SNR of 135 with  $\sigma < 2 \times 10^{-2}$ error in the log amplitude.

%Looking at \cref{fig:FOPT_prior} we again see that when we have a good knowledge of our instrument thus $\sigma_{inst}$ is small we have smaller error in the estimation of the slope and log amplitude. The errors increase with the increasing of uncertainties in the noise knowledge up to $\sigma_{inst} = 10^{-1}$ in this case where we see it becomes constant to $\sigma = 2 \times 10^{-1}$ the  log amplitude and to $\sigma = 5 \times 10^{-1}$ the slope. This means that we should know the noise with an accuracy better than $0.1\%$ to be able to estimate a FOPT signal with SNR of 149 with  $\sigma < 2 \times 10^{-2}$ error in the log amplitude.
%\todo[inline]{strange behavior for the FOPT with respect to the others...is this expected?}

%%%%%%%%%%%%%%%
\subsection{Signal reconstruction}\label{sec:sigrecon}

\begin{figure}
\centering
\includegraphics[width=0.4\textwidth]{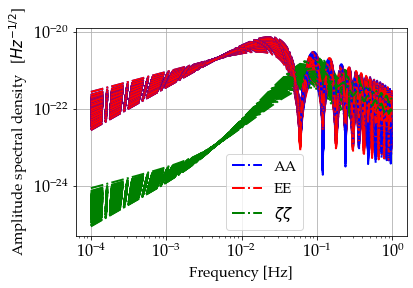}
\includegraphics[width=0.4\textwidth]{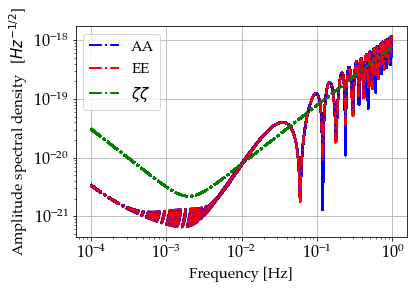}
\includegraphics[width=0.4\textwidth]{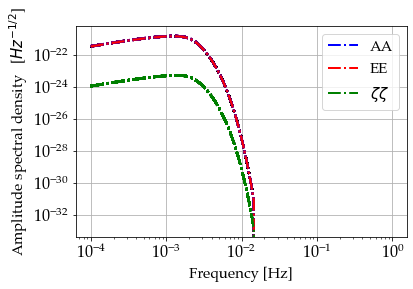}
\includegraphics[width=0.4\textwidth]{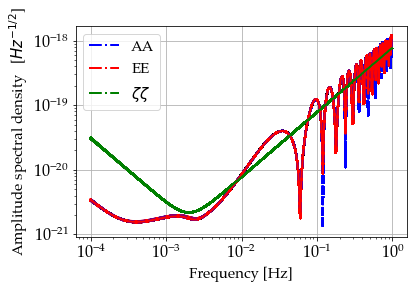}
\caption{\label{fig:rec_GB} We show the Power law signal, Galactic binaries and noise ASDs corresponding to random draws from the posterior, approximated using the Fisher matrix as described in the text. In each panel the curves correspond to the three TDI channels: A (blue), E (red) and $\zeta$ (green).  Upper first panel: reconstructed SGWB; upper second panel: reconstructed TM and OMS instrumental noise;  middle panel: reconstructed Galactic binaries; lower panel: total reconstructed ASD (signal + noise + GB).}
\end{figure}

To finish this section we will use our Fisher matrix results to illustrate how well we can reconstruct the Power law, the foreground, and the instrumental noise. To do this, we will approximate the posterior distribution on the model parameters using a multi-variate Gaussian with covariance matrix equal to the inverse of the Fisher matrix. We can then take random draws from this fake\footnote{Even if the measurements $x$ are Gaussian the posterior is not because is a function of $\theta$. A Gaussian with unknown variance $\sigma$ is not a Gaussian on that variance, $p(x | \theta) \propto \frac{1}{\sigma}\exp ( - \frac{x^2}{2 \sigma^2} )$, is a Gaussian in $x$, but not in $\sigma$} posterior distribution and plot the PSD of the SGWB, the foreground and instrumental noise corresponding to the drawn parameters. 
%Here we compare the uncertainty on the PSD as a function of frequency by taking random draws from the Fisher matrix posterior (i.e. a multivariate Gaussian with covariance equal to the inverse of the Fisher matrix) and then plotting the resulting PSD.  
In \cref{fig:rec_GB} we follow this procedure for a power law signal with an SNR of 872. The choice of the SNR is driven to avoid the breakdown of the Fisher matrix approximation for the SGWB parameters, because the SGWB parameter uncertainties are large and no longer in the linear signal regime as explained and, shown in details, in Appendix \ref{app:signa_rec}.\\

The three panels show the reconstructed ASDs for the Power law, for the foreground and for the instrumental noise and the total, which is the sum of the three.
%and we try to reconstruct either both the signal and noise PSD togheter or only the signal PSD or only the noise PSD. In \cref{fig:rec2} we inject a signal with bigger SNR and we perform the same exercise. \\

What we would expect is that our ability to measure the total spectral density is roughly independent of the relative  amplitudes of the two components, since this is what we actually see and measure in the data. Our model attempts to split that measurement into constituent components. If one of those components is much weaker than the other we would not expect to recover it as well as when the components are making comparable contributions to the data. Top panel of Fig. \ref{fig:rec_GB} (and the Figure~\ref{fig:rec} in the appendix \ref{app:signa_rec}) is consistent with this expectation. We see indeed in Fig. \ref{fig:rec_GB} that the galactic binaries are well recovered.
%\jg{Add a comment if we include the theoretical uncertainty on the direct measurement of the PSD in this figure.}
 Moreover, the only noise reconstruction for the TDI A and E suffers from the presence of the GW background and foreground in the regime $0.4mHz$ -- $4mHz$ where these two signals have the majority of power.  As a final point, it is clear that the SGWB is being best constrained around a frequency of 4mHz where the power of the GB is less and the uncertainty in the instrumental noise is also largest (although still small) at this point. This can be understood from Figure~\ref{fig:sgwbnoisepw}, which shows that the power law background is closest to the instrumental noise ASD at that frequency and the Galactic binaries pick at 1mHz,  and so this frequency range dominates the SNR in the signal. We expect to be able to measure the background best in the frequency range where it is most dominant relative to the instrumental noise and distinguishable from the galactic binaries.

%%%%%%%%%%%%

\section{Discussion and conclusion}\label{sec:nine}

%\todo[inline]{check the conclusion, insert comments also in case we include the priors and on future activity. I did not have time to write this section properly.. }
We have explored the impact of noise knowledge uncertainty on measuring the parameters of various modelled stochastic gravitational wave backgrounds. This was done by modelling instrumental noise uncertainties using cubic splines to represent deviations away from the design PSDs and CSDs for the three TDI channels $A$, $E$ and $\zeta$. We then used a Fisher matrix analysis to evaluate the expected uncertainties in the measurements of the model parameters when fitting a model including the instrumental noise uncertainties and compared it to fitting a model without those uncertainties. The degree of uncertainty was characterized by including a Gaussian prior on the instrumental noise parameters, allowing us to quantify the impact of imposing a requirement on our noise knowledge.

%%%%%%

 %%%%%
 %The increase in uncertainty varies according to the model and in particular to the number of knots we used.
 
This analysis showed that, for all SGWB models, allowing for instrumental noise uncertainties leads to a significant increase in the uncertainty in our measurements of the background parameters. The increase in uncertainty was a factor of $2-8$ for the Gaussian bump model, which reduces to $2-4$ when not including GB as foreground, $55-60$ for the power law ($15-30$ without GB as foreground), $20-35$ for the First order phase transition ($20-50$ without GB as foreground) and $30-75$ for the power law with running ($20-75$ without GB as foreground).
%Using the requirement that the uncertainty in log-amplitude is less than unity as a proxy for detectability, 
These increased uncertainties correspond to the threshold background energy density required for detection increasing by a factor of $10$ ($5$ without GB) for the Gaussian bump model, a factor of $60$ with and without GB for the Power-with-running and a factor of $20$ with and without GB for all other models ($50$ for the power law when including GBs). The threshold energy density at 1mHz at which the backgrounds start to be detectable are $4\times 10^{-13}$, $4\times 10^{-13}$, $2 \times 10^{-16}$ and $10^{-12}$
($10^{-13}$, $2.5\times 10^{-13}$, $8 \times 10^{-17}$ and $5 \times 10^{-13}$ if we do not include the GB foreground) for the power law, power law with running, Gaussian bump and FOPT models respectively. Comparing these to the reference
background amplitudes introduced in Section \ref{sec:sgwbmods} we see that the power law,  the Gaussian bump and FOPT backgrounds are detectable at the reference amplitudes, while the power law with running is not as the threshold energy density is $1.5$ times higher than the reference. However, for this latter background the amplitudes were specified based on the SNR and not on a physical model. The reference amplitudes for the power law and FOPT backgrounds are based on physical model predictions, so it is more important that these backgrounds are still detectable. We note that this result does depend on the particular choice of model we used for representing the instrumental noise uncertainty. If this model is made even more flexible, for example by increasing the number of spline knots used, the threshold would increase further and potentially also make the reference backgrounds undetectable.\\

When we vary our assumed level of knowledge of the instrumental noise, we find that the uncertainties on the SGWB parameters show a similar trend for all models, starting to degrade at relative small uncertainties, increasing and then saturating after a certain point. The point at which the sensitivity starts to degrade is when the uncertainty in the log-spectral density of the noise reaches $\sim 10^{-6}$--$10^{-5}$, depending on the SGWB model. %\LS{just to be precise: when the uncertainty in the power spectral density of the noise reaches $\sim 10^{-6}$--$10^{-5}$, right? the noise uncertainties have the same dimesnions of the PSDs.} 
The uncertainty saturates at log-spectral density uncertainties of $\sim 10^{-2}/10^{-1}/10^{-2}/10^{1}$ ($10^{-2}/10^{-1}/10^{-3}/10^{-1}$ in the absence of the GB foreground) for the power law/power-law with running/Gaussian bump/FOPT backgrounds respectively.
%when our uncertainty in the log-amplitude of the noise reaches $10^{-6}$ ($10^{-5}$ in the absence of the GB foreground) and saturates to its limiting value by the time this uncertainty reaches $10^{-2}$ for the power law, for the power-law-with running and Gaussian bump it starts degrading at $10^{-6}$ and saturates at $10^{-1}$ and $10^{-2}$ ($10^{-3}$ without GB) respectively, and it starts degrading at $10^{-5}$ ($10^{-6}$ without GB) and saturating at ($10^{-1}$ without GB) $10^{1}$  for the FOPT model. 
This means that if we wanted to limit the degradation in the science that arises from lack of noise knowledge we would have to impose a very stringent requirement on our knowledge of the noise. This is likely to be impossible to implement in practice, so we will have to accept that our ability to resolve SGWBs will not be as good as calculations that assume perfect noise knowledge predict.

%We see that these four SGWBs are mostly measurable, in particular the power law and the first order phase transition, when assuming that PSD variations are slowly varying, and we have a modelled SGWB. Conversely we measure with less precision the power law with running and the gaussian bump which show an error of about a factor 5 and 2  $\sigma$ respectivately, at the reference energy density.

%Generally speaking including noise knowledge uncertainties bring to a deterioration in the parameter estimation and this can be appreciated in all the plot but in presence of a modelled signal and sufficient SNR is still possible to recover the background. It is interesting to notice what happen when we vary the prior, in most case we notice that we need to know the noise better than 0.001, the only ecception is the FOPT where we required a knowledge better than 0.1 and this could be linked to the high SNR of the signal.. \todo[inline]{is this true? is the results correct?}

%It is important to underline that the models used here rely on assumption which could be proven to be not correct. So, claiming a detection of a SGWB is challanging. 
It is important to note that these results are based on some assumptions which might not hold in practice. In particular, we have considered only modelled SGWBs and we have assumed a particular form for variations in the PSD that forces variations to be smoothly varying as a function of frequency. If the number of knots is increased to obtain a more flexible instrumental noise model, with potentially faster variations of the PSD as a function of frequency,  we already see a degradation of 2 or 3 orders of magnitude in the estimation of the log-energy density. \\ It is the distinguishability of the models that allows us to measure the parameters of the SGWBs. In the extreme picture where we do not want to make any assumption at all about the form of the instrumental or SGWB spectral densities, then spectral separation will not be possible. We will be able to report measured power spectral density in all channels, and cross-spectral densities between them, and translate these into upper limits on the SGWB amplitude; but any interpretation of this as an actual detection will require independent confirmation from another detector~\cite{Muratore:2022nbh}. \\
%This is the . In that case, you can measure the PSD in each channel and the CSDs, but then that is all you know, since to make any other kind of assumption about how the PSD and CSD measurements relate relies specifying a model for the propagation of the noise through the instrument. Thus if we want to be agnostic on the noise model we can't go any further than characterising the total PSD and CSD in each channel. \\

All previous studies of the separation of instrumental noise and stochastic backgrounds have required assumptions: in \cite{Baghi:2023qnq} it was assumed that the instrumental uncertainty is a spline and that the SGWB has a power law spectrum; in \cite{Hartwig:2023pft} it was assumed that the instrumental noise is determined by 12 individual noise levels; and in this paper we are assuming something similar to \cite{Baghi:2023qnq}, although with a bit more flexibility, a wider variety of SGWB models and a different noise model for single satellite links. The SGWBinner \cite{Caprini_2019} is agnostic on the spectrum of the background but it can only work because it assumes a specific model for the instrumental noise. That is not going to be possible in practice. SGWBinner could be adapted to use a more flexible noise model, similar to the model used here, but the precision on the background recovery will be degraded. If we have a completely general instrumental noise model and a completely general SGWB model then we won't be able to separate them. In that case, the only hope would be that the SGWB is above the design sensitivity and we trust that the instrumental noise meets the mission requirements, in which case the best interpretation of such an observation would be a SGWB. However, even then an assumption would be made that the mission had met the design sensitivity requirements. An exploration of how our ability to separate instrumental noise as the spectral models of the SGWB and the instrumental noise are made more complicated should be the focus of future work.
% as well as introducing a more flexible noise models which allows mimic more features but without becoming degenerate with the signals.

\section{Acknowledgement}
M.M gratefully acknowledges the support of the German Space Agency, DLR.
    The work is supported by the Federal Ministry for Economic Affairs and Climate Action based on a resolution of the German Bundestag (Project Ref. No. FKZ 50 OQ 2301).  The authors thank Chiara Caprini for her review and suggestions regarding the use of SGWB templates in the text. Additionally, we acknowledge Mauro Pieroni for providing feedback on the presented results and the implementation of spline models. Furthermore, we thank Olaf Hartwig for discussions on the accurate computation of the GW transfer function and on the noise models and Marc Lilley for comparisons on the results obtained with the Fisher matrix formalism. We thank Lorenzo Sala for feedbacks on the recent analysis on the LISA Pathfinder performance results. The scientific discussions with Quentin Baghi, Jean Baptiste Bailey, Germano Nardini, Nikolaos Karnesis, Jesus Torrado, Nam Dam Quam, Henry Hinshauspe, Antoine Petiteau, and Riccardo Buscicchio have also been highly appreciated. Their feedback on the results and methodology presented in the paper has been valuable for the final outcome

%This will still rely on making some kind of assumptions about the scale over which the PSD varies. (ie. number of spline knots). 

%With the assumption made here of a slowly varying PSD uncertainties and that the noise is below the design target, it is possible to place constraints and we could detect a sufficiently bright background. the other thing that we should bear in mind is that we will have a good prediction for the background from SBHs from LIGO observations \cite{}. This gives us something to calibrate against in a certain frequency range. But if we can use this technique, the SOBH background has to be visible and so it must be above the noise/other backgrounds in the right range. 

%So the only way to use that information to provide sensitivity to other backgrounds is to make some assumptions about how the models vary with frequency. 

\appendix
\section{Likelihood derivation}\label{sec:appendix}

We derive the likelihood starting from the noise properties and explain why it takes the form shown in Sec.~\ref{sec:first}.
If we assume that the real time series $n(t)$ is a stationary, zero-mean, Gaussian and ergodic random process, then the Fourier transform of the noise $\tilde n_k = \tilde n(f_k)$ at each frequency $f_k$ is normally distributed with zero-mean and variance $\sigma_k ^2$. Thus, the natural log-likelihood at each frequency takes the form of a two-dimensional normal distribution
\begin{equation}
    \ln p(\tilde n _k) = -\frac{1}{2} \Big ( \frac{\Re [\tilde n_k]^2 }{ \sigma^2_k} + \frac{\Im [\tilde n_k]^2 }{ \sigma^2_k} \Big ) -\frac{1}{2} \ln [ (2 \pi \sigma_k ^2)^2 ]
\end{equation}
where we assumed that the real $\Re$ and imaginary $\Im$ part of the noise are not correlated and have the same variance. If we further assume that the variance of the noise at different frequencies follow a one-sided power spectral density $S_n(f)$ then :
\begin{align}
<\tilde{n}^{*}(f') \tilde{n}(f)> & = \frac{1}{2}S_n(f)\delta(f-f')\\
< \tilde{n}(f') \tilde{n}(f) > & = 0
\end{align}
where we used the expectation value of $<>$ over the data generating process. For a set of frequencies the first relation can be written as
\begin{align}\label{eqapp:variance_noise}
<\tilde{n}^{*}_k \tilde{n}_j> & = \frac{T}{2}S_n(f_k) \delta_{jk}
\end{align}
Therefore the variance of the real and imaginary part of the noise is given by
\begin{equation}\label{eqapp:variance_re_im_noise}
<\Re [\tilde{n}_k]^2>  = <\Im [\tilde{n}_k]^2> =  \frac{T}{4}S_n(f_k) = \sigma_k^2 \, .
\end{equation}
We can write the natural log-likelihood for all the measured frequencies as:
\begin{equation}\label{eqapp:loglike}
\sum _k \ln p(\tilde n_k) = - \sum_{k = 1}^{n} \ln[2 \pi \frac{T}{4} S_n(f_k)] - \frac{1}{2} \sum_{k = 1}^{n} \frac{|\tilde{n}(f)|^2}{\frac{T}{4} S_n(f_k)}
\end{equation}
which becomes in the continuum limit:
\begin{equation}\label{eqapp:loglike1}
\ln p(\tilde n) = - \ln\bigg{[}2 \pi \frac{1}{4} \det [S_n(f)\delta(f-f')] \bigg{]} - \frac{1}{2} 4 \int_{0}^{\infty} \frac{|\tilde{n}(f)|^2}{S_n(f)} \rm{d}f
\end{equation}
Note that in the continuum limit the variance of the noise can be thought as an operator. In fact, one can define the inner product
\begin{equation}
    (a(t)|b(t)) = 4 \Re \int_{0} ^{\infty} \int_{0} ^{\infty}  \tilde{a}^* (f) \, \Sigma^{-1}(f,f')\tilde b (f') \, \rm{d}f\, \rm{d}f'
\end{equation}
with $\Sigma^{-1} $ defined through the relation:
\begin{equation}\label{eq: int_sig}
    \int _0 ^\infty \Sigma^{-1}(f,f')\Sigma(f',f'') \rm{d}f' = \delta(f-f'') \, ,
\end{equation}
where if we set in Eq.\ref{eq: int_sig} that $\Sigma(f',f'')=\delta(f'-f'')S_n(f')$ we obtain
\begin{equation}
    \Sigma^{-1}(f,f'')S_n(f'') = \delta(f-f'') 
\end{equation}
and the inner product becomes
\begin{align}
    (a(t)|b(t)) &= 4 \Re \int_{0} ^{\infty} \int_{0} ^{\infty}  
    \frac{\tilde{a}^* (f) \, \delta(f-f')\tilde b (f')}{
    S_n(f')
    }
    \, \rm{d}f\, \rm{d}f' \nonumber \\
    &=4 \Re \int_{0} ^{\infty}
    \frac{\tilde{a}^* (f) \tilde b (f)}{
    S_n(f)
    }
    \, \rm{d}f
\end{align}

\section{Fisher matrix derivation}\label{app:fisher_derivation}
\subsection{Single detector}
To compute the Fisher matrix of Eq.~\ref{eq:fisher} we need the first derivative of the log-likelihood $l$ with respect to the parameters of the power spectral density.

Here we present the derivation of the Fisher matrix for the noise parameters $\vec\lambda$ affecting the one-sided spectral density $S_n(f|\vec\lambda)$, but this can be easily extended also including the gravitational wave background parameters $S_n (f|\vec \lambda)\rightarrow S_n (f|\vec \lambda) + S_{\rm GW} (f|\vec \theta)$. We differentiate the log-likelihood of equation~\ref{eqapp:loglike} with respect the parameters $\vec\lambda$:
\begin{align}
\frac{\partial l}{\partial \lambda^i} = & \sum_{k = 1}^{n} [  -  \frac{1}{S_n(f_k)} \frac{\partial S_n(f_k)}{\partial \lambda^i}  + \frac{1}{2}  \frac{|\tilde{n}(f_k)|^2}{\frac{T}{4} S_n(f_k)^2}  \frac{\partial S_n(f_k)}{\partial \lambda^i}  ], 
\end{align}
where we have omitted the dependency from $\vec\lambda$ to have a lighter notation. 
The second derivative of the likelihood is then
\begin{align}
\frac{\partial^2 l}{\partial \lambda^i \partial \lambda^j}  = &   \sum_{k = 1}^{n} [
\frac{1}{S^2_n(f_k)}\frac{\partial S_n(f_k)}{\partial \lambda^i}\frac{\partial S_n(f_k)}{\partial \lambda^j} \nonumber \\
&-  \frac{1}{S_n(f_k)} \frac{\partial^2 S_n(f_k)}{\partial \lambda^i \partial \lambda^j} \nonumber \\
&- \frac{1}{2} \frac{2 |\tilde{n}(f_k)|^2}{\frac{T}{4} S_n(f_k)^3} \frac{\partial   S_n(f_k)}{\partial \lambda^j}\frac{\partial S_n(f_k)}{\partial \lambda^i}\nonumber \\
&+ \frac{1}{2} \frac{|\tilde{n}(f_k)|^2}{\frac{T}{4} S_n(f_k)^2} \frac{\partial^2 S_n(f_k)}{\partial \lambda^j \partial \lambda^i}
]
\end{align}
using the definition of \cref{eqapp:variance_noise} the second and last term cancels and we get:
\begin{align}\label{fisher_single}
\Gamma_{ij} & =  \sum_{k = 1}^{n}  \frac{1}{S_n(f_k)^2} \frac{\partial  S_n(f_k)}{\partial \lambda^i } \frac{\partial  S_n(f_k)}{\partial \lambda^j} \, .
\end{align}
If we want to get the continuum limit we need to recast a factor of  $T \rm{d}f$:
\begin{equation}
\Gamma_{ij} = T \int_{0}^{\infty}   \frac{1}{S_n(f)^2} \frac{\partial  S_n(f)}{\partial \lambda^i } \frac{\partial  S_n(f)}{\partial \lambda^j} \rm{d}f    
\end{equation}

\subsection{Multiple detectors: real and imaginary part as separate random variables}\label{sec:fisher_detectors}
If we want to generalize our derivation to multiple detectors or channels we need to define the noise properties of each channel. For simplicity let us consider two channels $A$ and $E$ with 4 independent random variables $X(f_k) = X_k =  \{ \Re[\tilde X_k^A], \Im[\tilde X_k^A],\Re[\tilde X_k^E], \Im[\tilde X_k^E]\}$ at each frequency $f_k$. Since the final likelihood will be given by the product over all the frequencies, we consider only one frequency and we drop the subscript "$_k$". We can specify the spectral densities of each channel and the cross-spectral densities with:
\begin{subequations} \label{eqapp:conditions}
\begin{align}
<\tilde{X}^{c*}(f') \tilde{X}^c(f)> & = \frac{1}{2}S_c(f) \delta(f-f')\\
<\tilde{X}^{c}(f') \tilde{X}^c(f)> & = 0 \\
<\tilde{X}^{E*}(f') \tilde{X}^A(f)> & = \frac{1}{2}S^{*}_{AE}(f) \delta(f-f')\\
<\tilde{X}^{A*}(f') \tilde{X}^E(f)> & =\frac{1}{2}S_{AE}(f) \delta(f-f')\\
<\tilde{X}^{A}(f') \tilde{X}^E(f)> & = 0 
\end{align}
\end{subequations}
where the first two rows are valid for both channels $c=A,E$, and $S_c$ is real and $S_{AE}$ is complex.
From the above expression we can deduce:
\begin{subequations}
\begin{align}
<\Re[\tilde X_c]\Re[\tilde X_c]>  + <\Im[\tilde X_c]\Im[\tilde X_c]> & = \frac{S_{c}}{2}\\
<\Re[\tilde X_c]\Re[\tilde X_c] - \Im[\tilde X_c]\Im[\tilde X_c]> &=0\\
<\Re[\tilde X_c]\Im[\tilde X_c]>&=0\\
<\Re[\tilde X_A]\Re[\tilde X_E]>  + <\Im[\tilde X_A]\Im[\tilde X_E]> & = \frac{\Re[S_{AE}]}{2}\\
<\Re[\tilde X_A]\Re[\tilde X_E]>  - <\Im[\tilde X_A]\Im[\tilde X_E]> & = 0\\
<\Re[\tilde X_A]\Im[\tilde X_E]>  - <\Im[\tilde X_A]\Re[\tilde X_E]> & =\frac{ \Im[S_{AE}]}{2}\\
<\Re[\tilde X_A]\Im[\tilde X_E]>  + <\Im[\tilde X_A]\Re[\tilde X_E]> & =0 
\, ,
\end{align}
\end{subequations}
where in the last four rows $c=A,E$. Note that these are in total 10 independent conditions (3 equations for A, 3 equations for E and 4 equations for AE) that specify uniquely the 10 independent elements of a symmetric covariance matrix.\\
For a single frequency we can generalize the likelihood to two channels as:
\begin{equation} \label{Lik:im}
p(X) = \frac{1}{\sqrt{(2 \pi)^{2 \times N_c} \det(\Sigma)}} e^{-\frac{1}{2} X^T \Sigma^{-1} X  }
\end{equation}
where $N_c=2$ are the number of channels, X is a quadri-dimensional vector define above, $\Sigma$ is the multiple-channels covariance matrix:
\begin{equation}
\Sigma  = \left(
\begin{array}{cccc}
 \frac{{S_A}}{4} & 0 & \frac{\Re({S_{AE}})}{4} &
   \frac{\Im({S_{AE}})}{4} \\
 0 & \frac{{S_A}}{4} & \frac{\Im({S_{AE}})}{4} &
   \frac{\Re({S_{AE}})}{4} \\
 \frac{\Re({S_{AE}})}{4} & \frac{\Im({S_{AE}})}{4} & \frac{{S_E}}{4} &
   0 \\
 \frac{\Im({S_{AE}})}{4} & \frac{\Im({S_{AE}})}{4} & 0 &
   \frac{{S_E}}{4} \\
\end{array}
\right)
\label{Cov:im}
\end{equation}
where here each element is evaluated at the fixed frequency.
It can be shown that the expectation value of $X^T \Sigma^{-1} X$ equals the degrees of freedom, in this case 4. We have two channels, where each one has two degrees of freedom associated with the real and imaginary part of $\tilde X$.

We can then derive the Fisher matrix for the multiple channel case.
Taking the first derivative of the log-likelihood
\begin{align}
\frac{\partial \ln p(X) }{\partial \lambda ^i} & = -\frac{1}{2} \frac{1}{\det(\Sigma) } \frac{\partial \det(\Sigma)  }{\partial \lambda ^i} 
-\frac{1}{2} X^T \frac{\partial \Sigma^{-1}}{\partial \lambda ^i} X 
\end{align}
where we can use the following property of the determinant:
\begin{align}
\frac{ \partial  \det(\Sigma) }{\partial  \lambda^i } &= \det(\Sigma) \; \rm{Tr}[ \Sigma ^{-1}  \frac{ \partial \Sigma }{\partial  \lambda^i } ] \nonumber \\
&= \det(\Sigma) \; \Big [\Sigma ^{-1} \Big ]_{lm}  \Big [\frac{ \partial \Sigma }{\partial  \lambda^i } \Big ]^{ml}
\end{align}
to obtain
\begin{align}
\frac{\partial \ln p(X) }{\partial \lambda ^i} & = -\frac{1}{2} \Big [\Sigma ^{-1} \Big ]_{lm}  \Big [\frac{ \partial \Sigma }{\partial  \lambda^i } \Big ]^{ml} 
-\frac{1}{2} X^T \frac{\partial \Sigma^{-1}}{\partial \lambda ^i} X \, .
\end{align}
Then, the second derivative of the log-likelihood takes the form:
\begin{align}
\frac{\partial^2 \ln p(X) }{\partial \lambda^i \partial \lambda^j} & = - \frac{1}{2} \frac{\partial (\Sigma^{-1})^{lm}}{\partial \lambda^i} \frac{\partial \Sigma_{ml}}{\partial \lambda^j} \nonumber \\ &-
 \frac{1}{2} \Sigma_{lm}^{-1} \frac{\partial^2 \Sigma^{ml}}{\partial \lambda^i \partial \lambda^j}  
-\frac{1}{2}  X^T \frac{\partial^2 \Sigma^{-1}}{\partial \lambda^i \lambda^j} X
\end{align}
.

We can finally compute the Fisher matrix for a single frequency with:
\begin{align}
\Gamma_{ij} & = \frac{1}{2} \Big [\frac{\partial \Sigma_{lm}^{-1}}{\partial \lambda^i} \frac{\partial \Sigma^{ml}}{\partial \lambda^j} +
 \Sigma_{lm}^{-1} \frac{\partial^2 \Sigma^{ml}}{\partial \lambda^i \partial \lambda^j} + \Sigma_{ml} \frac{\partial^2 (\Sigma^{-1})^{lm} }{\partial \lambda^i \lambda^j} 
 \Big ],
\end{align} 
where we have considered $<X_l^T\frac{\partial^2 \Sigma_{lm}^{-1}}{\partial \lambda^i \lambda^j}X_m> =\frac{\partial^2 \Sigma_{lm}^{-1}}{\partial \lambda^i \lambda^j}\Sigma_{ml}$. If we use the property
\begin{equation}
    \frac{ \partial (\Sigma)^{-1}_{lm} }{\partial  \lambda } = - (\Sigma)^{-1}_{ln}  \frac{ \partial (\Sigma)_{nq}  }{\partial  \lambda } (\Sigma)^{-1}_{qm}
\end{equation}
we obtain the following expression: 
\begin{align}\label{eqapp:fisher_single_1}
\Gamma_{ij}  = & \frac{1}{2} \rm{Tr} \Big [ 
 - \Sigma^{-1}  \frac{ \partial \Sigma  }{\partial  \lambda^i } \Sigma^{-1} \frac{\partial \Sigma}{\partial \lambda^j} +  \Sigma^{-1} \frac{\partial^2 \Sigma}{\partial \lambda^i \partial \lambda^j}+  \Sigma \frac{\partial^2 \Sigma^{-1} }{\partial \lambda^i \partial \lambda^j}
 \Big ]
\end{align}
which can be further simplified if we use the following properties:
\begin{subequations}
\begin{align}
    \partial (\Sigma \Sigma^{-1}) = 0 \\ 
    \partial \Sigma \Sigma^{-1} + \Sigma \partial \Sigma^{-1} = 0 \\
    \partial^2 \Sigma \Sigma^{-1} + 2\partial \Sigma \partial\Sigma^{-1} + \Sigma \partial^2 \Sigma^{-1} = 0 \\
    \partial^2 \Sigma \Sigma^{-1} - 2\partial \Sigma \,  \Sigma^{-1} \, \partial \Sigma \, \Sigma^{-1} + \Sigma \partial^2 \Sigma^{-1} = 0 \label{sub6}
\end{align}
\end{subequations}
Note that in the above expression the first and last term of Eq.\ref{sub6} correspond to the last two terms in the Fisher matrix expression (\ref{eqapp:fisher_single_1}). The final expression for all frequencies can be easily obtained by taking the sum over all the frequencies:
\begin{equation}\label{fisher_2} 
\Gamma_{ij}  = \frac{1}{2}  \sum_{k=1}^{n}  [ (\Sigma_k ^{-1})_{lr } \frac{\partial \Sigma_k ^{rp}}{\partial \lambda^i}( \Sigma_k ^{-1})_{pm}\frac{\partial \Sigma_k ^{ml}}{\partial \lambda^j}  ] \, .
\end{equation}
Note that there is an additional factor of $1/2$ with respect to Eq. \ref{fisher_single}. If we insert only the first 2 columns and rows of $\Sigma$ we obtain the previous equation for the single channel as expected.

\subsection{Multiple detectors: complex random variables}
Equivalently, the likelihood can be written in terms of complex variables $\tilde X_A$ and $\tilde X_E$ \citep{complex_rnd_distr}
\begin{equation}
p(\tilde X_A,\tilde X_E) = \frac{e^{- [\tilde X_A,\tilde X_E]^{\rm H} \Sigma  ^{-1} [\tilde X_A,\tilde X_E] } }{(2 \pi)^{N_c} \det(\Sigma_k)}
\end{equation}
where "$^{\rm H}$" indicates the Hermitian conjugate, the factor of $1/2$ disappeared because it is a complex distribution and must match Eqs~(\ref{Lik:im})--(\ref{Cov:im}), and the new complex covariance matrix is defined as 
\begin{equation}
\Sigma  = \frac{1}{2}\begin{pmatrix}
 S_A &  S_{AE} \\
 S^*_{AE} &  S_E
\end{pmatrix}
\, ,
\end{equation}
where $\Sigma$ is now an Hermitian matrix and can be obtained from the conditions imposed in Eqs. \ref{eqapp:conditions}. The expectation value of $[\tilde X_A,\tilde X_E]^{\rm H} \Sigma  ^{-1} [\tilde X_A,\tilde X_E]$ over complex variable realizations $[\tilde X_A,\tilde X_E]$ is now 2. %, because the complex degrees of freedom are only 2. 
However, since the exponential does not have any factor of $1/2$ for a complex distribution, we recover the same number of degrees of freedom in the argument of the exponent as in the previous derivation, i.e. we got $\exp{[\frac{1}{2} \, 4]}$ for the case of multiple detectors with real and imaginary part as separate random variables' and $\exp{[2]}$ for the case considered here.\\

The derivation of the Fisher matrix differs from the previous one (Eq. \ref{fisher_2}) only by the factor $1/2$:
\begin{equation}
\Gamma_{ij}  = \sum_{k=1}^{n}  \left[ (\Sigma_k ^{-1})_{lr } \frac{\partial \Sigma_k ^{rp}}{\partial \lambda^i}( \Sigma_k ^{-1})_{pm}\frac{\partial \Sigma_k ^{ml}}{\partial \lambda^j}  \right],
\end{equation}
where the matrix $\Sigma_k$ is given by $\Sigma$ with spectral densities evaluated at given frequency $f_k$. \\
Note that we can recover the single channel realization by using the first element of $\Sigma$. \\ 
The continuum limit of the Fisher matrix in this formulation is given by
\begin{equation}
\Gamma_{ij}  =T \int_{0} ^\infty  (\Sigma_k ^{-1})_{lr } \frac{\partial \Sigma_k ^{rp}}{\partial \lambda^i}( \Sigma_k ^{-1})_{pm}\frac{\partial \Sigma_k ^{ml}}{\partial \lambda^j} \,\rm{d}f\, .\label{eq:fisher_mult}
\end{equation}

\subsection{Deterministic sources and noise cross-correlation}
\label{sec:detsrc_FM}
In the presence of a deterministic source, $\tilde{h}(f_k|\vec\mu)$, the derivative of the log-likelihood in Eq.~(\ref{eq:loglik}) with respect to the source parameters, $\vec\mu$, is:
\begin{equation}
\frac{\partial l}{\partial \mu^i} = - \sum_{k=1}^{n } \frac{|\tilde{s}(f_k) -  \tilde{h}(f_k|\vec\mu) | }{\frac{T}{4}S_n(f_k|\vec\lambda) }   \frac{\partial  \tilde{h}(f_k|\vec\mu)}{\partial \mu^i} \, , 
\end{equation}
and the derivative with respect to the parameters characterising the spectral density, $\vec\lambda$, is:
\begin{align}
\frac{\partial l}{\partial \lambda^i} =&  \sum_{k=1}^{n} [ -\frac{1}{2} \frac{T}{\det(S_n(f_k|\vec{\lambda}))}  + \nonumber \\  &+   \frac{|\tilde{s}(f_k) -  \tilde{h}(f_k|\vec\mu) |^2 }{\frac{T}{2}S_n(f_k|\vec\lambda)^2 }  ]\frac{\partial S_n(f_k|\vec{\lambda}) }{\partial \lambda^i}  \, .
\end{align}
The first of these expressions is odd in the noise component, $\tilde{n}(f_k) = \tilde{s}(f_k) - \tilde{h}(f_k|\vec\mu)$, while the second term is even. Since $\mathbb{E}[\tilde{n}(f_k)]=0$, from this we deduce that 
\begin{equation}
\mathbb{E}_{\cal L} \left[ \frac{\partial l}{\partial \mu^i} \frac{\partial l}{\partial \lambda^j} \right] = 0,
\end{equation}
i.e., at this level of approximation the terms in the Fisher matrix that mix signal and noise parameters vanish. We conclude that the estimation of the noise parameters and of the signal parameters is, at leading order, independent. Lack of knowledge of the noise should therefore not significantly affect measurements of the parameters of deterministic signals, except indirectly through the change in the spectral density that enters the likelihood for the deterministic sources.

\section{Impact of instrumental noise knowledge uncertainty on SGWB recovery in absence of galactic foreground}\label{sec:resuls no GB}
Below we show the same computations we did in Sec. \ref{sec:results} but in case we do not consider the presence of the foreground.
\subsubsection{Power law}
\begin{figure}
\includegraphics[width=0.45\textwidth]{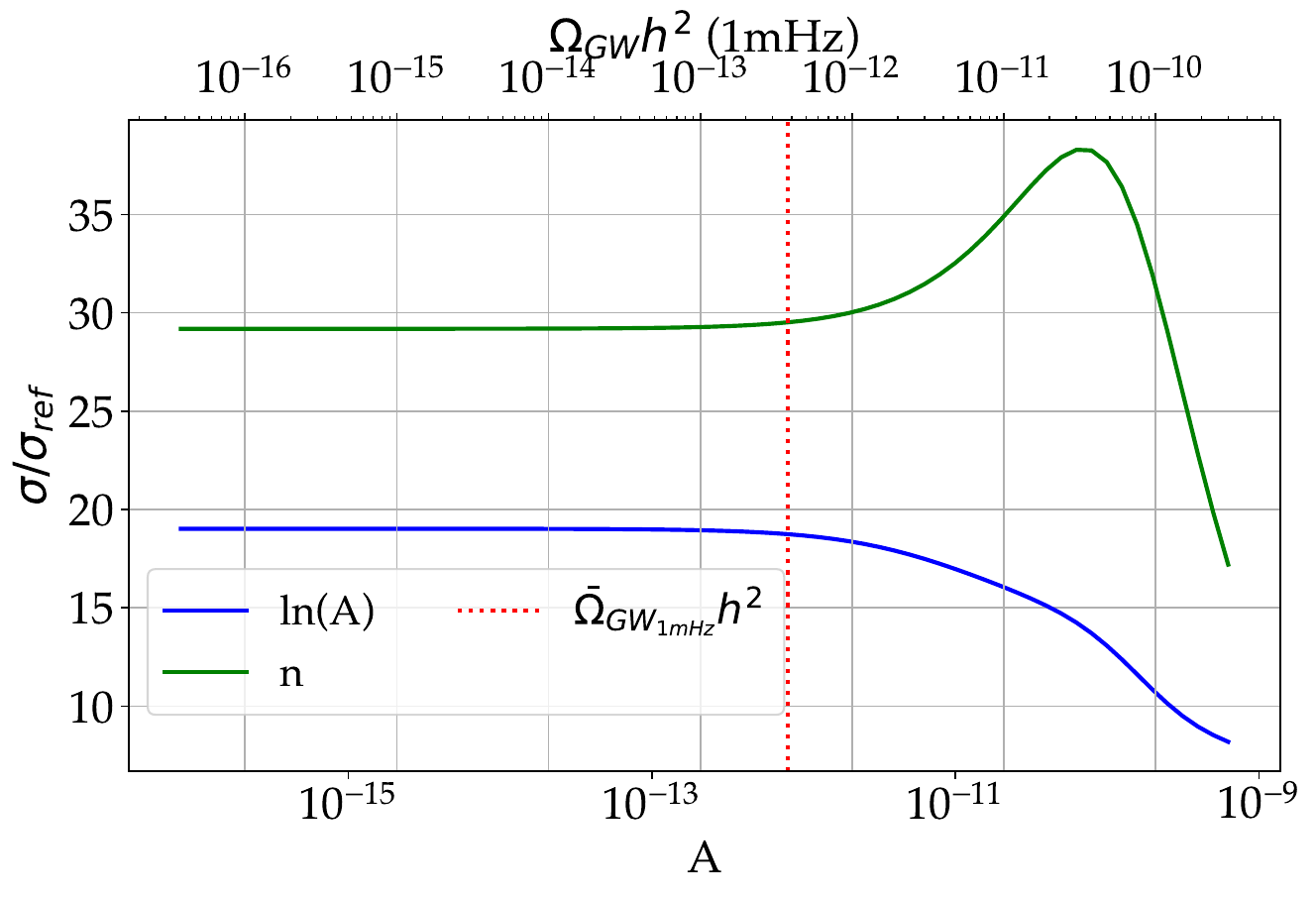}
\includegraphics[width=0.5\textwidth]{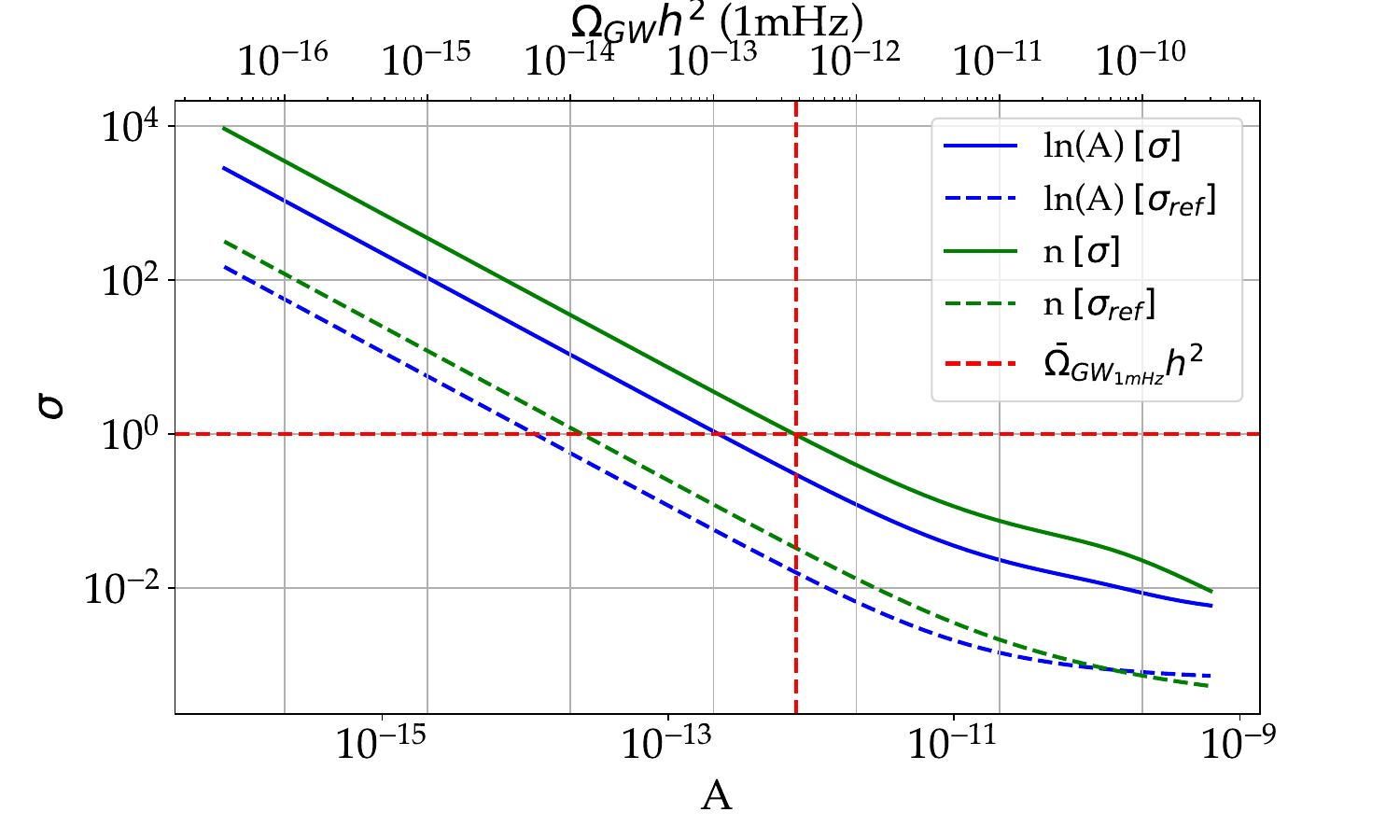}
\caption{\label{fig:power_law} Results for the power law SGWB model without foreground}
\end{figure}

Figure~\ref{fig:power_law} shows the results computed for the power law model. We see that in the presence of instrumental noise uncertainties, the uncertainty in the SGWB parameters increases by a factor of $\sim 19$--$36$, with the uncertainty in the slope being more affected than that of the amplitude.  Considering the raw uncertainties, to achieve the same measurement precision, the background energy density would have to be $\sim 33$ times larger than it would need to be in the absence of noise knowledge uncertainties. However, a background with amplitude equal to the reference value should be detectable even allowing for confusion with instrumental noise mis-modelling.

%%%%%%%%

%%%%%%
\subsubsection{Power law with running}
\begin{figure}
\includegraphics[width=0.45\textwidth]{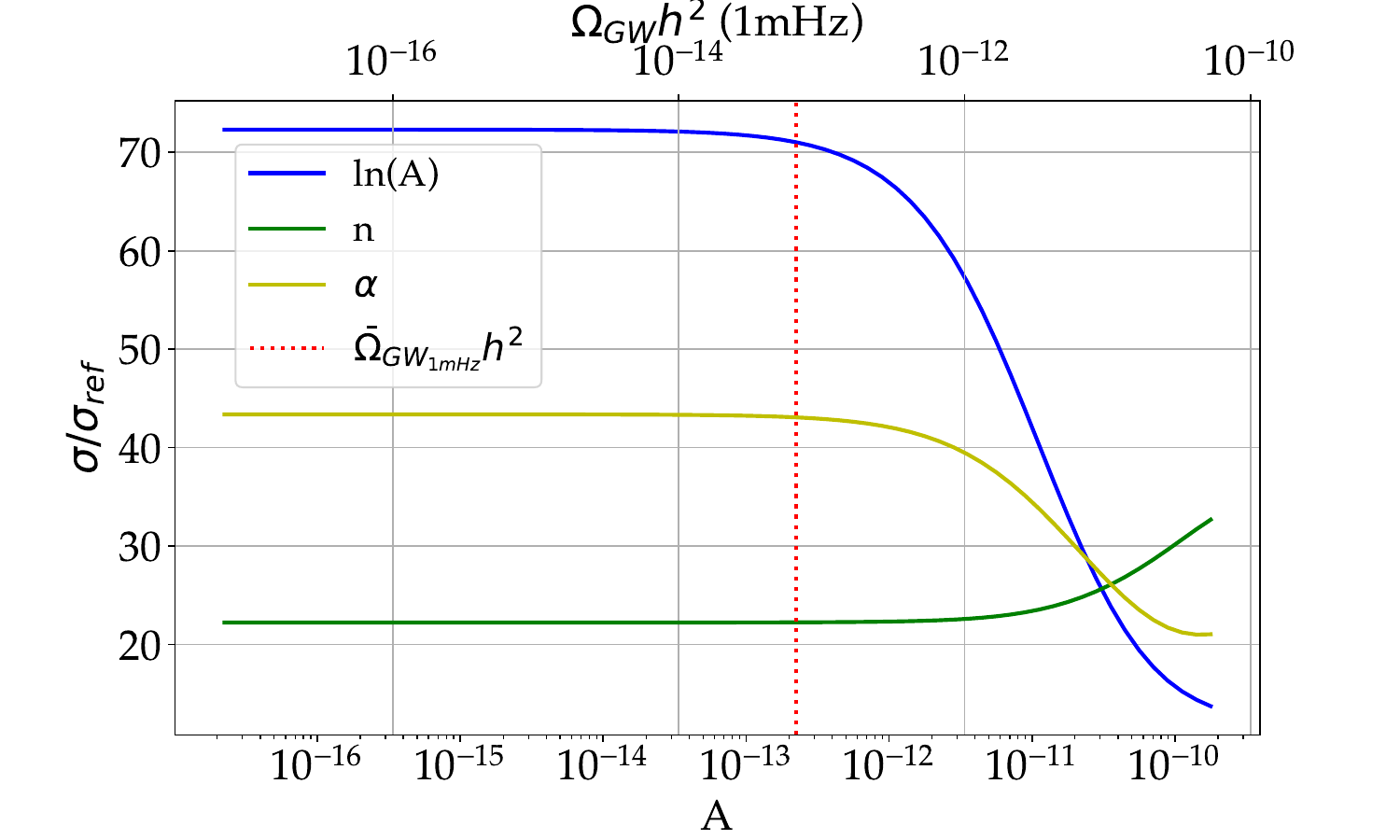}
\includegraphics[width=0.45\textwidth]{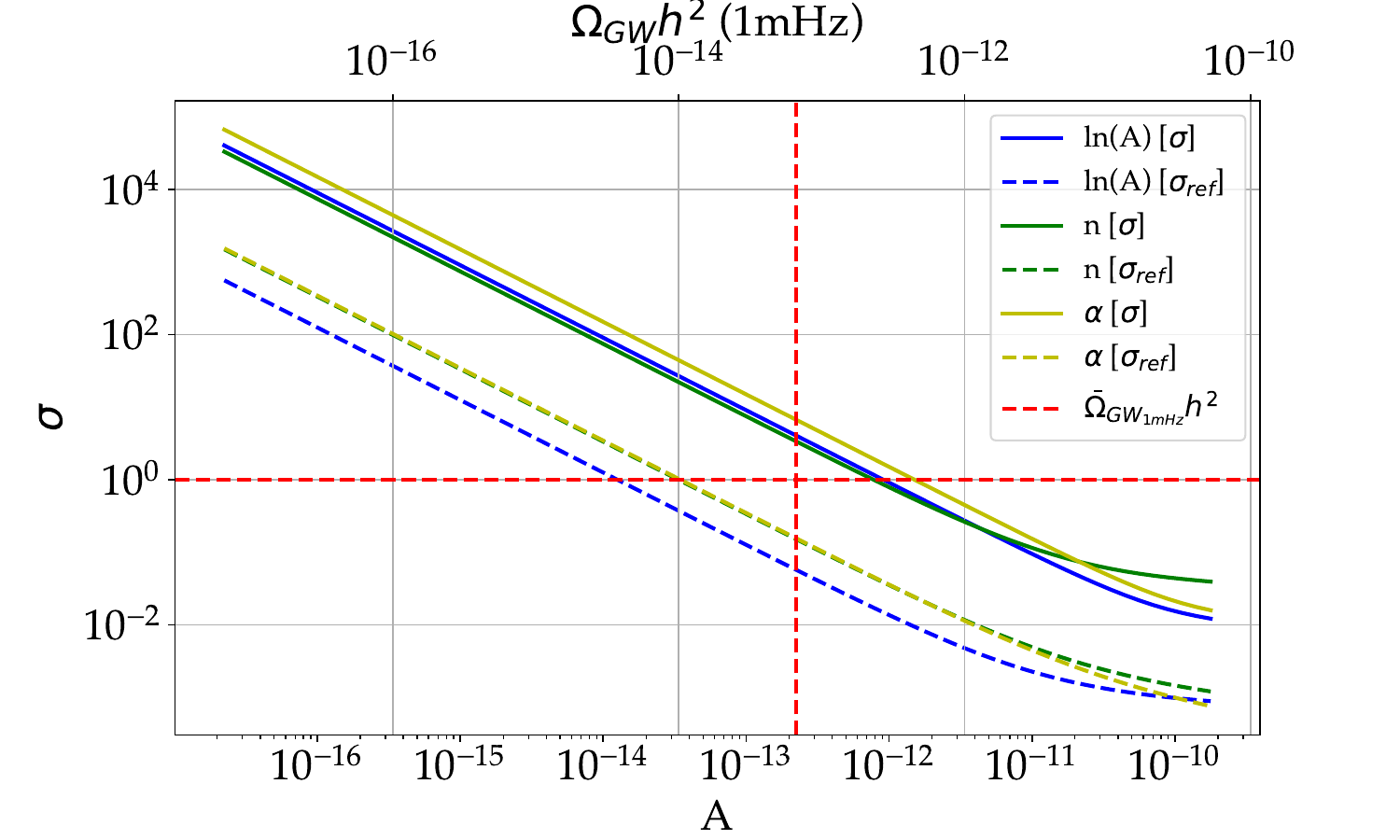}
\caption{\label{fig:power_law_wr} Results for the power-law with running SGWB model without foreground.}
\end{figure}
The results for this model are shown in Figure~\ref{fig:power_law_wr}. In this case we see that the uncertainties in the SGWB parameters increase by a factor of $\sim 21$--$72$, with the uncertainty on the log-amplitude being most affected in this case. Once again, the relative increase in the uncertainty is somewhat lower at higher background amplitudes. The lower panel of \cref{fig:power_law_wr} shows that the background is not detectable at the reference amplitude. An energy density $\sim 5$ times higher would be required for a detection. In general, the background again has to have an energy density $\sim 100$ times higher to be characterised with the same measurement precision when there is instrumental noise uncertainty as it could be without those uncertainties. 
\subsubsection{Gaussian bump}
\begin{figure}
\includegraphics[width=0.45\textwidth]{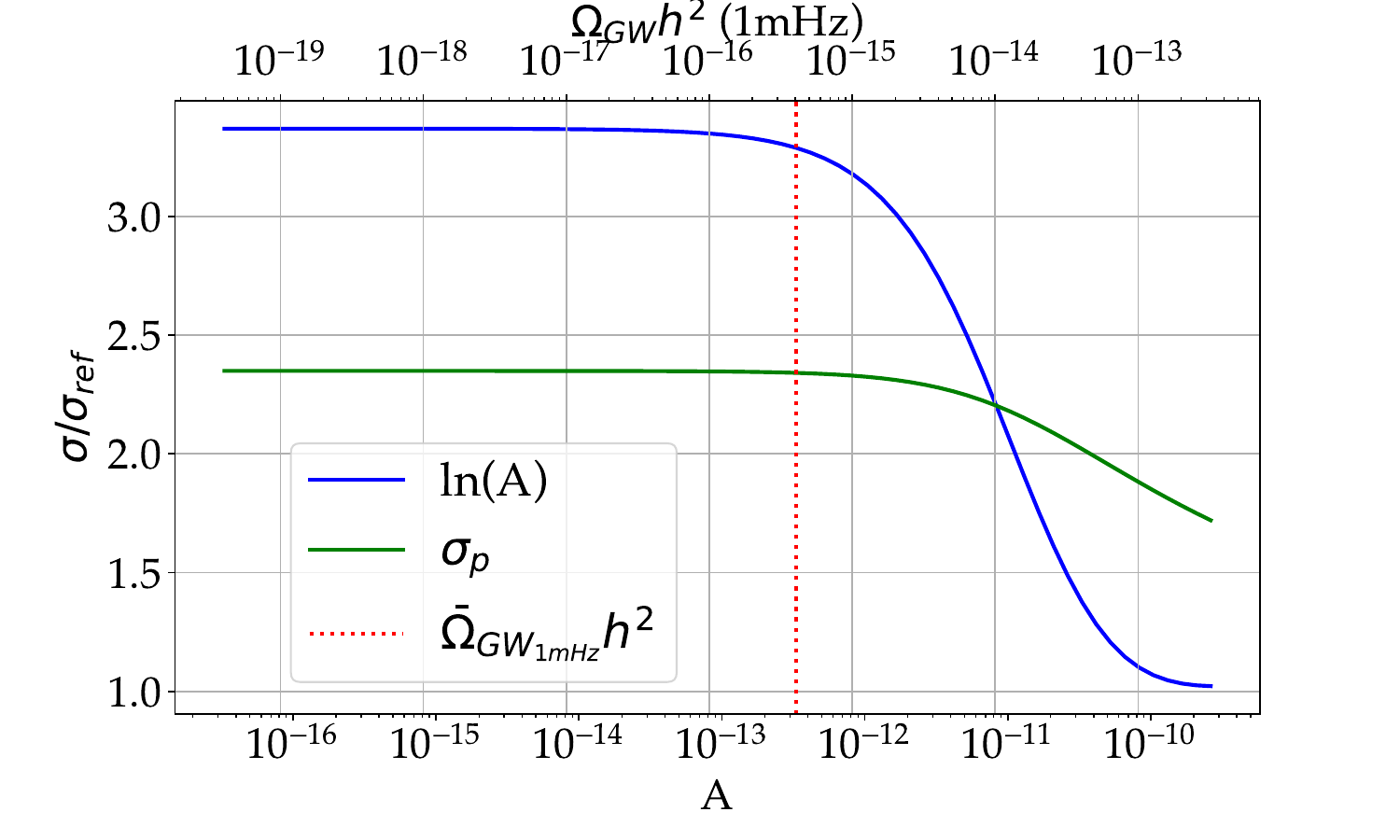}
\includegraphics[width=0.45\textwidth]{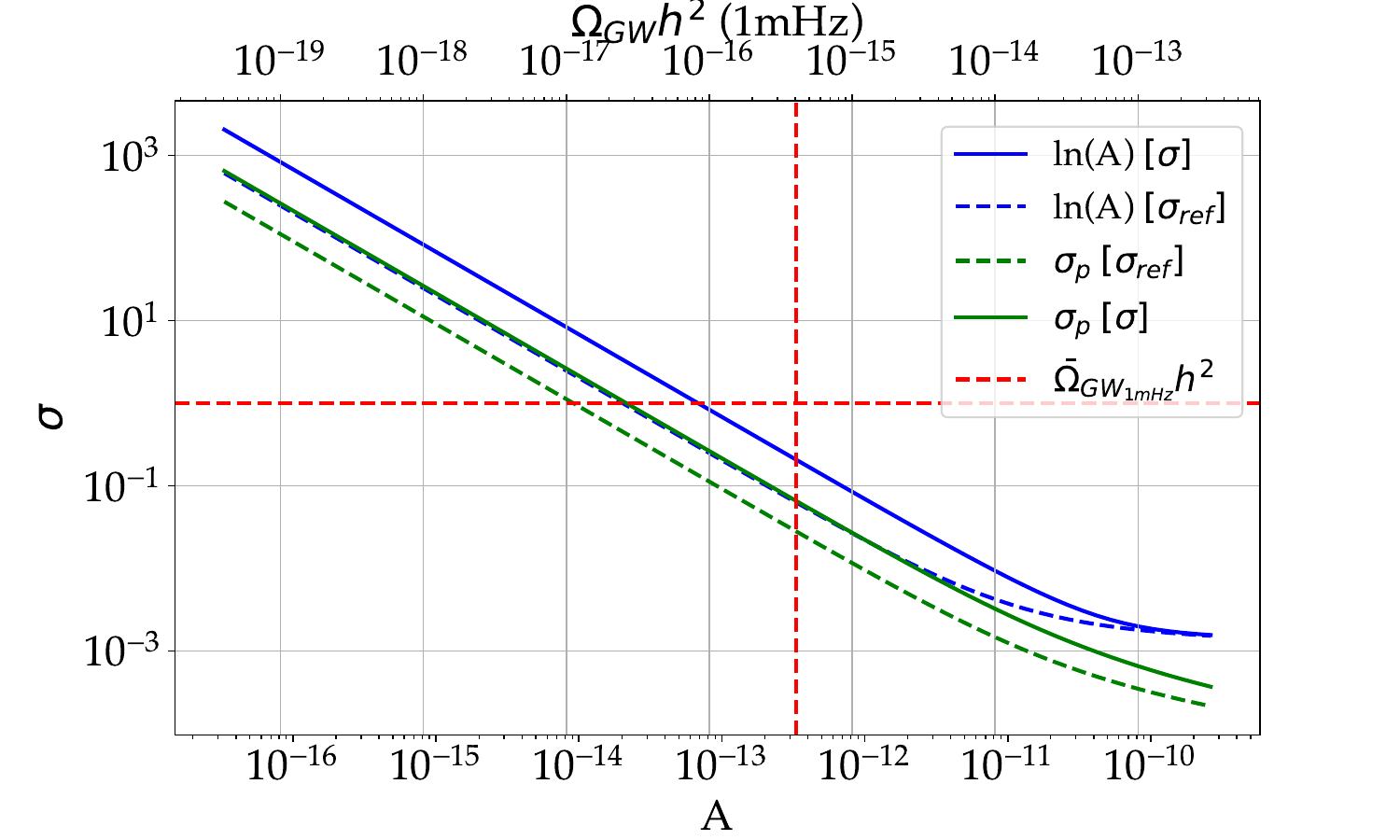}
\caption{\label{fig:gaussian} Results for Gaussian bump SGWB model without foreground
}
\end{figure}
The results for this model are shown in Figure~\ref{fig:gaussian}. In this case, the degradation in the precision of parameter measurement is a factor of $\sim2.3$--$4$ when allowing for lack of knowledge of the instrumental noise. From the lower panel of Figure~\ref{fig:gaussian} we see that the energy density in a Gaussian bump SGWB has to be just a small factor of $\sim 2.5$ times bigger to achieve the same measurement precision when the instrumental noise is not known perfectly. Moreover a Gaussian bump background at the reference amplitude can be measured to percent precision at the reference amplitude. The width of the Gaussian can be measured to a few tens of percent precision at the reference amplitude, improving approximately linearly with the background energy density.
%%%%%%%%%%%%%%%
\subsubsection{First order phase transition }
\begin{figure}
\includegraphics[width=0.45\textwidth]{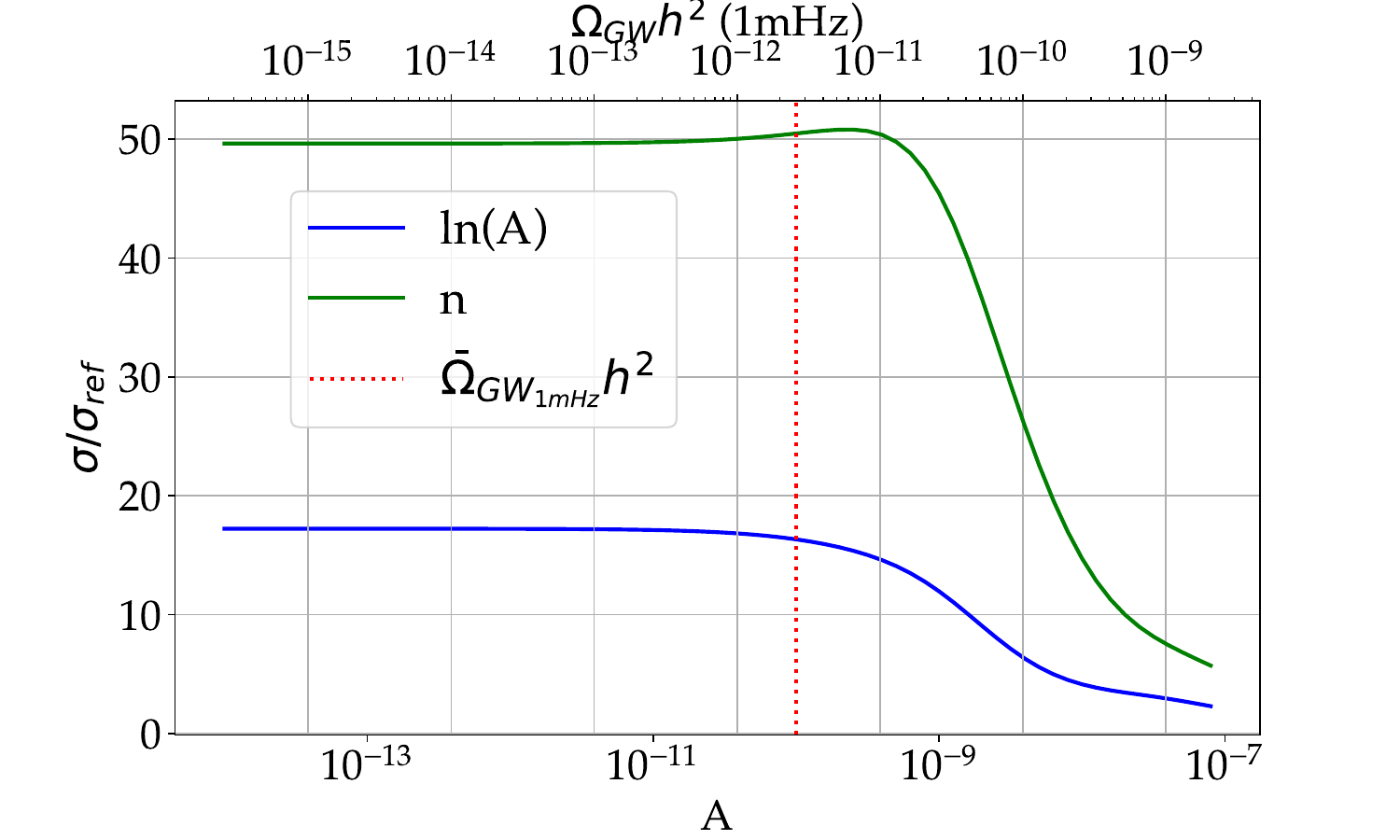}
\includegraphics[width=0.45\textwidth]{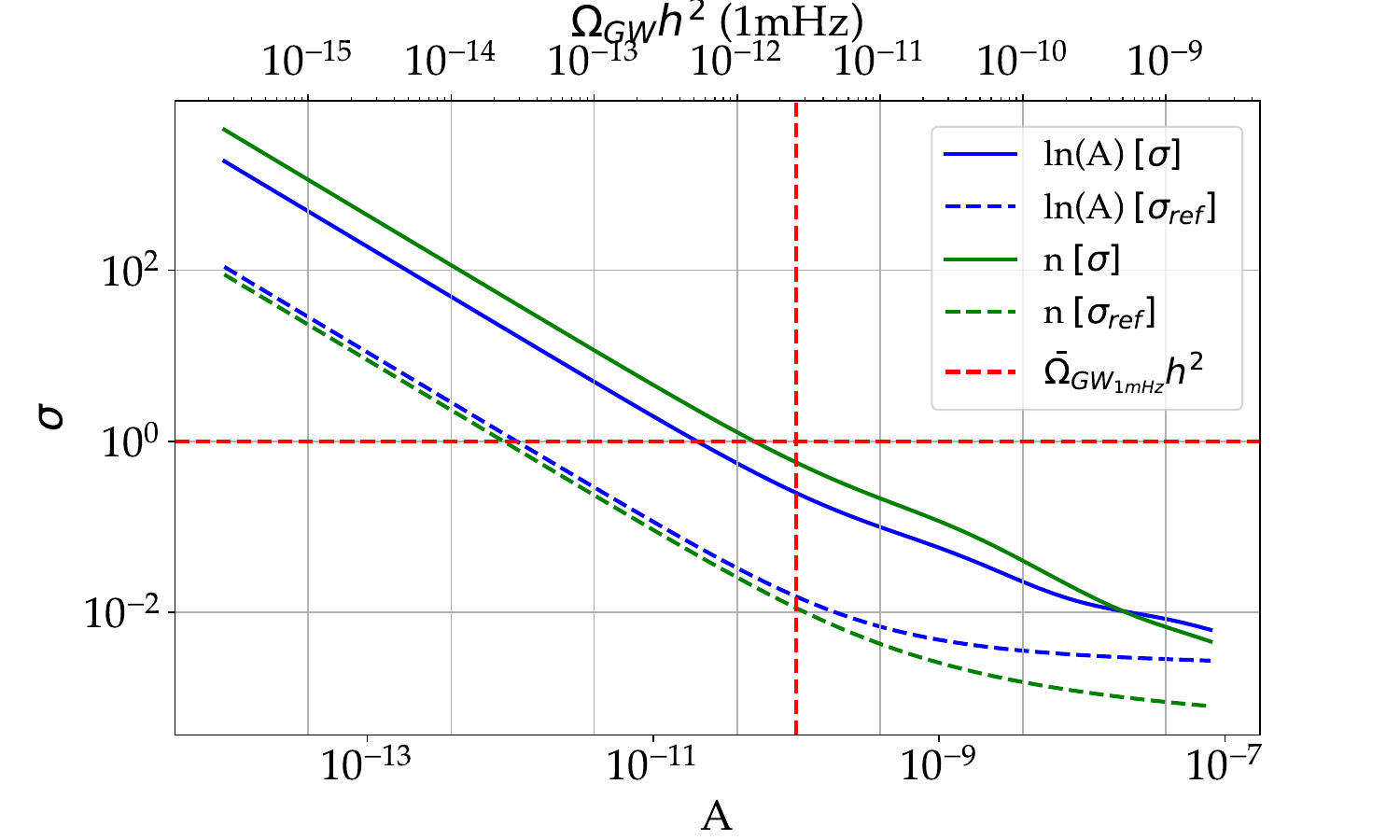}
\caption{\label{fig:FOPT} Results for first order phase transition SGWB model without foreground.}
\end{figure}
 The results for First order phase transition are shown in Figure~\ref{fig:FOPT}. The results for this SGWB model are quite similar to those for the power law background. When allowing for instrumental noise knowledge uncertainties, the precision with which the SGWB log-amplitude can be characterised degrades by a factor of $\sim 18$. The degradation in the determination of the spectral index is even larger, $\sim 50$. Once again, to achieve the same measurement precision, the background energy density would have to be $\sim 50$ times larger than it would need to be in the absence of noise knowledge uncertainties. Nonetheless, a FOPT background at the reference amplitude would still be detectable and provide a measurement of the spectral index at the level of about 0.5 percent.\\
 
\begin{figure}
\includegraphics[width=0.45\textwidth]{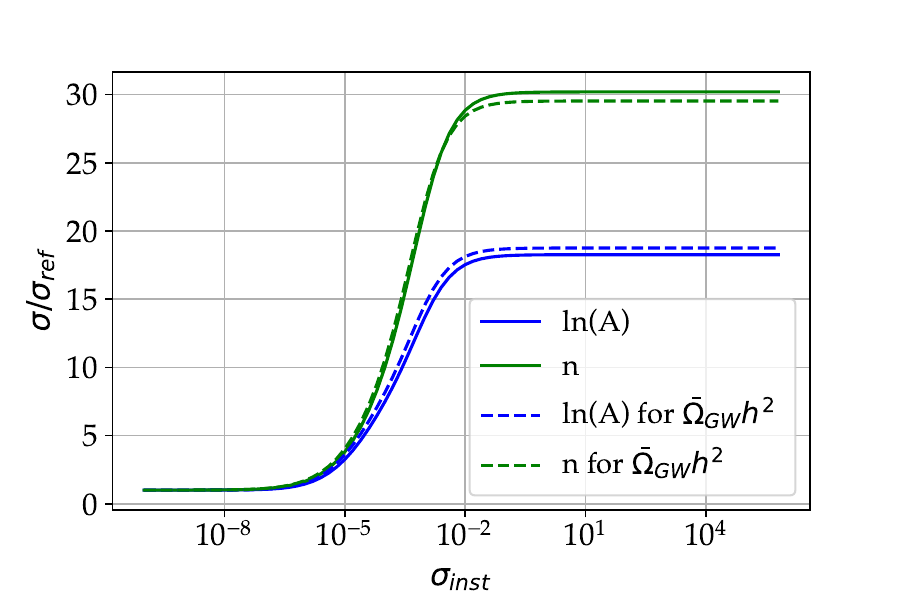}
\includegraphics[width=0.45\textwidth]{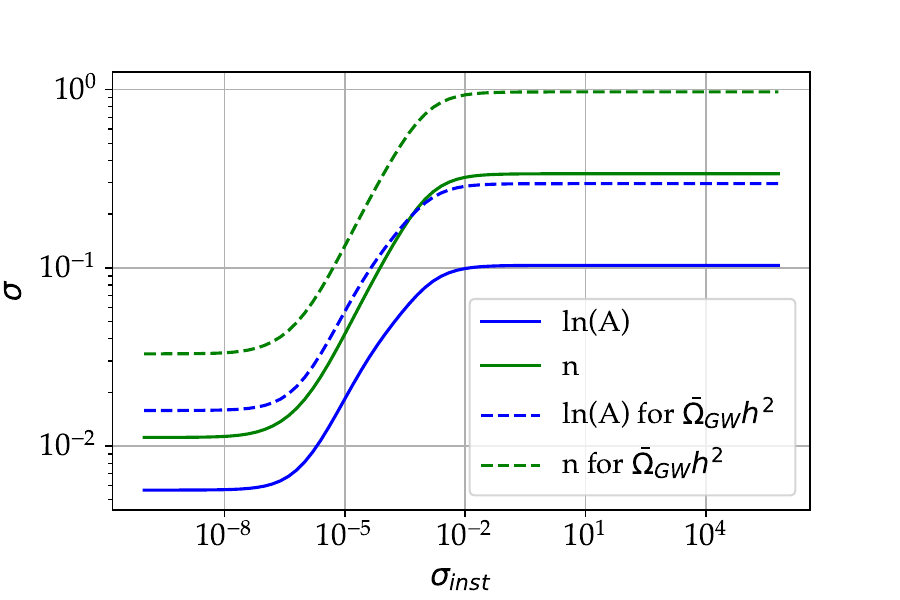}
\caption{\label{fig:PL_prior} As Figure~\ref{fig:power_law}, but now for fixed background amplitude and varying the variance of the Gaussian prior on the instrumental noise spline model. This plot is for a power law background, and the amplitude has been fixed such that the SNR in TDI channel A is 136 (continous lines) and 43 (dashed lines)
%Results for the power law model. Upper plot shows the ratio of the uncertainties of the amplitude and slope versus prior uncertainty in the instrumental noise. Lower plot shows the uncertainties of the amplitude and slope versus prior uncertainty in the instrumental noise.
}
\end{figure}

\begin{figure}
\includegraphics[width=0.45\textwidth]{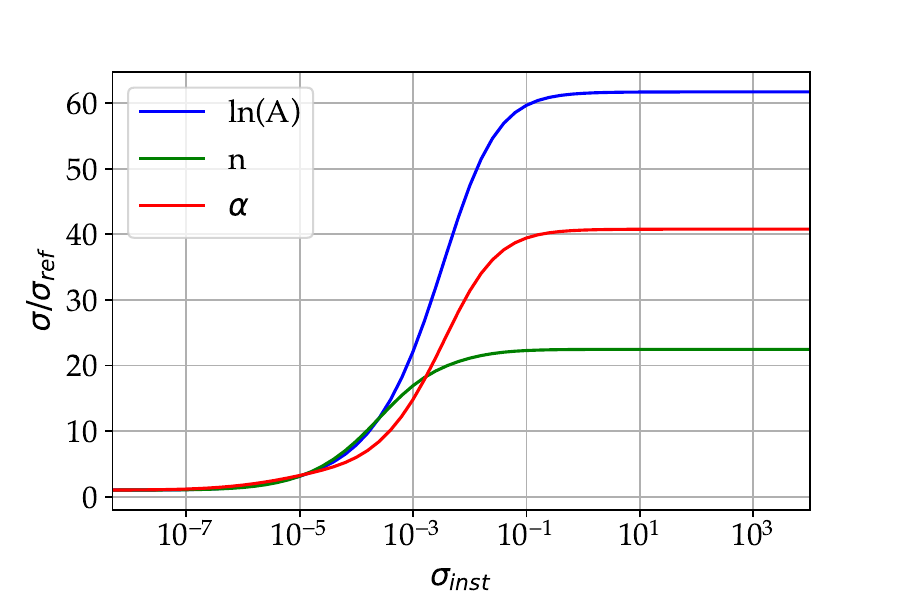}
\includegraphics[width=0.45\textwidth]{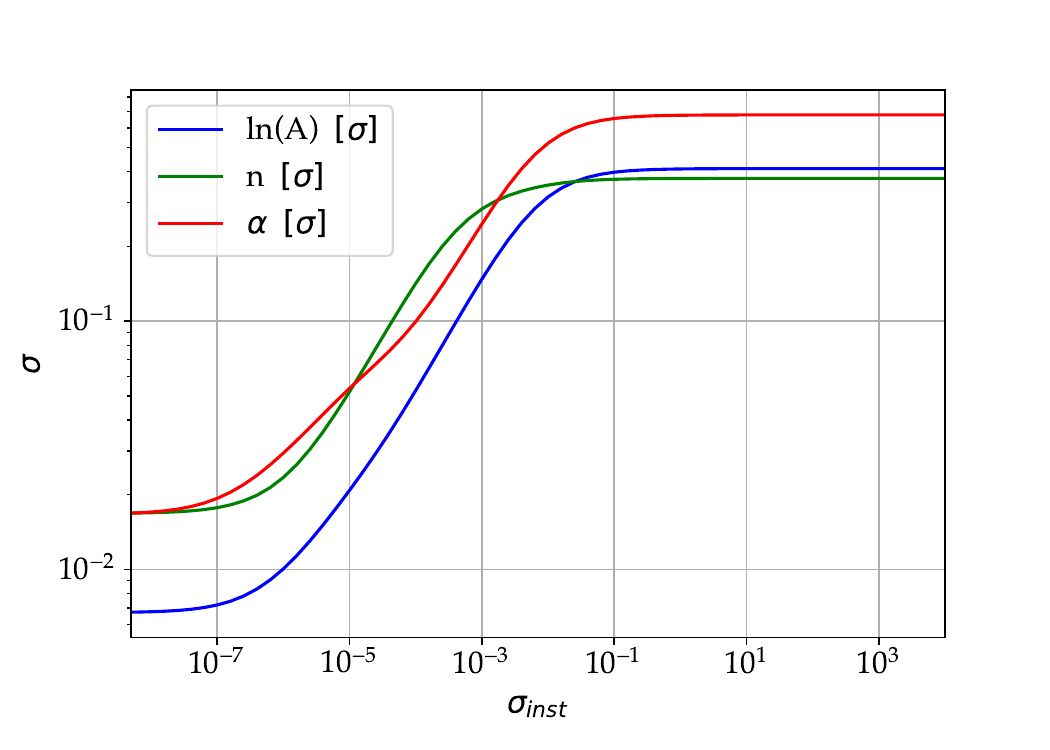}
\caption{\label{fig:PL_wr_prior} As Figure~\ref{fig:PL_prior} but now for the power-law-with-running model. The background amplitude has been fixed to give an overall SNR of $145$.
%Power law with running model.  Upper plot shows the ratio of the uncertainties of the amplitude, sloop and $\alpha$ versus prior uncertainty in the instrumental noise. Lower plot shows the uncertainties of the amplitude, sloop and $\alpha$ versus prior uncertainty in the instrumental noise
}
\end{figure}

\begin{figure}
\includegraphics[width=0.45\textwidth]{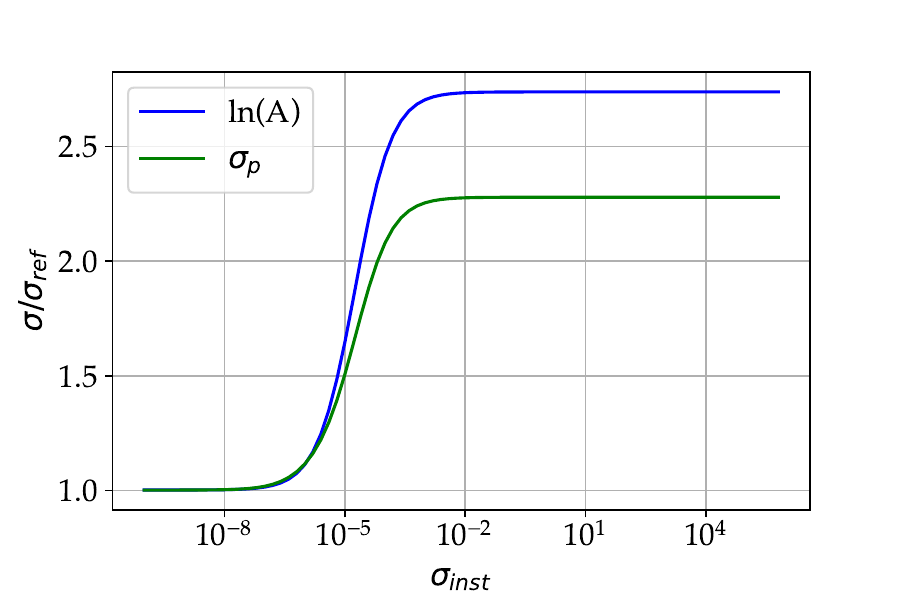}
\includegraphics[width=0.45\textwidth]{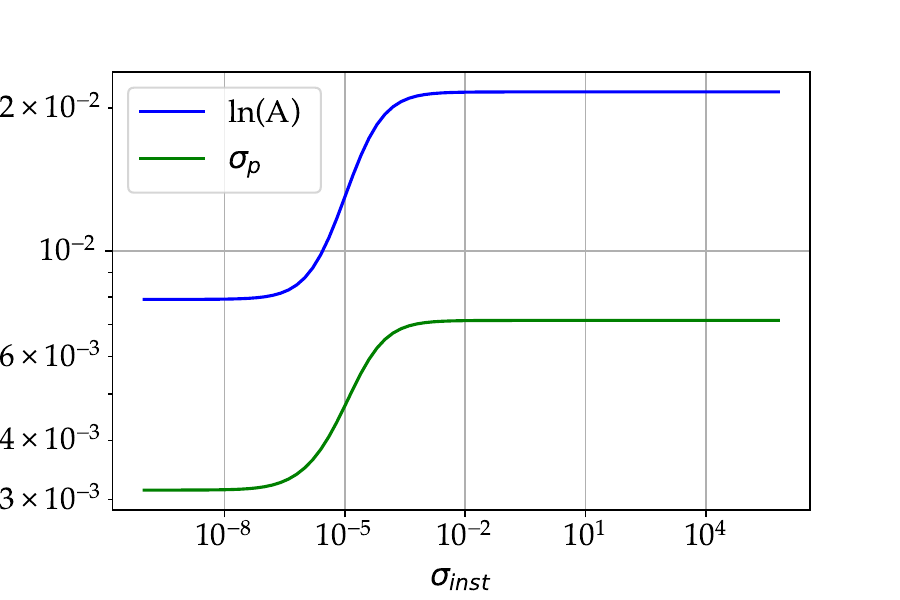}
\caption{\label{fig:gaussian_prior} As Figure~\ref{fig:PL_prior} but now for the Gaussian bump model. The background amplitude has been fixed to give an overall SNR of $135$.
%Gaussian bump model.  Upper plot shows the ratio of the uncertainties of the amplitude and Gaussian amplitude versus prior uncertainty in the instrumental noise. Lower plot shows the uncertainties of the amplitude and Gaussian amplitude versus prior uncertainty in the instrumental noise 
}
\end{figure}

\begin{figure}
\includegraphics[width=0.45\textwidth]{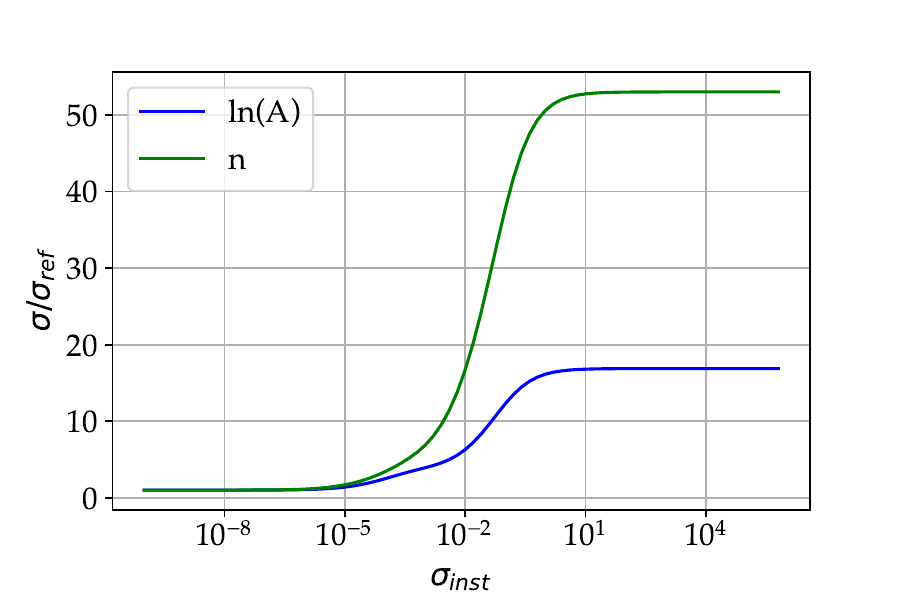}
\includegraphics[width=0.45\textwidth]{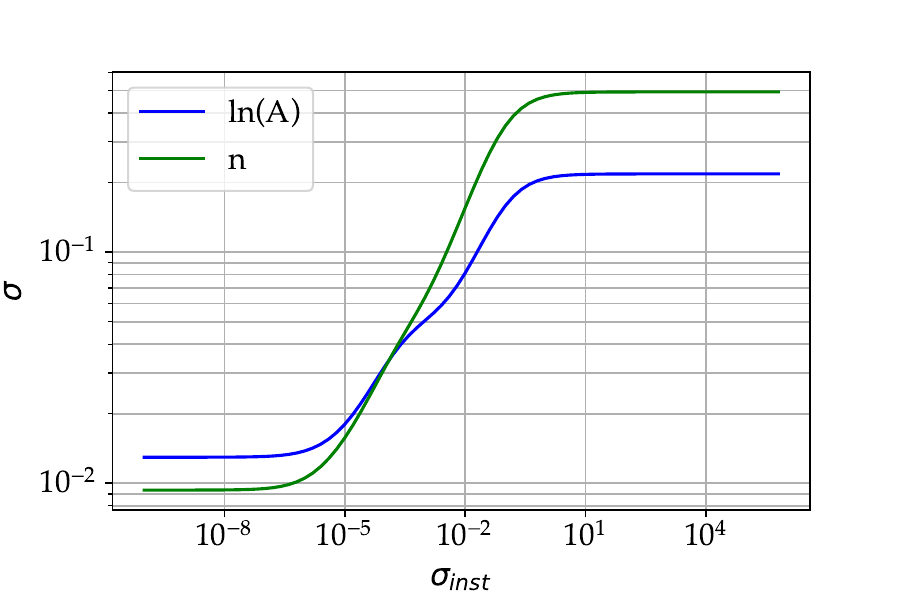}
\caption{\label{fig:FOPT_prior} As Figure~\cref{fig:PL_prior} but now for the FOPT model. The background amplitude has been fixed to give an overall SNR of $149$.
%FOPT law model. Upper plot shows the ratio of the uncertainties of the log amplitude and slope versus prior uncertainty in the instrumental noise. Lower plot shows the uncertainties of the log amplitude and slope versus prior uncertainty in the instrumental noise
}
\end{figure}

 We can do the same analysis we did in section \ref{sec:priordep} but without including the foreground. The results for all four SGWB models are qualitatively similar among themselves and also to the previous case with foreground in Sec. \ref{sec:priordep}. Figure \ref{fig:PL_prior} consider the power law case, Fig. \ref{fig:PL_wr_prior} the case of power law with running, Fig. \ref{fig:gaussian_prior} the case of gaussian model and  Fig. \ref{fig:FOPT_prior} the case of FOPT model.

%are included on the signal's parameters as a function of the magnitude of the allowed prior uncertainty in the instrumental noise, and the second one shows again the uncertainties on the signal's parameters as a function of the magnitude of the allowed prior uncertainty in the instrumental noise. 
%%%%%%%%%%%%%%%%%%%%%%%%%%

%For very low prior uncertainties the ratio of the uncertainties tends to unity. This is expected as this limit corresponds to the limit in which the instrumental noise is perfectly known. As the prior uncertainty is increased beyond $\sim 10^{-6}$ the measurement precision in the presence of noise knowledge uncertainties starts to increase. When the noise knowledge uncertainty reaches $\sim 10^{-2}$ (or $\sim10^{-1}$ for the FOPT model), the measurement precision ratio saturates. This final value reflects the expected uncertainty in the absence of any noise knowledge. The results given in Section~\ref{sec:ampdep} were all computed in this regime.

The main conclusion from these results is that again if we wanted to ensure that there was no degradation in LISA science due to lack of noise knowledge, the necessary requirement on the noise knowledge would be $\ll 10\%$. 

%Looking at \cref{fig:PL_prior} we see that when we have a good knowledge of our instrument thus $\sigma_{inst}$ is small we have smaller error in the estimation of the slope and amplitude. The errors increase with the increasing of uncertainties in the noise knowledge up to $\sigma_{inst} = 10^{-3}$ where we see it becomes constant to $\sigma \approx \times 10^{-1}$ for the slope and to $\sigma = 5 \times 10^{-1}$ for the log amplitude. This means that we need noise uncertainty accuracy smaller than $0.001\%$ to be able to estimate a power law signal with SNR of 136 with  $\sigma <  10^{-1}$ error in the amplitude. Also if we consider the reference energy density at $\sigma_{inst} > 10^{-3}$ we reach a constant error both for the amplitude and for the slope, but as expected given the law SNR we are able to estimate the power law signal amplitude with less accuracy.
%%%%%%%%%%%%
%%%%%%%%%%%
%Looking at \cref{fig:PL_wr_prior} we again see that when we have a good knowledge of our instrument thus $\sigma_{inst}$ is small we have smaller error in the estimation of the slope and log amplitude. The errors increase with the increasing of uncertainties in the noise knowledge up to $\sigma_{inst} = 10^{-2}$ where we see it becomes constant to $\sigma = 4 \times 10^{1}$ for the  log amplitude and  to $\sigma = 7 \times 10^{-1}$  for $\alpha$ , and up to $\sigma_{inst} = 10^{-3}$ for the slope where it becomes constant to  $4 \times 10^{2}$ . This means that we should know the noise with an accuracy better than $0.01\%$ to be able to estimate a power law with running signal with SNR of 145 with  $\sigma < 5 \times 10^{-1}$ error in the log amplitude.

%%%%%%%%%%%%
%%%%%%%%%%%%%
%Looking at \cref{fig:gaussian_prior} we again see that when we have a good knowledge of our instrument thus $\sigma_{inst}$ is small we have smaller error in the estimation of the Gaussian amplitude and log amplitude. The errors increase with the increasing of uncertainties in the noise knowledge up to $\sigma_{inst} = 10^{-3}$ where we see it becomes constant to $\sigma = 2 \times 10^{-2}$ for the amplitude and to $\sigma = 6 \times 10^{-3}$ for the $\sigma$. This means that we should know the noise with an accuracy better than $0.001\%$ to be able to estimate a Gaussian bump signal with SNR of 135 with  $\sigma < 2 \times 10^{-2}$ error in the log amplitude.

\subsection{Signal reconstruction without foreground of Galactic binaries}\label{app:signa_rec}
\begin{figure}
\centering
\includegraphics[width=0.4\textwidth]{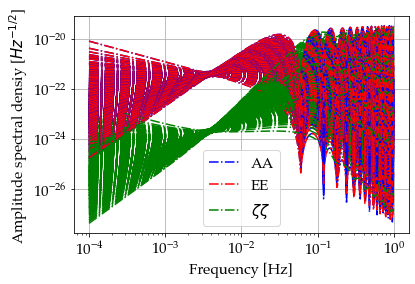}
\includegraphics[width=0.4\textwidth]{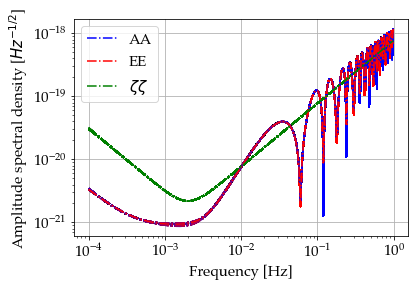}
\includegraphics[width=0.4\textwidth]{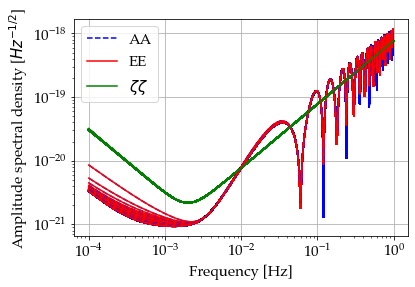}
\caption{\label{fig:rec} Signal and noise reconstruction for a power law SGWB with SNR 48.70. We show the signal and noise PSDs corresponding to random draws from the posterior, approximated using the Fisher matrix as described in the text. In each panel the curves correspond to the three TDI channels: A (blue), E (red) and $\zeta$ (green). Upper panel: reconstructed SGWB; middle panel: reconstructed instrumental noise; lower panel: total reconstructed ASD (signal + noise).}
\end{figure}

\begin{figure}
\centering
\includegraphics[width=0.4\textwidth]{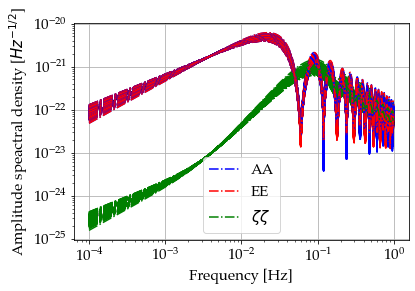}
\includegraphics[width=0.4\textwidth]{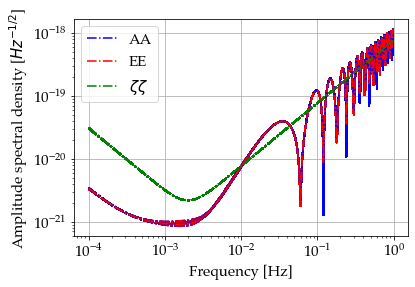}
\includegraphics[width=0.4\textwidth]{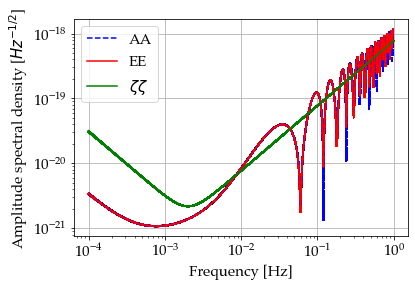}
\caption{\label{fig:rec2} As Figure~\ref{fig:rec}, but for a power law background with higher SNR, of 862. }
\end{figure}

We consider a power law signal with an SNR of 48.70 in Fig. \ref{fig:rec}. The three panels show the reconstructed ASDs for the SGWB and for the instrumental noise and the total, which is the sum of the three. No foreground has been considered in this case. In Fig \ref{fig:rec2} we show corresponding results for a power law with a higher SNR, of 862.
%%%%%%%%%%%%

We see that our ability to reconstruct the signal component of the data stream is poor when the SNR is low. However, we are able to obtain good measurements of the instrumental noise and the total spectral density. We note that the total ASD reconstruction in Figure~\ref{fig:rec} is somewhat poorer than the noise-only component, which does not fit with the expectation that we are actually measuring the total. This happens due to the breakdown in the Fisher matrix approximation for the SGWB parameters in this case, because the SGWB parameter uncertainties are large and no longer in the linear signal regime. 
At higher SNR, we start to be able to reconstruct the SGWB more precisely, shown by a reduction in the scatter in Figure~\ref{fig:rec2}. As the SNR is increased we would expect the scatter to reduce further. The reconstruction of the noise spectral density is comparable to what is seen in the lower SNR case, but we would eventually expect it to degrade as the SGWB becomes more dominant in the data. The reconstruction of the total spectral density is similar to the low SNR case, as expected. However, this higher SNR case does not show the noise at low frequency that arises from the breakdown of the Fisher matrix approximation, presumably because the measurement uncertainties are within the linear regime in this case. although, what is interesting to notice in the total reconstruction is that the channels A and E are affected by the SGWB where between $1mHz$ and $4mHz$ the noise level deviated from the expected one shown in Fig \ref{fig:noises}.

%\begin{align}
% S_{EA}(\omega)   \approx  & 16 \left(\delta _5-\sqrt{3} \delta _6\right) T \omega  \sin ^3\left(\frac{T \omega }{2}\right) \cos \left(\frac{T \omega }{2}\right)  \nonumber   \\ & \left(2 S_g^{disp} (2 \cos (T
%   \omega )+\cos (2 T \omega )+3)+S_{\text{oms}}\right)
%\end{align}
%\begin{align}
% S_{\zeta A}(\omega)   \approx  & \sqrt{2} \left(\sqrt{3} \delta _5-3 \delta _6\right) T \omega  e^{-\frac{3}{2} i T \omega } \sin \left(\frac{T \omega }{2}\right)   \\ &  \left(S_g^{disp} \left(4 e^{i T
%   \omega }+3 e^{-2 i T \omega }+e^{2 i T \omega }+4\right) \nonumber  \right.  \\ & \left.  +S_{\text{oms}} (1+2 i \sin (T \omega ))\right)\nonumber
 %  \end{align}
%\begin{align}
% S_{E\zeta}(\omega)   \approx  & \sqrt{6} \left(\sqrt{3} \delta _5+\delta _6\right) T \omega  e^{-\frac{3}{2} i T \omega } \sin \left(\frac{T \omega }{2}\right)    \\ & % \left(S_g^{disp} \left(4 e^{i T
%   \omega }+3 e^{-2 i T \omega }+e^{2 i T \omega }+4\right)  \nonumber  \right.  \\ & %\left. +S_{\text{oms}} (1+2 i \sin (T \omega ))\right)\nonumber
%\end{align}
\newpage
\bibliographystyle{plain}
\bibliography{bibliography.bib}

\end{document}